\documentclass[11pt]{article}
\usepackage{hyperref}
\usepackage{lscape,longtable}
\usepackage{multirow}
\usepackage{tabularx} 
\usepackage{amsmath}  
\usepackage{amssymb}
\usepackage{comment}
\usepackage{float}
\usepackage{listings}
\usepackage{xcolor}
\usepackage{eso-pic}
\usepackage{enumitem}
\usepackage{booktabs}
\usepackage{pdfpages}
\usepackage{authblk}
\let\address\affil
\usepackage[export]{adjustbox}
\usepackage{hyperref}
\usepackage{siunitx}
\usepackage{epsfig}
\usepackage{graphicx}
\usepackage{caption}
\usepackage{subcaption}
\usepackage[margin=1in,letterpaper]{geometry} 
\usepackage{cleveref}
\usepackage{blindtext}
\usepackage{epstopdf}
\usepackage{afterpage}
\usepackage{fancyvrb}
\usepackage{fancyhdr}
\usepackage{chngcntr}
\usepackage{xfrac}
\usepackage{lineno}
\usepackage{tikz}
\usepackage{bm}
\usepackage[numbers,sort&compress]{natbib}
\usepackage{tikz}
\usetikzlibrary{decorations.pathmorphing,decorations.markings,calc,arrows.meta}
\usepackage{array}
\newcolumntype{L}[1]{>{\raggedright\arraybackslash}m{#1}}
\usepackage[utf8]{inputenc}
\usepackage[T1]{fontenc}

\usepackage[protrusion=true,expansion=false]{microtype}
\usepackage{parskip}
\usepackage{xcolor}
\usepackage{array}
\usepackage{enumitem}
\usepackage{tcolorbox}
\hypersetup{bookmarksdepth=3}
\tcbuselibrary{skins}
\begin{document}
%
\begingroup
  \thispagestyle{empty}
  \AddToShipoutPictureBG*{%
    \put(0,0){%
      \includegraphics[width=\paperwidth,height=\paperheight]%
                      {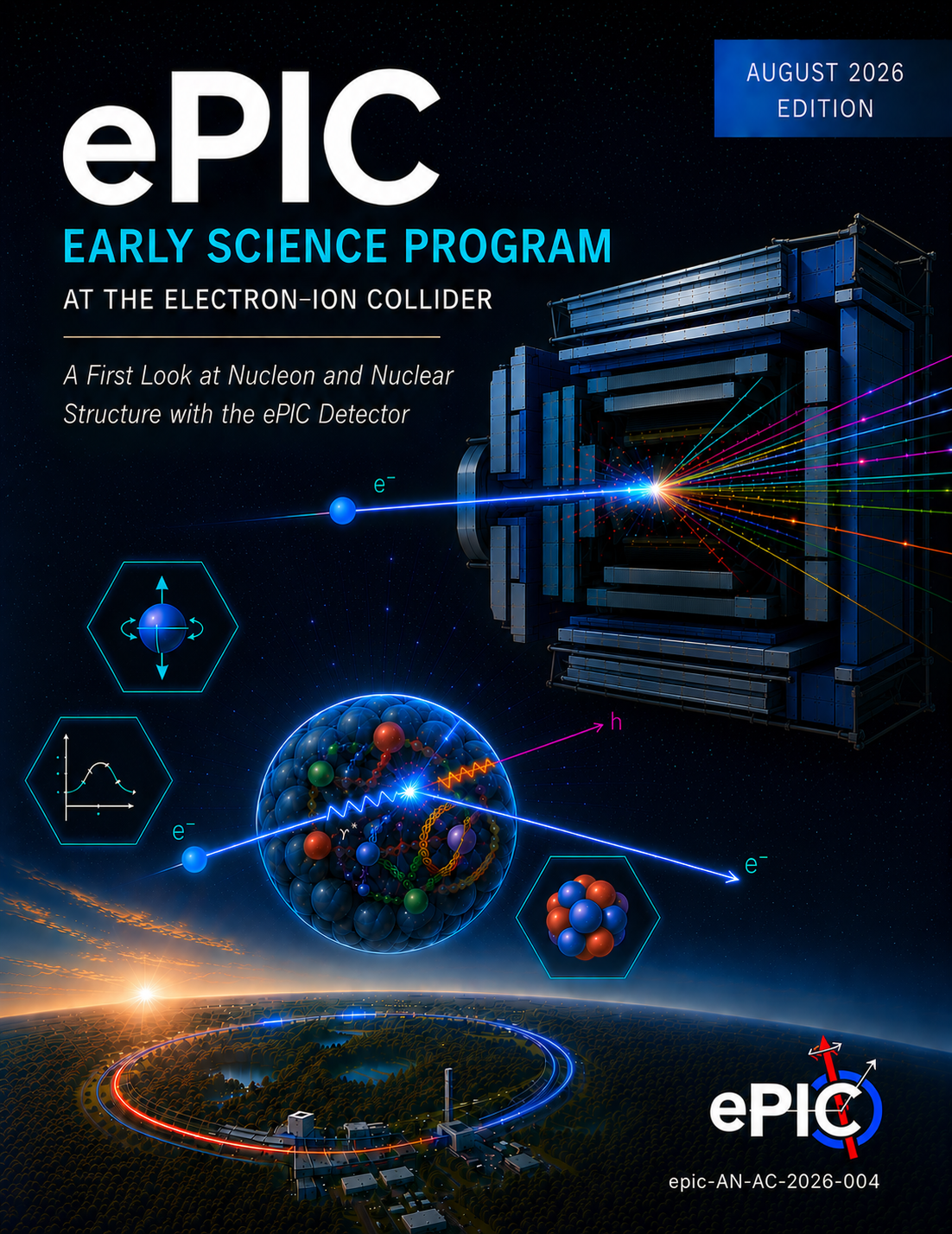}%
    }%
  }%
  \null            
  \clearpage       
\endgroup
\pagestyle{empty}
\vspace*{0.1cm}
\begin{center}
{\bf\Large epic-AN-AC-2026-004 } \\
\vspace*{1.0cm}
{\bf\Huge ePIC Early Science Report} \\
\vspace*{0.3cm}
\end{center}
\vspace*{0.2cm}
\begin{center}
\end{center}
\vspace*{1cm}
\begin{abstract}
This Early Science Report from the ePIC Collaboration outlines the compelling physics program achievable during the first years of operation of the Electron-Ion Collider (EIC), prior to the establishment of the full design luminosity and energy range. The analyses are based on realistic early-running beam configurations and detailed Geant4 ePIC detector simulations, hit digitization and data reconstruction.
The projected studies from the physics working groups of ePIC span inclusive, semi-inclusive, exclusive, diffractive and tagging, as well as jet and heavy flavor measurements in both electron\,–\,proton and electron\,–\,ion collisions. Even before the collider reaches its full design performance, these measurements will constrain parton distribution functions in nucleons and nuclei, access transverse-momentum-dependent and spin-dependent observables, probe gluon dynamics in nuclei, and initiate a program of imaging of quarks and gluons.
Each measurement is directly connected to the core science pillars of the EIC, identified in the 2018 report by the National Academy of Sciences: understanding the origin of the nucleon mass, unraveling the spin structure of the nucleon, and exploring the emergent properties of dense gluonic matter. The results presented here provide examples that demonstrate that the early years of EIC running with ePIC will deliver novel world-leading insights into Quantum Chromodynamics. In addition, the early science program will establish measurement and analysis methodologies that will pave the way to the subsequent full EIC physics program.
\end{abstract}
\vfill
\begin{center}
{\includegraphics[width=0.4\linewidth]{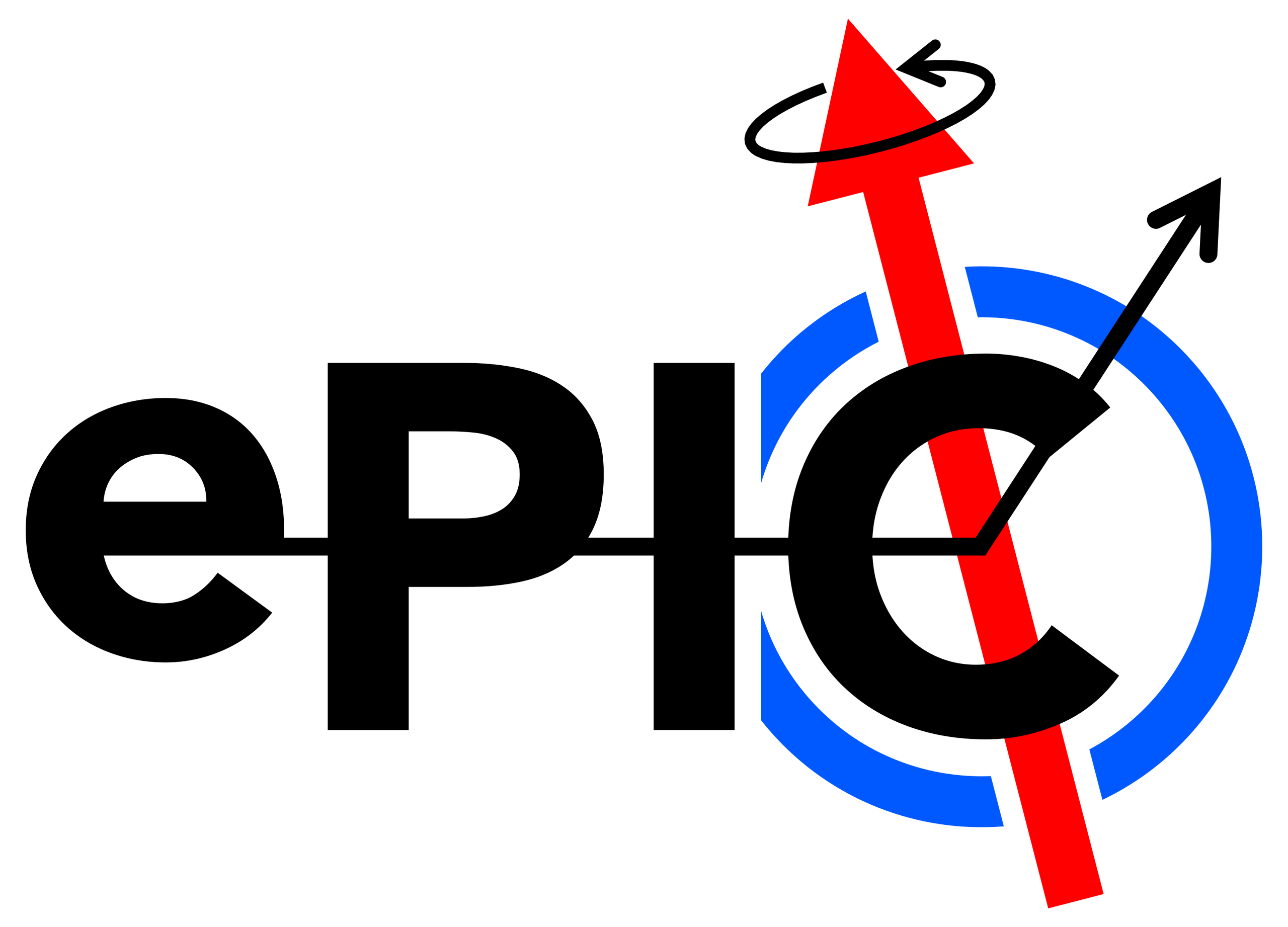}}
\end{center}
\vfill
\pagebreak
\thispagestyle{empty}

\clearpage
\begingroup
\renewcommand\Authfont{\small}
\renewcommand\Affilfont{\footnotesize}

\begin{center}
{\bfseries\Large The ePIC Collaboration}
\end{center}
\vspace*{0.4cm}


\begin{center}
\makeatletter
\Authfont \AB@authlist
\makeatother
\end{center}

\vspace*{0.6cm}
\begin{center}
{\bfseries\Large and the following external collaborators}
\end{center}
\vspace*{0.2cm}

\makeatletter
\gdef\AB@authlist{}
\setcounter{authors}{0}
\expandafter\gdef\csname @sep1\endcsname{}%
\expandafter\gdef\csname @sep2\endcsname{\Authand}%
\makeatother

%

\author[1,161]{A.~Accardi}
\author[162]{I.~Borsa}
\author[42]{M.~Cerutti}
\author[31]{C.~Cocuzza}
\author[155]{J.~Jim\'{e}nez-L\'{o}pez}
\author[163]{K.~Kumericki}
\author[164]{R.~Martinez}
\author[165]{L.~Rossi}
\author[166]{P.~Sznajder}
\author[167]{A.~Vladimirov}
\author[168]{K.~Wichmann}
\author[169] {S.~Yadav}
\author[167]{P.~Zurita}

\address[161]{School of Engineering and Computing, Christopher Newport University, Newport News, VA 23606, USA}
\address[162]{Institute for Theoretical Physics, University of T\"ubingen, 72076 T\"ubingen, Germany}
\address[163]{Department of Physics, Faculty of Science, University of Zagreb, HR-10000 Zagreb, Croatia}
\address[164]{Universidad de Buenos Aires, FCEyN, Departamento de F\'isica and IFIBA-CONICET, Argentina}
\address[165]{Dipartimento di Fisica, Universit\`a di Milano and INFN - Sezione di Milano, 20133 Milano, Italy}
\address[166]{National Centre for Nuclear Research (NCBJ), Pasteura 7, 02-093 Warsaw, Poland}
\address[167]{Departamento de F\'isica Te\'orica \& IPARCOS, Universidad Complutense de Madrid, E-28040 Madrid, Spain}
\address[168]{Deutsches Elektronen-Synchrotron DESY, Germany}
\address[169]{Dr. Hari Singh Gour University, Sagar, Madhya Pradesh, India}

\begin{center}
\makeatletter
\Authfont \AB@authlist
\makeatother
\end{center}

\vspace*{0.6cm}

\begin{center}
\makeatletter
\enlargethispage{-1\baselineskip}
\Affilfont \AB@affillist
\makeatother
\end{center}

\endgroup

\thispagestyle{empty}
\pagenumbering{roman}
\pagestyle{plain}
\nolinenumbers
\begingroup
\setlength{\parskip}{0pt}
\newpage
\setcounter{tocdepth}{4} 
\tableofcontents
\enlargethispage{1.8\baselineskip}
\thispagestyle{empty}
\endgroup
\cleardoublepage
\cleardoublepage
%
\cleardoublepage
%
%
%

\section*{Executive Summary}
\addcontentsline{toc}{section}{Executive summary}
\setcounter{subsection}{0}
\renewcommand{\thesubsection}{\Alph{subsection}}
\pagenumbering{roman}
\label{sec:executive-summary}

The Electron-Ion Collider (EIC) will begin producing world-leading QCD measurements from its very first years of operation, well before reaching its full design capabilities. This next major facility for studying the strong interaction will be realized at Brookhaven National Laboratory (BNL) in partnership with Thomas Jefferson National Accelerator Facility (JLab).
Its general-purpose detector, ePIC, will be operational from the start of EIC data taking. As is typical for a new collider, the EIC will approach its full design luminosity, beam energies, and polarization configurations over a multi-year period.
This early science report was prepared by the ePIC Collaboration in response to a charge letter from the Associate Lab Directors of BNL and JLab, reported in the \hyperref[app:charge-letter]{Appendix}. It demonstrates that, even during these first years, ePIC will deliver world-leading measurements that address the three central science questions identified by the 2018 National Academy of Sciences (NAS) assessment of the EIC science case: the origin of the nucleon mass, the origin of the nucleon spin, and the emergent properties of dense systems of gluons.
Based on realistic early running beam configurations and full ePIC detector Geant4 simulations, hit digitization and data reconstruction, the report shows that impactful, globally competitive measurements will be possible from day one of operations. The analysis methods developed during this phase will provide an essential foundation for the full EIC physics program that follows.

\subsection*{Early Running Conditions}

The early years of EIC operations will use a 9~GeV electron beam colliding with a sequence of hadron and ion beams. 
The start of operations will prioritize technical stability with unpolarized medium-mass ions (e.g., $e+\mathrm{Ag}$ at 9 GeV on 115 GeV). Following this, longitudinally polarized electron beams and transversely polarized hadron beams will be available. Next, hadron spin rotators will facilitate the transition to longitudinal proton polarization. This stage will also include progressive commissioning of the far-forward detectors, including the Roman Pots, which are needed for proton and spectator tagging. Beam species across the early program include polarized protons (at both 130 and 275~GeV), polarized $^{3}$He, and unpolarized deuterons and Au, with per-species integrated luminosities of roughly $1.0$--$2.5~\mathrm{fb}^{-1}$. Despite this being only an early operational phase, the projected polarized $e+p$ yearly integrated luminosities already exceed the total lifetime integrated luminosity of the Hadron-Electron Ring Accelerator (HERA), providing immediate access to a broad Quantum Chromodynamics (QCD) program.

\subsection*{Scientific Highlights by NAS Pillar}
 
\paragraph*{Origin of Nucleon Mass: Gluon Dynamics and the Hadron Structure}
Inclusive deep-inelastic scattering (DIS) provides access to the proton structure functions $F_2$ and $F_L$. Even at early luminosities, planned measurements will deliver constraints on the valence up-quark and gluon distributions, with particularly strong improvements for $x > 0.3$. The parton distribution function (PDF) extractions will be performed simultaneously with the strong coupling $\alpha_S$, with projected experimental uncertainty 
at the level of both the current world average and lattice QCD determinations. 
Spectator-tagged measurements on deuterium will also provide the first collider extraction of the free-neutron structure function $F_2^n$ with minimal nuclear corrections, strengthening the isospin decomposition of nucleon PDFs.
Exclusive channels, notably deeply virtual Compton scattering (DVCS) and vector-meson production, provide direct access to generalized parton distributions (GPDs), and therefore to the transverse spatial structure of quarks and gluons. Projected DVCS measurements at early EIC kinematics will already improve the extraction of Compton form factors, the experimentally accessible convolutions of GPDs entering the DVCS amplitude. While a precise extraction of GPDs requires fine multi-dimensional binning, and thus the need for a large amount of data, the EIC GPD program can already be initiated during the early years.
Forward-tagged exclusive meson production will extend this program to the structure of the pion, providing a direct early handle on how QCD dynamics generate the structure of light mesons. Semi-inclusive DIS (SIDIS) will extend nucleon tomography to multi-dimensional momentum space through transverse-momentum-dependent (TMD) PDFs, while measurements of fragmentation functions (FFs), including nuclear FFs, will probe how observed hadrons emerge from struck quarks and gluons. Jet and heavy-flavor measurements, including open charm production and the $\Lambda_c/\mathrm{D}^0$ production ratio, will add complementary handles on the gluon distributions and hadronization mechanisms that constrain the nucleon mass budget.

\paragraph*{Origin of Nucleon Spin: Helicity, Orbital Motion, and Spin--Orbit Correlations}
The availability of polarized electron and proton beams (and polarized $^{3}$He towards the end of the early years of EIC operation) enables immediate progress on the nucleon spin puzzle. The measurement of the double-helicity asymmetry $A_1$ in inclusive DIS, using longitudinally polarized electron and proton beams, will allow one to constrain the gluon helicity distribution in the nucleon. The extension of this measurement to SIDIS will provide quark-flavor sensitivity, in particular to the so-far poorly constrained sea-quark helicity PDFs. 
The inclusion of the momentum dependence of the produced hadron and measurements with different beam-polarization states, including transversely polarized proton beams, allow one to probe correlations between the transverse momentum and/or spin of the parton and the spin of the nucleon. In parallel, GPD-sensitive measurements from exclusive reactions provide constraints relevant to partonic orbital angular momentum (OAM). For all processes, measurements using polarized $^{3}$He with spectator-proton tagging in the far-forward region provide direct access to various distributions inside the neutron. Collectively, the early polarized program will begin to disentangle quark, gluon, and orbital contributions to nucleon spin across an unprecedented kinematic domain.

\paragraph*{Emergent Properties of Dense Gluonic Matter}
The flexibility in beam species during early running -- deuterons, $^{3}$He, Ag, and Au -- opens a new regime for studying dense gluonic matter. Inclusive $e+\mathrm{A}$ DIS will extend the reach in nuclei by roughly an order of magnitude in $x$, down to $x \sim 10^{-3}$, allowing the first extraction of the nuclear structure function $F_2^A$ at low $x$ in heavy nuclei and precise measurements of nuclear $F_2$ ratios.
Impact studies show that with data from the early running we can constrain nuclear PDFs for quarks and gluons at intermediate-to-low $x$.
For sea and valence up-quark distributions, the EIC-only fits already outperform global fits across most of the $x$ range. Diffractive vector-meson (VM) production off nuclei is particularly powerful for studying dense gluonic matter, because its cross section scales quadratically with the gluon density. Simulations of coherent $\phi \to K^{+}K^{-}$ production in $e+\mathrm{Au}$ show that the ePIC resolution in four-momentum transfer between the initial- to final-state ion is sufficient, provided adequate control of the incoherent background contribution, to resolve the diffractive pattern in its cross section and to recover the associated transverse spatial gluon distribution via Fourier--Bessel transformations, even with the moderate integrated luminosity available. Coherent and incoherent VM production  provide access, respectively, to the average spatial gluon distribution and to event-by-event gluon-density fluctuations in nuclei. Di-hadron correlations in SIDIS will provide complementary sensitivity to the transition toward the saturation regime through azimuthal decorrelations. Charged-jet and $D^{0}$-in-jet nuclear modification ratios, $R_{eA}$, will probe parton energy loss and modifications of heavy-quark hadronization in cold nuclear matter, with projected statistical precision sufficient to resolve theoretical predictions in multiple rapidity intervals.
Taken together, all the measurements described above have the potential to facilitate the search for the onset of the long-conjectured gluon saturation regime by ePIC.
 
\subsection*{Cross-Cutting Measurements}
 
The studies presented in this report, performed by the four ePIC physics working groups (Inclusive DIS, Semi-Inclusive DIS, Exclusive, Diffractive and Tagging, and Jets \& Heavy Flavor), are linked to all three NAS pillars. Each measurement campaign with different beam species, polarization and energies will give access to physics results addressing several or all pillars of the EIC physics scope. The inclusive measurements require only precision reconstruction of the scattered electron with central tracking and electromagnetic calorimetry, and no far-forward instrumentation or robust particle identification beyond electrons. 
They therefore offer the quickest route to early science results.
As detector commissioning progresses, hadron particle identification (PID) enables jets and heavy flavor measurements, including hadron-in-jet Collins observables (sensitive to Collins FFs), jet fragmentation functions, and charm-quark hadronization measurements. PID also enables a SIDIS program: multiplicities, unpolarized TMD PDFs, longitudinal double-spin asymmetries, and Sivers and Collins asymmetries.
This is followed by far-forward instrumentation, which then unlocks the exclusive, diffractive, and tagged program (DVCS, exclusive vector mesons, spectator-tagged neutron structure in deuterium and $^{3}$He, coherent and incoherent vector-meson production in nuclei, and meson-structure measurements using forward tagging, including the pion form factor and pion/kaon structure functions).


\providecolor{massHead}{HTML}{BDD7EE}
\providecolor{massCell}{HTML}{DEEBF7}
\providecolor{massDark}{HTML}{1F4E79}
\providecolor{massMid}{HTML}{2E75B6}
\providecolor{massLite}{HTML}{9DC3E6}

\providecolor{spinHead}{HTML}{C5E0B4}
\providecolor{spinCell}{HTML}{E2EFDA}
\providecolor{spinDark}{HTML}{375623}
\providecolor{spinMid}{HTML}{548235}
\providecolor{spinLite}{HTML}{A9D18E}

\providecolor{gluHead}{HTML}{F8CBAD}
\providecolor{gluCell}{HTML}{FCE4D6}
\providecolor{gluDark}{HTML}{C55A11}
\providecolor{gluMid}{HTML}{ED7D31}
\providecolor{gluLite}{HTML}{F4B183}

\providecolor{epicblue}{HTML}{1F3864}
\providecolor{subtitlegray}{HTML}{595959}

\providecommand{\strength}[2]{%
  \tcbox[on line, colback=#1, colframe=#1, coltext=white,
         boxrule=0pt, arc=2pt, left=4pt, right=4pt, top=1pt, bottom=1pt,
         boxsep=0pt]{\scriptsize\bfseries #2}%
}

\makeatletter

\@ifundefined{pillcellbox}{%
\newtcolorbox{pillcellbox}[2]{%
  colback=#1, colframe=#1, boxrule=0pt, arc=4pt,
  left=5pt, right=5pt, top=4pt, bottom=4pt, boxsep=1pt,
  width=\linewidth, before skip=0pt, after skip=0pt,
  valign=top, equal height group=#2,
  before upper={\raggedright\hyphenpenalty=10000\exhyphenpenalty=10000\sloppy}
}
}{}

\@ifundefined{headerbox}{%
\newtcolorbox{headerbox}[1]{%
  colback=#1, colframe=#1, boxrule=0pt, arc=4pt,
  left=5pt, right=5pt, top=5pt, bottom=5pt,
  width=\linewidth, before skip=0pt, after skip=0pt,
  halign=center, valign=center, equal height group=hdr
}
}{}

\@ifundefined{cellitems}{%
\newenvironment{cellitems}{%
  \begin{itemize}[leftmargin=10pt, itemsep=1pt, topsep=2pt, parsep=0pt, label=\textbullet]\footnotesize
}{%
  \end{itemize}
}
}{}

\@ifundefined{labelbox}{%
\newtcolorbox{labelbox}[1]{%
  blank, boxrule=0pt, boxsep=0pt,
  left=0pt, right=0pt, top=4pt, bottom=4pt,
  width=\linewidth, before skip=0pt, after skip=0pt,
  valign=center, equal height group=#1,
  before upper={\raggedright\hyphenpenalty=10000\exhyphenpenalty=10000}
}
}{}

\makeatother

\begin{table}[!b]
\centering

{\color{epicblue}\bfseries\large NAS Science Pillars and ePIC early science Measurements}\\[2pt]
{\color{subtitlegray}\small Examples inside each cell indicate how ePIC early physics measurements connect to the three NAS pillars}\\[10pt]

\setlength{\tabcolsep}{3pt}
\renewcommand{\arraystretch}{1.0}

\begin{tabular}{@{}>{\raggedright\arraybackslash}p{0.18\linewidth}
                  p{0.255\linewidth}
                  p{0.255\linewidth}
                  p{0.255\linewidth}@{}}

\begin{labelbox}{hdr}
\centering
\textbf{\footnotesize ePIC early science}\\[-1pt]
\textbf{\scriptsize measurement type}
\end{labelbox}
&
\begin{headerbox}{massHead}\color{epicblue}\bfseries\small Origin of\\nucleon mass\end{headerbox}
&
\begin{headerbox}{spinHead}\color{epicblue}\bfseries\small Origin of\\nucleon spin\end{headerbox}
&
\begin{headerbox}{gluHead}\color{epicblue}\bfseries\small Emergent properties\\of dense gluon matter\end{headerbox}
\\[8pt]

\begin{labelbox}{rowA}\centering\textbf{\small Inclusive DIS}\end{labelbox}
&
\begin{pillcellbox}{massCell}{rowA}
\strength{massLite}{Supporting}
\begin{cellitems}
  \item $F_2$ and $F_L$ structure functions
  \item Proton PDFs and $\alpha_S$
\end{cellitems}
\end{pillcellbox}
&
\begin{pillcellbox}{spinCell}{rowA}
\strength{spinDark}{Strong}
\begin{cellitems}
  \item $A_1^{p,n}$ and $g_1^{p,n}$
  \item Gluon helicity PDFs
\end{cellitems}
\end{pillcellbox}
&
\begin{pillcellbox}{gluCell}{rowA}
\strength{gluMid}{Important}
\begin{cellitems}
  \item Nuclear $F_2$ ratios
  \item Nuclear PDF constraints
\end{cellitems}
\end{pillcellbox}
\\[8pt]

\begin{labelbox}{rowB}\centering\textbf{\small Semi-inclusive DIS}\end{labelbox}
&
\begin{pillcellbox}{massCell}{rowB}
\strength{massLite}{Supporting}
\begin{cellitems}
  \item Unpolarized TMD PDFs
  \item Nuclear fragmentation functions
\end{cellitems}
\end{pillcellbox}
&
\begin{pillcellbox}{spinCell}{rowB}
\strength{spinDark}{Strong}
\begin{cellitems}
  \item Sivers $A_{UT}$
  \item Collins asymmetries
  \item Quark helicity PDFs 
\end{cellitems}
\end{pillcellbox}
&
\begin{pillcellbox}{gluCell}{rowB}
\strength{gluDark}{Strong}
\begin{cellitems}
  \item Di-hadron correlations
  \item Nuclear hadronization / fragmentation
\end{cellitems}
\end{pillcellbox}
\\[8pt]

\begin{labelbox}{rowC}\centering\textbf{\small Exclusive /} \newline
\textbf{\small Diffractive /} \newline \textbf{\small Tagged}\end{labelbox}
&
\begin{pillcellbox}{massCell}{rowC}
\strength{massDark}{Strong}
\begin{cellitems}
  \item DVCS / GPD imaging
  \item Quarkonium production
  \item Meson form factors / structure functions
\end{cellitems}
\end{pillcellbox}
&
\begin{pillcellbox}{spinCell}{rowC}
\strength{spinMid}{Important}
\begin{cellitems}
  \item Spectator-tagged $A_1(n)$
  \item GPD constraints on OAM
\end{cellitems}
\end{pillcellbox}
&
\begin{pillcellbox}{gluCell}{rowC}
\strength{gluDark}{Strong}
\begin{cellitems}
  \item Coherent/incoherent VM in $eA$
\end{cellitems}
\end{pillcellbox}
\\[8pt]

\begin{labelbox}{rowD}
\centering
\textbf{\small Jets \& Heavy}\\
\textbf{\small Flavor}
\end{labelbox}
&
\begin{pillcellbox}{massCell}{rowD}
\strength{massMid}{Important}
\begin{cellitems}
  \item Open charm production
  \item $\Lambda_c / \mathrm{D}^0$ ratio
\end{cellitems}
\end{pillcellbox}
&
\begin{pillcellbox}{spinCell}{rowD}
\strength{spinLite}{Supporting}
\begin{cellitems}
  \item $\Delta g$ from charm/jets
  \item Hadron-in-jet Collins FF
\end{cellitems}
\end{pillcellbox}
&
\begin{pillcellbox}{gluCell}{rowD}
\strength{gluMid}{Important}
\begin{cellitems}
  \item Charged-jet $R_{eA}$
  \item $\mathrm{D}^0$-in-jet $R_{eA}$
\end{cellitems}
\end{pillcellbox}
\\

\end{tabular}

\caption{Mapping of example ePIC early-science measurements to the NAS science pillars. The table is intended as a compact guide to the cross-cutting structure of the early science program rather than as an exhaustive list of all measurements discussed in the report. The strength label in each cell indicates the directness of the connection: ``Strong'' denotes measurements providing a direct handle on the underlying physics of a given pillar, ``Important'' denotes a substantial but less direct sensitivity, and ``Supporting'' corresponds to a complementary or more indirect role.}
\label{tab:nas-pillars-exe}
\end{table}

The same measurement categories contribute to more than one NAS pillar. For example, inclusive DIS constrains proton and nuclear PDFs, while polarized inclusive DIS accesses helicity structure; SIDIS provides both momentum-space imaging and spin--orbit observables; exclusive and diffractive reactions provide spatial imaging and nuclear gluon-density information; and jets and heavy flavor probe gluon dynamics, hadronization, and cold nuclear matter effects. Table~\ref{tab:nas-pillars-exe} summarizes this cross-cutting structure. 
These same measurements also provide important benchmarks for the broader nuclear-physics program: early $e+\mathrm{A}$ constraints on nuclear PDFs, gluon imaging, heavy-flavor production, jet observables, and cold-nuclear-matter effects will help interpret heavy-ion measurements at the Relativistic Heavy Ion Collider (RHIC) and the Large Hadron Collider (LHC), where initial-state nuclear structure and cold-matter effects must be separated from final-state hot-QCD dynamics.
The entries are examples rather than an exhaustive list, and the labels indicate the directness of the connection: ``Strong'' denotes measurements providing a direct handle on the underlying physics of a given pillar, ``Important'' denotes a substantial but less direct sensitivity, and ``Supporting'' denotes a complementary or more indirect role.

The staged structure of these measurements is naturally matched to the staged availability of beam species, polarization, and luminosity. Details of the individual measurements are given in the running-configuration tables in the body of the report.

\subsection*{Conclusions and Outlook}
 
Even before the EIC reaches its full design parameters, ePIC will produce high-precision results that directly address the fundamental questions of QCD identified in the 2018 NAS report. Projected early-year integrated luminosities for polarized $e+p$ running exceed HERA's lifetime total, and the broad range of nuclear targets will enable the first collider DIS measurements on nuclei. Whilst the present report only assumes the running of a 9~GeV electron beam, the addition of a 5~GeV electron beam energy configuration is possible. 

At the same time, the early science program must be understood as the first stage of the EIC mission, not its fulfillment. A dedicated section of this report (Sec.~\ref{sec:limitations}) details the limits of the initial configuration: a lower-energy electron beam and reduced integrated luminosities compared to the EIC science program goals. The 9~GeV electron beam during the early running phase does not reach the low-$x$ domain where clearer evidence for the gluon saturation regime is expected to emerge. Firmly entering the onset to the saturated regime requires the 18~GeV beam of the upgraded EIC. Early luminosities, while exceeding the HERA lifetime total, remain an order of magnitude below those needed for quantitative GPD tomography and for closing the spin sum rule at low $x$; charged-current DIS and key hard-probe and near-threshold channels are deferred entirely.
Realizing the full NAS science case -- the questions this facility is built to answer -- therefore depends on realizing the full EIC design capabilities and preserving the planned upgrade path.

The early science program demonstrates not only the readiness and excitement of the ePIC Collaboration, but also the extraordinary discovery potential of the EIC from its very first years of operation. The measurements presented here will deliver world-leading insights into the quark and gluon structure of nucleons and nuclei, the nucleon spin decomposition, and the emergence of dense gluonic matter---while simultaneously establishing the validated analysis framework on which the full EIC physics program will then be built.

\newpage
\pagenumbering{arabic}
\renewcommand{\thesubsection}{\thesection.\arabic{subsection}}
\setcounter{subsection}{0}

\newpage

\section{Introduction}
\label{sec:introduction}

As is typical for new particle colliders, the Electron–Ion Collider (EIC) will not initially operate at its full design specifications. A staged commissioning plan, covering instantaneous luminosity, beam energies and beam polarization configurations is being carefully developed.
The general-purpose detector, the electron-Proton and Ion Collider (ePIC) experiment, will be operational during the EIC commissioning period. All subsystems of the ePIC detector are foreseen to be ready by the beginning of data taking. However, some of the far-forward subsystems close to the beamline will only be operated once the beam conditions are fully under control.    

This document has been prepared by the ePIC Collaboration in response to a charge from the Associate Lab Directors of Brookhaven National Laboratory (BNL) and Thomas Jefferson National Accelerator Facility (JLab), the two EIC host laboratories. The request, sent to ePIC management in June 2025 and reported on the \hyperref[app:charge-letter]{Appendix}, called for a summary of the scientific opportunities accessible with early ePIC data, before the ramp-up to the full EIC machine capabilities. The goal of this document is to address that charge and provide the broader nuclear physics community, funding agencies, and laboratory management with a clear view of the exciting physics results that will be achievable during the initial years of EIC operations, prior to the ramp-up to full machine capability. This period is  referred to as the ``early science'' years in this document.

This report details predictions for the impact ePIC will have on the field of nuclear physics in its first years of operation based on simulation studies performed using the ePIC Collaboration’s software and simulation stack. This continuously validated, modular, community-developed stack is built around monthly tagged releases and simulation campaigns, which serve both as production infrastructure and as recurring quality-control mechanisms for detector design, reconstruction development, and physics analysis readiness. 
The simulation chain begins with physics event generators for the processes of interest, 
followed by a full Geant4~\cite{Agostinelli:2002hh, Allison:2006ve, Allison:2016lfl} detector simulation, digitization, and event reconstruction.

The report is structured as follows. We begin with a section that provides links between the science pillars of the EIC physics case and the physical processes that will be measured with ePIC.  The following section summarizes the expected beam conditions and machine capabilities during the early years of running. Subsequent sections highlight selected measurements within each of the pillars of the EIC science program. These discussions focus on physics channels that offer particularly clean signatures, sizable cross sections, or unique sensitivities to hadron and nuclear structure, demonstrating the high-impact science that can be achieved during the early operating phase. The last section summarizes our findings.

\section{The NAS Science Pillars and the Physics Processes}
\label{sec:pillars}

In 2017, the National Academy of Sciences (NAS) assessed the physics case for an EIC as ``compelling, fundamental, and timely''. 
Quoting from the 2018 NAS report~\cite{NAP25171}, the EIC can uniquely address three profound and central questions in nuclear physics through deep-inelastic scattering (DIS):
\begin{itemize}
  \item How does the mass of the nucleon arise?
  \item How does the spin of the nucleon arise?
  \item What are the emergent properties of dense systems of gluons?
\end{itemize}

We do not attempt to re-state the science case for the EIC here, which has been extensively reported elsewhere~\cite{NAP25171, AbdulKhalek:2021gbh, Accardi:2012qut}. However, to provide some context for the above topics within our planned measurements and the following sections of this document, we here link the science pillars laid out in the NAS report to some examples of physical processes which will be measured at the EIC.

The vast majority of the nucleon mass arises from Quantum Chromodynamics (QCD). To better understand QCD dynamics, it is crucial to access quark and gluon distributions through exclusive and tagged measurements (Sec.~\ref{secmain:exclusive}). For example, exclusive vector-meson production and deeply virtual Compton scattering (DVCS) provide access to the spatial distributions of gluons and quarks, while forward-tagged meson-structure measurements can further elucidate how QCD dynamics generate the structure of light hadrons.

To further explore the nucleon mass enigma, it is also important to perform multi-dimensional tomography of the distributions of quarks and gluons inside the nucleon and nuclei, in terms of their transverse position and momentum. Generalized Parton Distributions (GPDs) and transverse-momentum-dependent (TMD) distribution functions are important tools for this, and they are accessible via several exclusive (Sec.~\ref{secmain:exclusive}) and semi-inclusive (Sec.~\ref{secmain:sidis}) processes respectively.


The polarization of beams at the EIC will enable us to investigate how the nucleon's spin originates from quarks, gluons, and their interactions, 
via measurements of Semi-Inclusive DIS (SIDIS) processes (Sec.~\ref{secmain:sidis}). 
Furthermore, exploiting spin-polarization observables in inclusive DIS (Sec.~\ref{secmain:inclusive}) also provides key insights into quark helicity and gluon contributions to the nucleon spin.

The information on the quark and gluon content of the nucleon is encoded in Parton Distribution Functions (PDFs), which are extracted via inclusive DIS (Sec.~\ref{secmain:inclusive}). Data at the EIC will enable the extraction of PDFs for free protons and a large variety of nuclear species. Nuclear gluon distributions are poorly constrained, and the ability to compare nucleon versus nuclear PDFs will enable new studies of the dense gluonic regime in hadron structure.
The production of vector mesons through diffractive scattering off nuclei in exclusive processes (Sec.~\ref{secmain:exclusive}) will also be essential for studying dense systems of gluons, as the cross sections for diffractive processes are quadratically dependent on the gluon density.
Studying the limit of high gluon density will allow us to map a potential transition to a gluon saturation regime, where the rate of gluon splitting and recombination reaches an equilibrium within a hadron.

The production of heavy flavor in $e + p$ and $e + \mathrm{A}$ scattering (Sec.~\ref{secmain:hf}) will be used to study gluon distributions in hadrons and hadronization mechanisms of heavy quarks (e.g., charm quarks). Jets (Sec.~\ref{secmain:hf}) are another key tool to address many NAS science pillars, in particular topics that are linked to the gluon structure of hadrons.

These measurements also have broader impact beyond the EIC science case itself. Nuclear PDFs, cold-nuclear-matter effects, gluon imaging, heavy-flavor production, and jet observables provide essential inputs and benchmarks for interpreting heavy-ion measurements at the Relativistic Heavy Ion Collider (RHIC) and the Large Hadron Collider (LHC), where signatures of quark-gluon plasma formation must be separated from initial-state nuclear structure and other cold-matter effects. By providing clean electron–nucleus constraints on gluon distributions, parton propagation, hadronization, and nuclear modifications in cold QCD matter, early ePIC data will complement the hot-QCD program and strengthen the role of the EIC within the wider nuclear-physics landscape.

The full realization of the scientific program at the EIC, as outlined by the NAS report~\cite{NAP25171}, will require the machine to reach its full capabilities, both in terms of beam energies and luminosity. Nevertheless, even as the machine progresses towards its full capabilities, early measurements will already yield scientific impact unique to the EIC. 
In the next section, we will briefly discuss the capabilities of the collider during the initial years. The measurements discussed throughout this report should always be interpreted in light of the machine constraints laid out in the next section.


\section{Early Physics Running at the EIC}
\label{sec:early_running}

The early years of the physics collisions delivered to ePIC will be driven by the commissioning and optimization of the machine. Some machine capabilities will be gradually made available over time while the luminosity will progressively increase to eventually reach the design peak. The EIC Project has provided an outline for the first years. The following assumptions are considered realistic at present: 
\begin{itemize}
  
  \item During the early-science era of the EIC, the electron beam energy will be 9~GeV, with a later upgrade path to 18~GeV. Operation at 10~GeV is also possible, but 9~GeV has been chosen to optimize luminosity in this initial phase. 

  
  \item Initial commissioning will use beams of medium-mass ions, such as Ag, at the energy of 115 GeV, to ensure technical stability during the commissioning period.
  Medium-mass ions are better suited for achieving desired, high center-of-mass energies while easing operational constraints during the initial phase. The ion energy will be chosen to match the bunch repetition rates for different orbit lengths in the hadron and electron storage rings. No beam polarization will be available during the first year for either the electron or the hadron beam.
  
  \item After completion of the initial beam commissioning, the longitudinal electron polarization (achieved via electron spin rotators) becomes a core capability for data taking. During this phase, electron beams can be polarized longitudinally, while transversely polarized hadron beams will also become available. During this period, the ePIC far-forward detectors might be commissioned, depending on the stability of the running conditions.

  \item After completion of the initial electron beam polarization commissioning, the commissioning of the hadron spin rotators is planned. This allows the experiment to transition from transversely polarized protons to longitudinal proton polarization.
 
\end{itemize}

Current assumptions for the collected luminosities per year, based on the EIC Project estimates, are listed in Table~\ref{tab:early-science-matrix}. 
The estimates for the luminosity per year-of-running during the early period are based on the following assumptions:
\begin{itemize}
  \item Electron bunch charge of 10\,nC (compared to 28\,nC in the final design).
  
  \item Constant proton beam Interaction Point (IP) divergences are maintained throughout the store by gradual increase of proton IP beta-functions (which control the beam size and focusing at the collision point) as the beam emittance (the spread of particles within the beam) increases.

  \item The electron beta-functions at the IP are adjusted accordingly to match electron and proton transverse beam size.

  \item Neither low-energy cooling of ion and proton beams, nor stochastic cooling of the heavy-ion beam is used in the store.

  \item A single Run is 30 weeks of operation at $80~\%$ uptime (assuming a 2~h store turnaround and 30 min for injection and detector turn-on).

  \end{itemize}

The aim of the present report is to show how, notwithstanding the stringent conditions listed above, a carefully optimized running plan would still allow ePIC to deliver an impactful physics output well before the collider reaches its full design capabilities. 

The beam species, energy configurations, polarization modes, and expected collected luminosities anticipated within the first five years of data taking are summarized in Table~\ref{tab:early-science-matrix}. 
The ion beam energies and luminosities are given per nucleon. This convention is followed throughout the document. Note that the degree of polarization of the beams is expected to reach 70\,\% for both electron and hadron beams.
These assumptions are expected to evolve as the accelerator design is refined, and commissioning experience accumulates. This table should therefore be regarded as a realistic, yet provisional, basis for early ePIC operations.

\setlength{\tabcolsep}{8pt} 
\renewcommand{\arraystretch}{1.2}
\begin{table}[htbp]
\centering
\small
\begin{tabular}{
>{\centering\arraybackslash}m{1.1cm} 
>{\centering\arraybackslash}p{2.5cm} 
>{\centering\arraybackslash}p{3.5cm} 
>{\centering\arraybackslash}p{2.6cm} 
>{\centering\arraybackslash}p{3.6cm}
}
\toprule
\textbf{Species} & \textbf{Beam energy (GeV)} &  \textbf{Integrated luminosity (fb$^{-1}$)} &\textbf{Electron-beam polarization} & \textbf{Hadron-beam polarization}\\

\midrule
 $e+$Ag & $9 \times 115$ &  1.0  & NO & N/A  \\
 
 $e+$D & $9 \times 130$ &  {1.5 } & LONG & NO \\
 
 $e +p$ & $9 \times 130$ &  {1.0} & LONG &TRANS and/or LONG\\

 $e+p$  & $9 \times 275$ &  {2.5 } & LONG & TRANS and/or LONG \\
 
 $e+$Au & $9 \times 100$ &  {1.0 } & LONG & N/A \\
 
 $e+{}^3$He & $9 \times 166$ &  {1.5 } & LONG &TRANS and/or LONG \\
 
\bottomrule
\end{tabular}
\caption{Proposed beam configurations and corresponding luminosities for the EIC early running. The $e+\mathrm{A}$ energy and luminosity are per nucleon.}
\label{tab:early-science-matrix}
\end{table}

This report uses 9~GeV, the nominal upper electron beam energy during early EIC running, as its baseline; lower electron beam energies remain technically feasible. Some measurements can benefit from such configurations. 
The studies presented here, together with future work, will help guide the optimization of the early running program.

Despite being part of the early operational phase, the projected yearly integrated luminosities are substantial. These values already exceed, within a single year, the total integrated luminosity accumulated during the full operational lifetime of the Hadron-Electron Ring Accelerator (HERA). The EIC will also be the first $e+\mathrm{A}$ collider with polarized protons and ion beams. Combined with the flexibility of beam species and energies, this provides immediate access to a broad range of QCD observables, including DIS structure functions, heavy--flavor production, diffractive and exclusive channels, and nuclear structure measurements.

Together, these capabilities form a robust foundation for the early scientific program of the EIC. Although the precise evolution of the running plan will depend on machine performance and operational experience, the overall strategy clearly positions the EIC and its detector ePIC to deliver impactful physics results from the outset. With an operational ePIC detector available from the start, the collider will enter operations with broad kinematic coverage and strong reconstruction performance, enabling precision studies in both proton and nuclear systems and providing a pathway toward the full realization of the EIC's long--term physics potential.

The following four sections will discuss the foreseen scientific impact driven by measurements of the different processes at the ePIC detector during the first years of EIC running. Studies in this report were performed using the April 2026 ePIC simulation campaign, which assumed a 10~GeV electron-beam energy and a maximum of 250~GeV for protons, consistent with the accelerator parameters available at that time. The physics results are not expected to change significantly when using the beam energies quoted in Table~\ref{tab:early-science-matrix}. 


\section{Inclusive measurements}
\label{secmain:inclusive} 
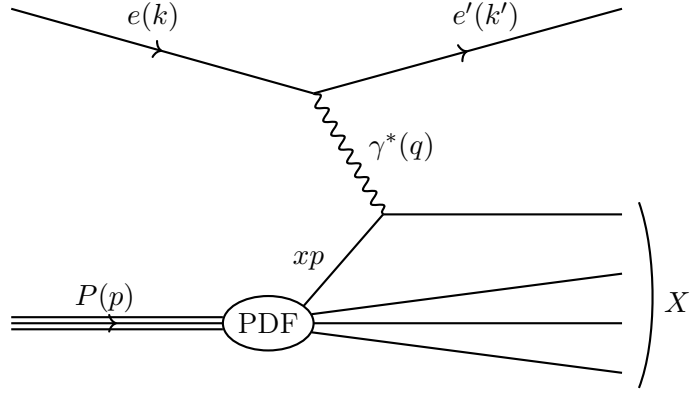
\begin{figure}[h]
    \centering
    \begin{tikzpicture}[scale=0.8, line width=0.8pt]

\tikzset{
  fermion/.style={
    postaction={decorate},
    decoration={markings, mark=at position 0.5 with {\arrow{>}}}
  },
  photon/.style={
    decorate,
    decoration={snake, amplitude=2pt, segment length=6pt}
  }
}

\coordinate (A) at (-0.2,1.55);      
\coordinate (B) at (0.95,-0.45);     

\coordinate (PDF) at (-0.95,-2.25);  

\def\rpdfx{0.75}                     
\def\rpdfy{0.45}                     

\coordinate (eL) at (-5.2,2.95);
\coordinate (eR) at ( 4.9,2.95);
\coordinate (pL) at (-5.2,-2.25);

\coordinate (xTop)  at (4.9,-0.45); 

\coordinate (xMidU) at (4.9,-1.43);
\coordinate (xMid)  at (4.9,-2.25);
\coordinate (xBot)  at (4.9,-3.07);

\coordinate (PDFW)  at ($(PDF)+(-\rpdfx,0)$);       
\coordinate (PDFNE) at ($(PDF)+(0.58,0.28)$);       

\coordinate (PDFEu) at ($(PDF)+(0.72,0.15)$);
\coordinate (PDFE)  at ($(PDF)+(\rpdfx,0)$);
\coordinate (PDFEd) at ($(PDF)+(0.72,-0.15)$);

\draw[fermion] (eL) -- (A);
\draw[fermion] (A) -- (eR);

\node at (-2.85,2.82) {$e(k)$};
\node at ( 2.65,2.82) {$e'(k')$};

\draw[photon] (A) -- (B);
\node[right] at ($(A)!0.50!(B)+(0.16,0.12)$) {$\gamma^{*}(q)$};

\draw ($(pL)+(0,0.10)$) -- ($(PDFW)+(0,0.10)$);
\draw[fermion] (pL) -- (PDFW);
\draw ($(pL)+(0,-0.10)$) -- ($(PDFW)+(0,-0.10)$);

\node at (-3.65,-1.82) {$P(p)$};

\draw (PDFNE) -- (B);
\node[right=-23pt, yshift=-0pt] at ($(PDFNE)!0.50!(B)$) {$xp$};

\draw (B)      -- (xTop);
\draw (PDFEu) -- (xMidU);
\draw (PDFE)  -- (xMid);
\draw (PDFEd) -- (xBot);

\draw[fill=white] (PDF) ellipse ({\rpdfx} and {\rpdfy});
\node at (PDF) {PDF};

\draw
  (5.18,-0.25)
  .. controls (5.48,-1.10) and (5.48,-2.55) ..
  (5.18,-3.32);

\node[right] at (5.42,-1.90) {$X$};

\end{tikzpicture}
    \caption{Neutral current DIS of an electron $e(k)$ from a proton $P(p)$, mediated by the exchange of a virtual photon $\gamma^{*}(q)$, in the one-photon exchange approximation. In this example only the scattered beam electron $e'(k')$ is reconstructed, without reconstructing the hadronic final state $X$.}
\label{fig:InclusiveDIS}
\end{figure}

Inclusive DIS measurements at the EIC will play a central role in determining the quark-gluon structure of nucleons and nuclei. Neutral-current (NC) DIS reduced cross sections provide the primary experimental input for the extraction of proton and nuclear structure functions (SFs) and, ultimately, for the determination of parton distribution functions (PDFs) and nuclear PDFs (nPDFs). Figure~\ref{fig:InclusiveDIS} illustrates an NC DIS process in which an electron scatters from a proton through the exchange of a virtual photon.
The focus in the early science running will be NC DIS, and charged current (CC) DIS is not discussed in this report.

The measurement of inclusive DIS cross sections relies primarily on the precise reconstruction of the scattered electron, complemented by information from the hadronic final state for kinematic reconstruction and systematic cross-checks. Importantly, these measurements do not require advanced particle identification (PID) or far-forward instrumentation, which will take longer to commission, making inclusive DIS one of the most accessible, high-precision, and impactful components of the early science program. Some of the physics topics related to these measurements and featured in this report are discussed  below.

Precise measurements of inclusive DIS reduced cross sections provide essential constraints for global analyses of collinear PDFs, and enable the extraction of the structure functions $F_{2}$ and $F_{L}$. PDFs give a one-dimensional description of the longitudinal momentum carried by quarks and gluons in the proton~\cite{Abramowicz:2015mha, Rezaei:2011zzb}.
Neutral current DIS is the cleanest probe of unpolarized collinear PDFs. A benchmark inclusive DIS dataset is provided by the combined H1 and ZEUS measurements from HERA, complemented by fixed-target measurements from other facilities, including JLab. The combined HERA data span several orders of magnitude in both $x$ and $Q^{2}$, and correspond to an integrated luminosity of order $1\,\mathrm{fb}^{-1}$. They form the basis of the HERAPDF2.0 set, extracted at leading order (LO), next-to-leading order (NLO), and next-to-next-to-leading order (NNLO) from $e+p$ data alone~\cite{Abramowicz:2015mha}. Although percent-level precision is reached at intermediate $x$, uncertainties remain larger at very small and large $x$. Existing inclusive DIS studies show that the full EIC dataset will substantially improve proton PDF precision~\cite{PhysRevD.109.054019} and further constrain the strong coupling constant~\cite{Cerci:2023StrongCoupling}.

Modern PDF phenomenology relies on global analyses performed at NNLO and approximate next-to-next-to-next-to-leading order ($\mathrm{N^{3}LO}$) accuracy in perturbative QCD, often supplemented by quantum electrodynamics (QED) corrections, on a broad range of precision measurements from HERA, fixed target and hadron collider experiments. Contemporary determinations, such as CT18/25~\cite{Hou:2021efy, Ablat:2025ct25}, CJ26~\cite{Accardi:2026hdv}, MSHT20~\cite{Bailey:2020ooq}, NNPDF4.0~\cite{Ball:2022qks}, ABMP~\cite{Alekhin:2017kpj}, and JAM~\cite{Cocuzza:2026highx} analyses, incorporate precision measurements from DIS, Drell–Yan production, electroweak boson production, inclusive jets, heavy-flavor production, and top-quark processes. 
These datasets are complementary: HERA dominates small and moderate $x$, fixed-target data constrain the large $x$ valence region, and LHC measurements add flavor and high-scale sensitivity. 
The comparison of PDFs accessed from DIS with those accessed from hadron-hadron collisions allows us to check theoretical assumptions entering the extraction, which is essential for the percent-level accuracy required for particle searches beyond the Standard Model. HERA will remain the primary small $x$ reference for proton PDFs until the full EIC, whose high-luminosity polarized and unpolarized DIS data will reduce sea-quark and gluon uncertainties, test dataset tensions, and enable a more unified determination of proton structure.

The EIC will also provide the first precision and high-energy collider DIS measurements on nuclei. Nuclear PDFs encode the partonic structure of bound nucleons, while nuclear modification ratios quantify deviations from the simple scaling of free-nucleon PDFs with atomic mass $A$.
Such deviations from an incoherent superposition of free-nucleon PDFs, especially in the gluon sector, signal high-density QCD dynamics.
The EIC reach and precision will transform nPDF extractions~\cite{PhysRevD.109.054019}: inclusive $e+\mathrm{A}$ DIS at low $x$ will enable $F_{L}$ measurements connected to the gluon density~\cite{Armesto:2010NuclearEffects}, clarify nuclear modifications of gluon distributions~\cite{Cazaroto:2008NuclearGluon}, and allow combined global QCD analyses of proton, deuteron and nPDFs~\cite{AbdulKhalek:2021gbh}, as well as potentially offering relevant information for heavy-ion physics at RHIC and the LHC.

Current nPDFs remain significantly less well constrained than their proton counterparts, particularly in the sea-quark and gluon sectors. Existing global sets and progress are reviewed in~\cite{EthierNocera:2020PDFreview, Klasen:2024nPDFreview}. Modern global analyses, such as EPPS21~\cite{Eskola:2021nhw}, nNNPDF~\cite{AbdulKhalek:2022ihb}, nCTEQ~\cite{Risse:2023ncteq}, and TUJU~\cite{Helenius:2022bws}, rely on fixed-target lepton–nucleus DIS measurements together with Drell–Yan and selected hard-scattering data from proton–nucleus and nucleus–nucleus collisions. Despite substantial progress, significant uncertainties remain, especially for gluons at low $x$.
Additional constraints have been obtained from jet and heavy-flavor production at both RHIC and the LHC, electroweak boson production at the LHC, as described in~\cite{Klasen:2024nPDFreview},  quarkonium production and ultra-peripheral collisions at RHIC and the LHC~\cite{Adare:2009xb, Abdulhamid:2024ups, Abelev:2007nb, Abelev:2012ba, Acharya:2019vlb, Acharya:2021khy}. However, these observables generally involve larger theoretical uncertainties than inclusive DIS.
Further constraints are expected from high-luminosity LHC nuclear running~\cite{Apollinari:2015HL-LHC} and from the JLab 12\,GeV program in the valence region~\cite{Dudek:2012vr}. The EIC will, however, uniquely address existing limitations through high-precision $e+\mathrm{A}$ measurements spanning a broad kinematic range in $x$ and Q$^{2}$, providing direct constraints on quark and gluon distributions in nuclei and enabling a new generation of global nPDF analyses.

The decomposition of the nucleon spin into quark and gluon spin contributions, as well as orbital angular momentum contributions also remains a central open problem in QCD. Polarized beams at the EIC will enable precision measurements of spin asymmetries, providing access to spin structure functions, particularly $g_{1}$, constraining polarized PDFs and the quark and gluon contributions to nucleon spin~\cite{Boer:2019fpa, Hughes:1999wr, Aidala:2012mv, Deur:2018roz}. In particular, the EIC will extend polarized DIS measurements into the previously unexplored low-$x$ region, where a substantial fraction of the nucleon spin may reside. The existing experimental foundation includes polarized DIS measurements from SLAC~\cite{Alguard:1976DISspin, Alguard:1978Bjorken, Anthony:1996E142, Abe:1995E143gp1, Abe1997E154, Abe:1997NLOQCD, Anthony:1999E155deuteron, Anthony:2000E155Q2, Anthony:1999E155g2, Anthony:2003E155g2}, EMC~\cite{Ashman:1988EMCg1}, HERMES~\cite{Ackerstaff:1997ELECTg1n, Airapetian:1998ELECTg1p}, SMC~\cite{Adeva:1993SMCg1d, Adams:1994SMCg1p, Adams:1994SMCg1pErratum, Adams:1995SMCg1d, Adeva:1997SMCg1d, Adams:1997SMCg1pFinal}, COMPASS~\cite{Ageev:2005COMPASSg1d}, and JLab~\cite{Amarian:2002GDHneutron, Amarian:2004MomentsNeutron, Amarian:2004Polarizabilities, Slifer:2008SumRules, Yun:2003CLASdeuteron, Fatemi:2003CLASg1p, Wesselmann:2007RSS, Slifer:2010RSStransverse, Dharmawardane:2006CLASA1, Guler:2015CLASdeuteron, Prok:2009CLASmoments, Bosted:2007CLASduality, Fersch:2017CLASg1p, Zheng:2004HallAAn1, Zheng:2004HallAStructure, Kramer:2005CLASg2n, Solvignon:2015E01012g2n, Solvignon:2008DualityNeutron, Sulkosky:2020E97110, Adhikari:2018CLASg1d, Prok:2014CLASg1, Armstrong:2019SANEcolor, Posik:2014d2n, Flay:2016d2nA1n, Qian:2011SSApi, Huang:2012ALT, Allada:2014SSAincl, Zhang:2014Pretzelosity, Zhao:2014KaonSSA, Zhao:2015DSA3He, Katich:2014TNSA}, which together have established much of our current understanding of nucleon spin structure.
Complementary polarized proton--proton measurements at RHIC, from STAR~\cite{Adamczyk:2015GluonPolarization, Adam:2019Wspin, Abelev:2009Spin} and PHENIX~\cite{Adare:2006GluonSpin}, provide direct sensitivity to the gluon helicity distribution and evidence for a non-zero gluon spin contribution for $x\gtrsim 0.05$, but leave large uncertainties at smaller $x$. Current polarized PDF fits, including DSSV~\cite{deFlorian:2008mr, deFlorian:2009vb}, JAM~\cite{Sato:2016tuz, Ethier:2017zbq, Adamiak:2023spin, Cocuzza:2025spin} and NNPDFpol~\cite{NNPDFpol:2025}, therefore remain limited by the kinematic reach of existing data, as summarized in~\cite{Deur:2018roz}. Further results are expected from JLab, CERN and RHIC before EIC operations, with possible complementary studies at NICA~\cite{Savin:2015NICA}. The EIC will, however, provide the decisive high-luminosity polarized DIS data at unprecedentedly small $x$.

Finally, inclusive diffractive DIS measurements offer a complementary probe of the low-$x$ structure of protons and nuclei, with sensitivity to gluon dynamics and diffractive parton distributions. Diffractive PDFs, as previously studied at HERA~\cite{Aktas:2006hy, Aktas:2006hx, Chekanov:2009aa, H1ZEUS:2012mb}, will also be extracted at the EIC and will be accessible in the early years of running.
Diffractive measurements are discussed further in Sec.~\ref{secmain:exclusive}.

\begin{table}[t]
\centering
\footnotesize
\renewcommand{\arraystretch}{1.8}
\begin{tabular}{L{0.15\textwidth} L{0.40\textwidth} L{0.35\textwidth}}
\hline
\textbf{Configuration} &
\textbf{Representative measurements} &
\textbf{Enabling capabilities} \\
\hline

$e+\mathrm{A}$ running with medium- and light-mass nuclei &
Reduced cross sections for nuclear $F_{2}$ extraction; $A$-dependence of nuclear effects; nPDFs; diffractive nPDFs &
No polarization needed, moderate luminosity; tracking, calorimetry, PID; rapidity gap selections for diffractive nPDFs \\

$e+p$ running at multiple energies, with polarization &
Reduced cross sections for $F_2$ and $F_L$ extractions, collinear and diffractive PDFs and $\alpha_S$ extractions; spin asymmetries for proton spin SFs and polarized PDFs &
 Lever arm in center-of-mass energies; proton polarization; electron polarization; far-forward instrumentation; and previous capabilities\\

$e+\mathrm{A}$ running with heavy nuclei &
Reduced cross sections for nuclear modifications of $F_2$, $A$-dependence  of nuclear effects in SF and nPDFs, diffractive nPDFs &
Requires only the previous capabilities \\

$e+\mathrm{A}$ running with polarized $^3$He &
Spin asymmetries for neutron spin SFs and polarized PDFs & Spectator tagging and previous capabilities\\

 \\

\hline
\end{tabular}
\caption{Overview of inclusive measurements in the ePIC early-science years grouped by the approximate running conditions and capabilities that enable them. 
}
\label{tab:inclusive_staging}
\end{table}

The measurements discussed above require different beam species, polarization configurations, and levels of detector commissioning. In the early science years ePIC will utilize these tools to build upon a rich legacy of inclusive measurements.
Table~\ref{tab:inclusive_staging} gives an estimate of how the early years may evolve and the approximate running conditions under which the principal representative inclusive DIS measurements become accessible during the early science program.

This program is accessible from the very start of operations, as laid out in Table~\ref{tab:inclusive_staging}, with the earliest measurements being inclusive DIS reduced cross sections.
Although the luminosities are modest compared to the ultimate EIC design goals, they are higher than previously available at HERA.
Combined with the large inclusive DIS cross sections, this makes these measurements ideal for impactful results in the initial years. Predicted statistical precision for inclusive measurements in early running is excellent, except at the highest $Q^{2}$ and lowest $x$, and will be much better than the anticipated systematic uncertainties.
The measurements will also mostly rely on precision reconstruction of the scattered electron.
There is no requirement for robust PID or exclusivity. Far-forward instrumentation, although not needed for generic inclusive DIS measurements, will enable the inclusive diffractive program.  Beyond the physics outputs, inclusive DIS will be key for ePIC to establish tracking, electromagnetic calorimetry, and detector alignment.

The earliest $e+\mathrm{A}$ running configurations, with medium and light nuclei, will enable measurements of reduced cross sections and nuclear SFs, providing the foundation for subsequent nPDF analyses.
Running heavy nuclei later in the early years will supplement the DIS program, enabling further independent nPDF fits, in addition to providing sensitivity to the onset of gluon saturation.

Reduced cross sections in inclusive $e+p$ DIS will yield $F_{2}$ extractions and impact global PDF fits for the proton and $\alpha_S$ extractions.
$F_{L}$ extractions will also be made, although they are more challenging.
Adding more beam energy configurations, for example a lower electron energy setting, would assist the $F_{L}$ program. However, the planned proton energies remain very useful, offering wide and overlapping ($x, Q^{2}$) coverage, including low-$x$ and high-$Q^2$ reach.
Spin polarization of the electron and proton beams will enable double spin asymmetries for proton spin SF and polarized PDFs.

Spin-polarized electron and $^{3}$He beams, later in early running, would facilitate neutron spin structure studies analogous to proton ones, providing a wealth of new information on the spin decomposition of the nucleon.
Although these measurements can be performed in a fully inclusive manner, the measurements will benefit from light spectator-ion tagging capabilities of the far-forward instrumentation. Tagging is also discussed further in Sec.~\ref{secmain:exclusive}.

Measurements for inclusive diffractive PDFs will be possible from the first beams and continue throughout the period. The configurations are powerful because they include protons spanning different center-of-mass energies, nuclei, polarization, and high luminosities compared to previous experiments. Rapidity gap selections early on will facilitate the measurements, with the measurements becoming even more established as the far-forward instrumentation is commissioned.

Studies of some of the above topics using the full ePIC detector simulation are presented below.

\paragraph{Proton Structure Functions and Parton Distribution Functions}
Inclusive $e+p$ DIS reduced cross sections will constitute one of the cornerstone measurements of the early science program. These measurements provide access to the proton PDFs and SFs $F_{2}$ and $F_{L}$. 
The projected impact of early science data on SF measurements and PDF extractions has been studied using the full ePIC detector simulation; full details may be found in~\cite{jimenezlopez2026inclusiveelectronprotonmeasurementprospects}.

\begin{figure}
    \centering
        \includegraphics[origin=c,width=0.48\linewidth]{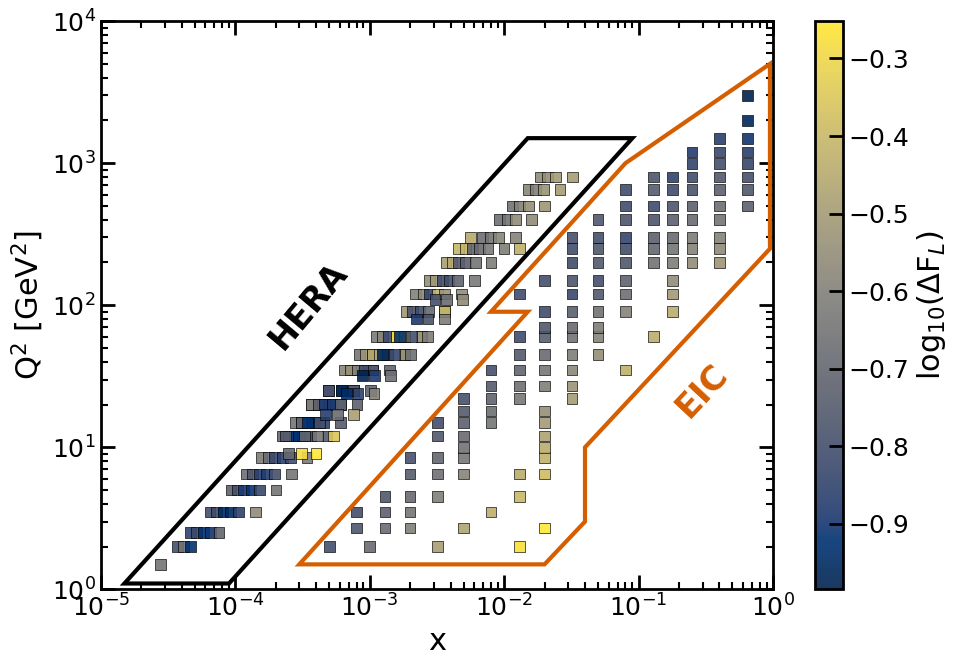}
        \includegraphics[origin=c,width=0.48\linewidth]{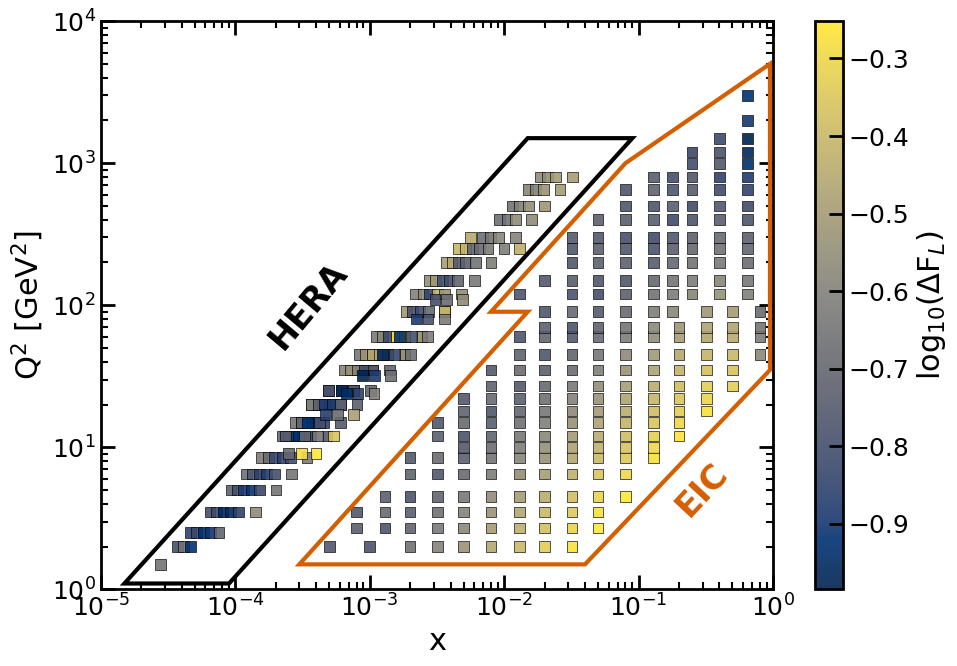}
    \caption{Kinematic coverage of EIC early science pseudo-data compared with HERA data for the proton F$_{L}$~\cite{Abramowicz:2015mha}, demonstrating complementarity of the EIC data and the expanded phase space for measurements. The color of the points indicates the absolute uncertainties. The uncertainties are at a comparable level to HERA but over a much larger phase space. The left panel shows the phase-space for the two $e+p$ configurations given in Table~\ref{tab:early-science-matrix}. Right panel shows the benefit of adding a third lower energy configuration (5\,GeV\,$\times$\,130\,GeV).}
    \label{inclusive-phase-space-FL}
\end{figure}

Figure~\ref{inclusive-phase-space-FL} shows the kinematic coverage of early science pseudo-data compared with world data from HERA for F$_{L}$, demonstrating an extremely complementary kinematic reach and expanded phase space of the early science data.
Two configurations are shown in Fig.~\ref{inclusive-phase-space-FL}. The left panel shows the phase space obtained with the two $e+p$ running configurations given in Table~\ref{tab:early-science-matrix}. The right panel demonstrates the benefit of adding a third lower energy configuration (5\,GeV\,$\times$\,130\,GeV), which has also been studied. 
In addition to significantly expanding the phase space, the addition of a third configuration is expected to improve the precision and uncertainties on $F_{L}$ points. It has been demonstrated~\cite{jimenezlopez2026inclusiveelectronprotonmeasurementprospects} that the ratio of uncertainties on the $F_{L}$ points for a three-beam-energy configuration to a two-beam-energy configuration typically falls between 10~\% and 20~\%, and never exceeds 32~\%. More details on the anticipated precision can be found in~\cite{jimenezlopez2026inclusiveelectronprotonmeasurementprospects}.

\begin{figure}[h]
    \centering
    \includegraphics[width=0.4\linewidth]{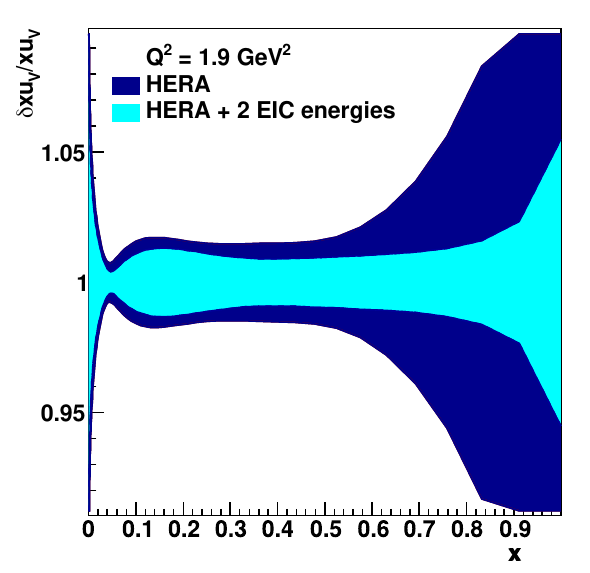}
   \includegraphics[width=0.4\linewidth]{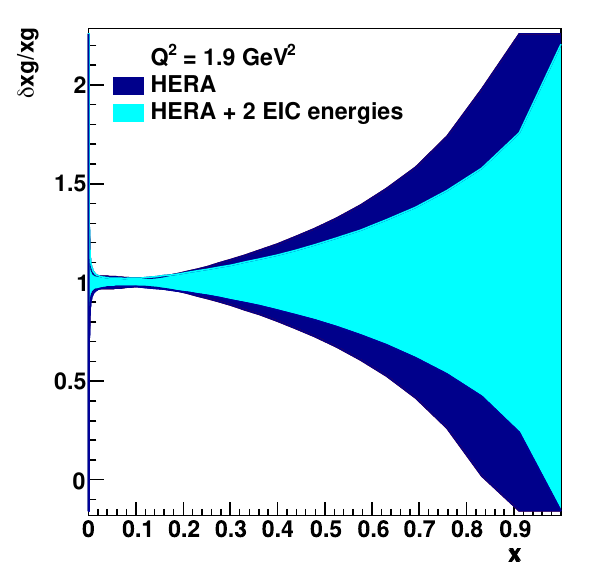}
    \caption{Impact of early science data on the NNLO collinear parton distributions of the proton, shown on a linear $x$ scale and performed using the HERAPDF2.0 NNLO fitting framework~\cite{h12015combinationmeasurementsinclusivedeep} and xFitter fitting platform~\cite{Alekhin2015HERAFitter}. The bands show relative experimental uncertainties for the up-valence (left) and gluon (right) distributions.}
    \label{fig:inclusive_pdf}
\end{figure}

Early science $e+p$ data and SF extractions will have a significant  impact on proton PDFs extracted from fits to globally available DIS data, as shown in Fig.~\ref{fig:inclusive_pdf}. This is particularly true for intermediate and high $x$, where we can expect to obtain significant impact in proton PDF fits for improving our knowledge of the up-valence quark and gluon distributions within the proton.
The PDF impact demonstrated in Fig.~\ref{fig:inclusive_pdf} has been evaluated using PDF fits at NNLO in perturbative QCD (pQCD) using the HERAPDF2.0 Dokshitzer–Gribov–Lipatov–Altarelli–Parisi (DGLAP) framework~\cite{Abramowicz:2015mha} and xFitter fitting platform~\cite{Alekhin2015HERAFitter}. For the eventual early science data collected, it is expected that the DGLAP fits will be performed at N$^{3}$LO in pQCD.

The SF data points for the above early science studies are assigned systematic uncertainties derived from the EIC Yellow Report~\cite{AbdulKhalek:2021gbh}, plus small statistical uncertainties. A point-to-point uncorrelated systematic uncertainty of 1.9~\% stems from the estimated contributions of radiative corrections, bin migration, detector efficiency, and charge symmetric background effects. A normalization uncertainty of 3.4~\%, which is uncorrelated between different beam configurations, is also included for each data set, for a total systematic uncertainty of 3.9~\% on each data point. The normalization uncertainties include contributions from the luminosity measurement (1.5~\%), as well as electron purity and detector efficiency effects. In the PDF fits, the correlated uncertainties are treated as ``nuisance parameters'' that are determined during the fitting, a method which is described further in~\cite{Abramowicz:2015mha}.

\begin{figure}[thb!]
	\centering
        \includegraphics[height=0.55\textwidth]{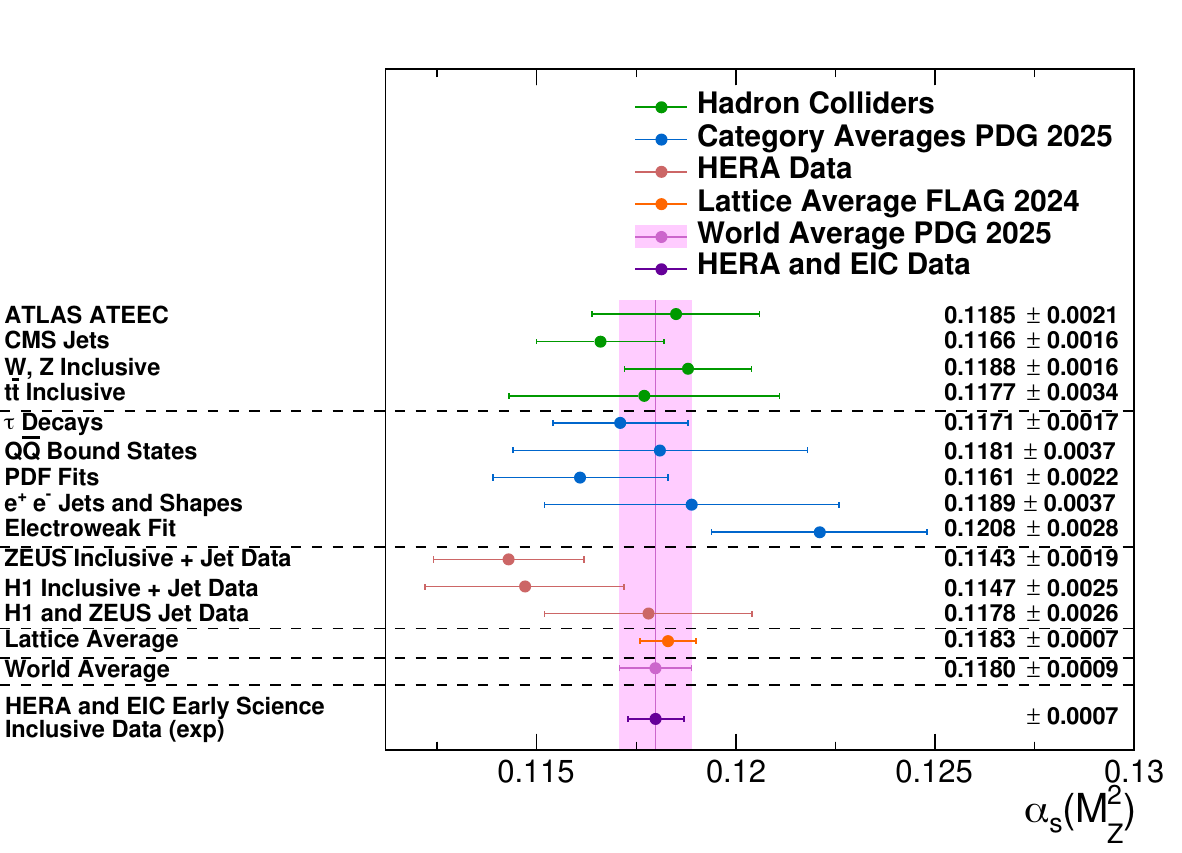}

  	\caption{Comparison of the early science projected $\alpha_S$ accuracy with existing measurements and the world average from the data linked in the 2026 PDG review on QCD~\cite{ParticleDataGroup:2026}.}
	\label{fig:inclusive-alphaS}
\end{figure}

Since PDFs evolve with scale via DGLAP equations that depend directly on the strong coupling $\alpha_{S}$, the PDF fits may be accompanied by a simultaneous extraction of $\alpha_{S}$ (rather than fixing it externally).
A projection of the strong coupling extracted from a simultaneous NNLO fit for the PDFs and $\alpha_{S}(M_{Z}^{2})$ using the early science pseudo-data together with the HERA data has been performed. A projected value of $\alpha_{S}(M_{Z}^{2}) = 0.1157 \pm 0.0007$ is obtained.
The quoted central value depends on the specific pseudo-data used  and should not be interpreted as a prediction; only the projected uncertainty is relevant for assessing the expected sensitivity.
The quoted uncertainty is the total experimental uncertainty; it does not include model and parameterization, or theory uncertainties. Model and parameterization uncertainties are expected to be small, while theoretical uncertainties (for example from higher orders beyond NNLO) may be understood further in the future by comparisons of NNLO versus N$^{3}$LO fits.
The experimental uncertainty demonstrates that a very precise result is achievable, reaching the level of the present world average or lattice QCD determinations~\cite{ATLAS2025StrongCoupling, ParticleDataGroup:2026, Cerci:2023StrongCoupling}.
Figure~\ref{fig:inclusive-alphaS} compares the projected uncertainty with existing measurements and the world average from the data linked in the 2026 Particle Data Group (PDG) review on QCD~\cite{ParticleDataGroup:2026}.

\paragraph{Nuclear Structure Functions and Parton Distribution Functions}
In the early running years the reach of $e+\mathrm{A}$ DIS data on nuclei in the low-$x$ region will be extended by an order of magnitude, down to $x \sim 10^{-3}$. Reduced cross section measurements will provide the opportunity to extract the $F_2^\mathrm{A}$ SF at low-$x$ for the first time in heavy nuclei, and to explore the collinear partonic structure of nuclei with an unprecedented level of detail. 

\begin{figure}[thb!]
	\centering
        \includegraphics[height=0.6\textwidth]{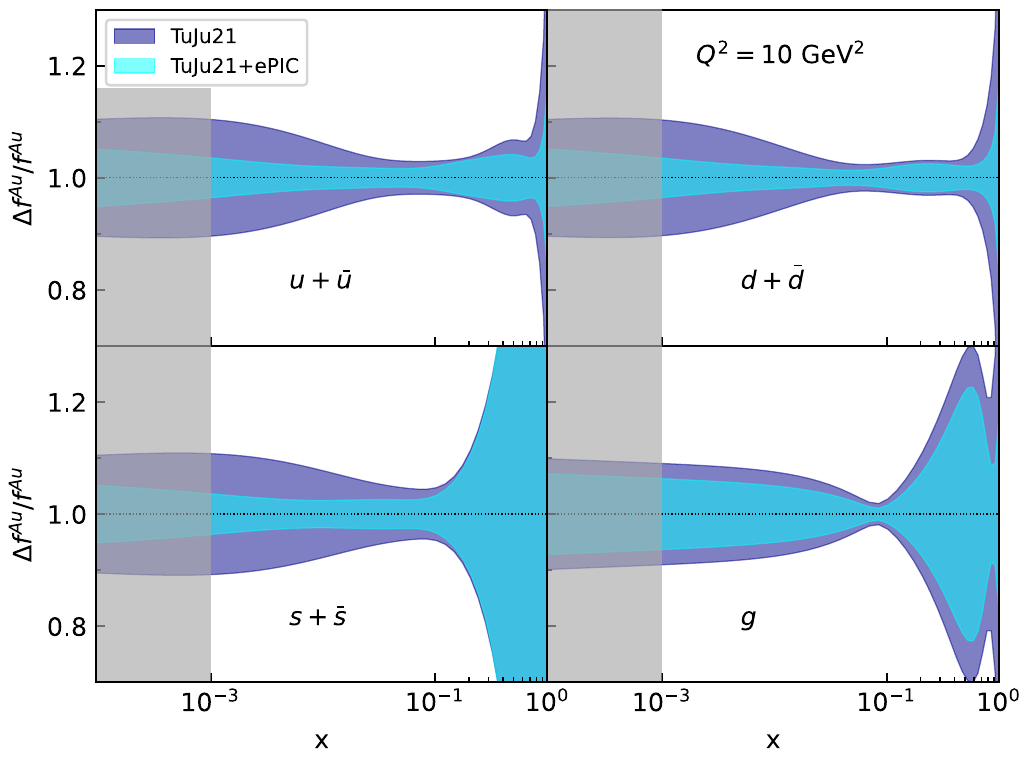}
  	\caption{Impact of early science inclusive data on nPDFs for a gold nucleus. The relative uncertainties of the nPDFs before and after reweighting of a global nPDF set~(TUJU21~\cite{Helenius:2022bws}) with EIC e+Au pseudo-data are shown as a function of $x$ at $Q^2 = 10$~GeV$^2$. The region of $x$ extending below 10$^{-3}$ is shaded, since it will not be directly constrained by experimental data. }	\label{fig:inclusive_R_Au}
\end{figure}

The potential impact of inclusive DIS data from the early science runs on nPDFs has been evaluated by the reduction in uncertainties when using a reweighting procedure with a set of global nPDF fits (TUJU21~\cite{Helenius:2022bws}), which includes data from fixed target DIS on nuclei and $W^\pm$ and $Z$ production data from $p+\mathrm{Pb}$ collisions at the LHC. A Bayesian reweighting procedure~\cite{Armesto:2013kqa} is used to update the TUJU21 nPDFs in accordance with the simulated early running $e+\mathrm{Au}$ data (TUJU21+EIC). The projected impact of the early-science EIC data
is summarized in Fig.~\ref{fig:inclusive_R_Au}, which shows the relative uncertainties on the nPDFs for parton species before and after reweighting of the PDFs.

There are significant improvements in the uncertainties for all parton species in the low-$x$ region, where the EIC will provide the first-ever data for DIS on nuclei. This paves the way to understanding gluons at high density (low $x$), which is a key goal of the full EIC program, offering the potential to discover the long-predicted saturation of the gluon density. Data for hard processes at the LHC, which also play a significant part in constraining gluon densities at low $x$, have been included in more recent global fits~\cite{Eskola:2021nhw,AbdulKhalek:2022ihb}. The DIS process is well understood, so nuclear DIS measurements over a large range of center-of-mass energy at the EIC offer superb constraining power. It has been shown that EIC inclusive data alone can constrain the nuclear modification factors with unprecedented precision~\cite{PhysRevD.109.054019}.

\paragraph{Nucleon Spin Structure Functions}
Polarized electron, proton, and $^3$He beams will provide unique opportunities to investigate the spin structure of the nucleon.
The measurement of observables such as double spin asymmetries for different electron/hadron polarization states would be possible. This will provide access to the spin dependent structure function $g_{1}$, which can elucidate how the spins of the quarks and gluons inside a nucleon are aligned compared to the nucleon's spin. Studying this across different ranges in $x$ will offer insights into quark and gluon spin dynamics inside the nucleon.

\begin{figure}[thb]
    \centering
    \includegraphics[width=0.45\linewidth]{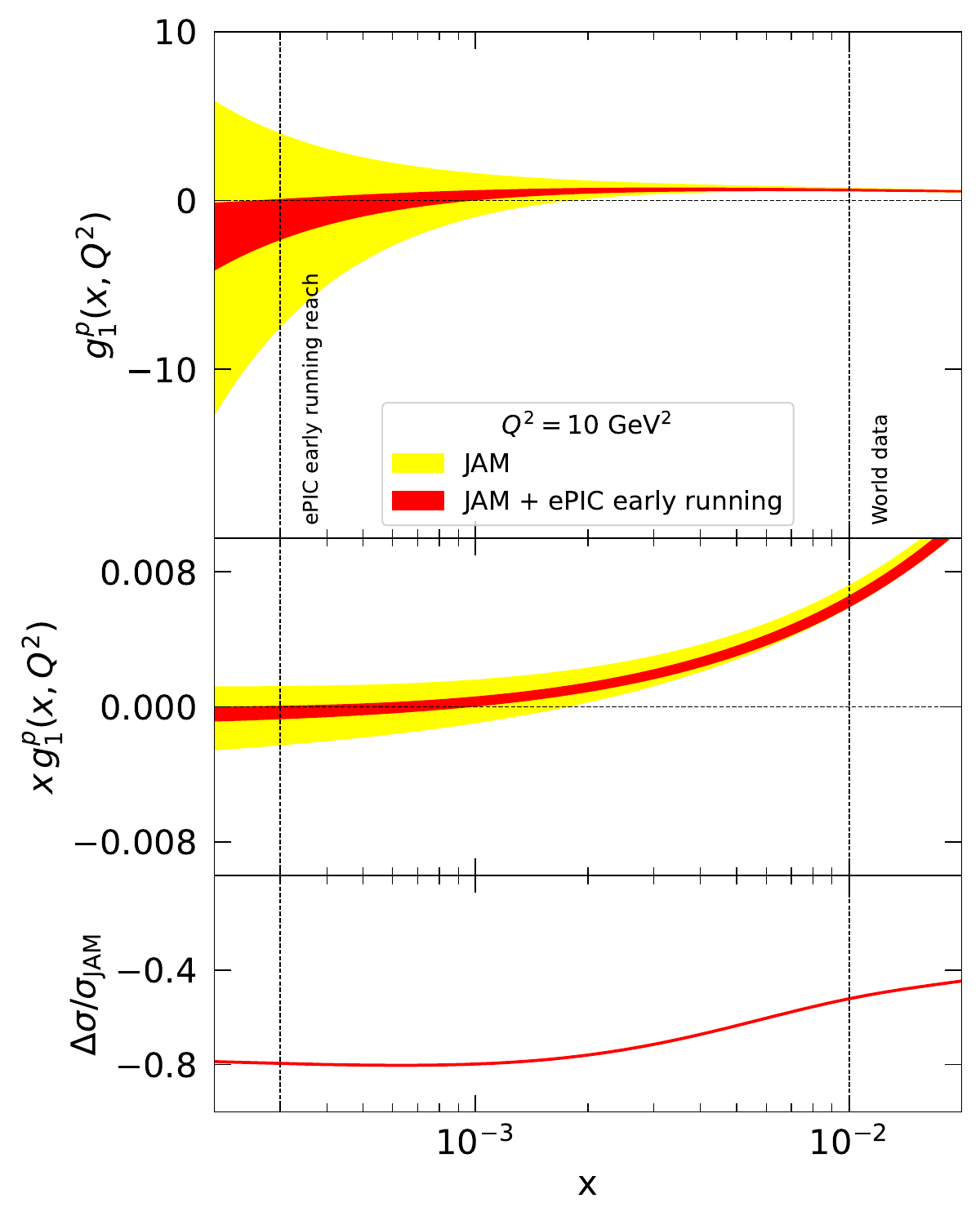}
    \includegraphics[width=0.45\linewidth]{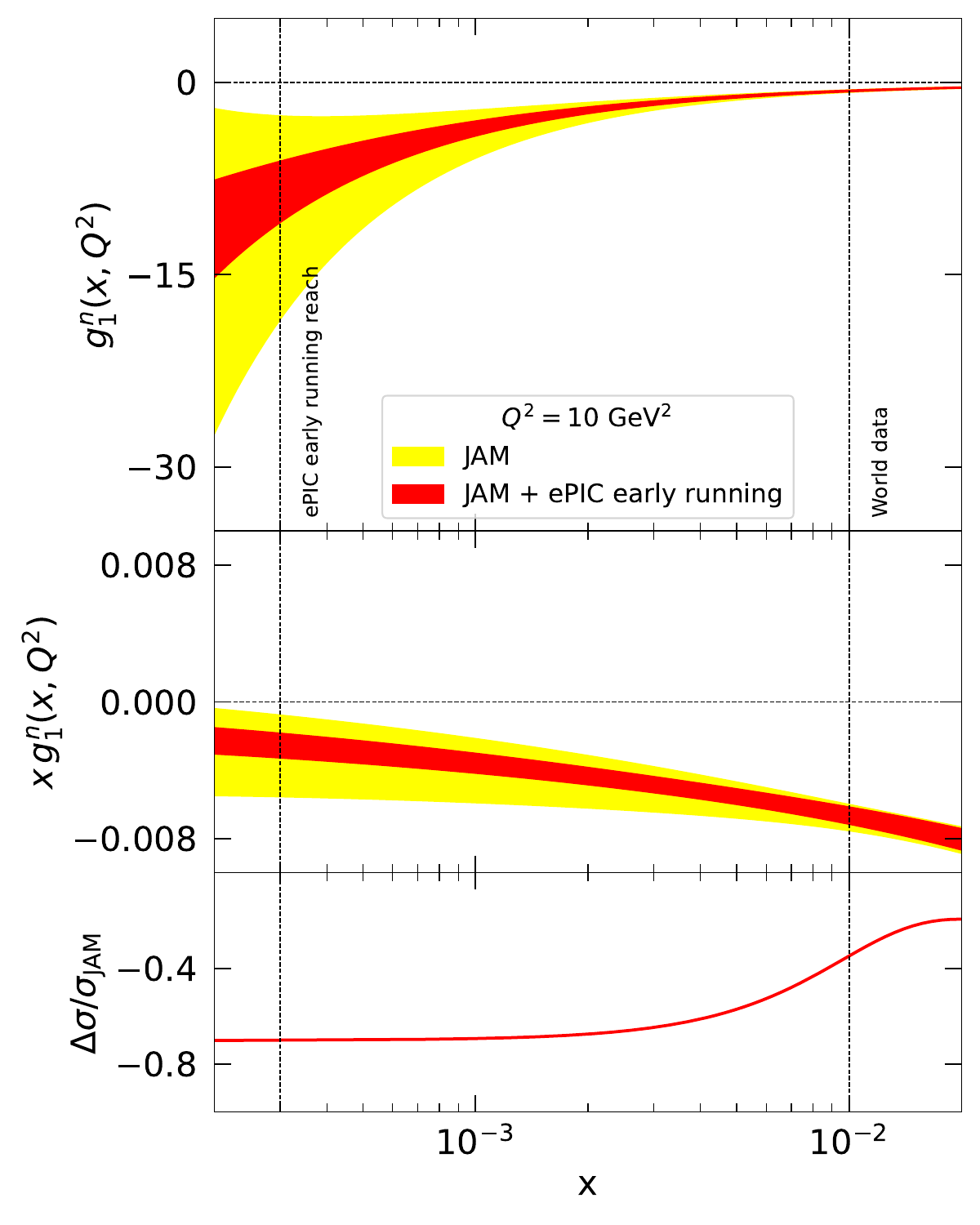}
    \caption{Left: impact of projected $A_1$ data on $g_1^p$. Right: impact of projected $A_1$ data on $g_1^n$. The top and middle panels show uncertainties from JAM global fits~\cite{6fn9-1wqb} for $g_1$ and $xg_1$, respectively, at $Q^2=10\;\rm{GeV^2}$ with (red bands) and without (yellow bands) projected early science ePIC data. The bottom panels show the corresponding fractional uncertainty reductions relative to current JAM results. The vertical dashed lines indicate the regions directly constrained by data.}
    \label{fig:inclusive_g1-main}
\end{figure}

Early running measurements of the double-helicity
asymmetry with a proton, A$_{1}^{p}$, or a neutron, A$_{1}^{n}$, vastly expand the landscape into an unexplored phase space compared to previous data, particularly into the low-$x$ region. Early science data will also partially overlap with some previous measurements for comparisons and consistency checks. The low-$x$ reach of early science data will be very important for exploring $g_{1}$ and polarized PDFs in the high gluon density region and to help pin down the gluonic contribution to the nucleon spin. This is demonstrated in Fig.~\ref{fig:inclusive_g1-main}, which shows the impact of early science data on $g_{1}^p$ and $g_{1}^n$ uncertainties from global QCD analyses performed by the JAM collaboration~\cite{6fn9-1wqb}. Taking advantage of the extended low-$x$ reach of the EIC, the behavior of $g_{1}$ as $x \to 0$ is expected to be much better understood, and the uncertainties associated with extrapolating values of $g_{1}$ to lower values of $x$ are significantly reduced. 
The double spin asymmetry measurements are expected to be statistically dominated. The projection study includes a total of 3.4~\% uncorrelated and 2.9~\% correlated uncertainties to account for systematic effects. These uncertainties are much smaller than the statistical uncertainty projected by the simulations.
Eventually, with the full running capacity of the EIC, inclusive measurements will provide even further constraints on the polarized SFs and polarized PDFs.
In addition to these improvements, by measuring $g_1^p$ and $g_1^n$ over a new, larger kinematic phase space, the EIC will test the Bjorken sum rule with higher precision than before, and can simultaneously extract $\alpha_S$ through the $Q^2$ dependence of the sum rule. This will provide a complementary determination of $\alpha_S$ to the aforementioned extraction (Fig.~\ref{fig:inclusive-alphaS}).

\paragraph{Summary}
Inclusive DIS measurements will constitute one of the earliest and most impactful components of the ePIC early science program. Relying primarily on precise reconstruction of the scattered electron and the hadronic final state, these measurements can be performed without advanced PID capabilities or fully commissioned far-forward instrumentation, making them ideally suited to the initial years of EIC operation.

The projected luminosities and broad kinematic coverage available during early science running will enable precise measurements of inclusive reduced cross sections in both $e+p$ and $e+\mathrm{A}$ collisions. These data will provide important constraints on proton and nuclear SFs, improve determinations of proton and nuclear PDFs, and significantly enhance our knowledge of the poorly constrained gluon distributions at low $x$. In addition, inclusive DIS measurements offer sensitivity to the strong coupling constant, providing a competitive and complementary determination within a global QCD analysis framework.

The availability of polarized electron, proton, and $^3$He beams will further enable precision measurements of spin-dependent SFs and polarized PDFs. By extending polarized DIS studies into the previously unexplored low-$x$ region, the EIC will substantially improve constraints on the quark and gluon contributions to nucleon spin already in the early years of running.

Overall, the inclusive DIS program will deliver a rich set of high-impact early science measurements while establishing the experimental and analysis foundations for the broader ePIC physics program.

\section{Semi-inclusive measurements}
\label{secmain:sidis}

\begin{figure}[htbp]
    \centering
    \begin{tikzpicture}[scale=0.9, line width=0.8pt]

\tikzset{
  fermion/.style={
    postaction={decorate},
    decoration={markings, mark=at position 0.5 with {\arrow{>}}}
  },
   doublefermion/.style={
    double,
    double distance=1.5pt,
    postaction={decorate},
    decoration={markings, mark=at position 0.5 with {\arrow[scale=0.65]{>}}}
  },
  photon/.style={
    decorate,
    decoration={snake, amplitude=2pt, segment length=6pt}
  }
}

\coordinate (A) at (-0.2,1.55);     
\coordinate (B) at (0.95,-0.45);    
\coordinate (PDF) at (-0.95,-2.25); 
\coordinate (FF)  at (2.35,-0.45);  

\def\rpdfx{1.05}
\def\rpdfy{0.70}
\def\rff{0.75}

\coordinate (eL) at (-5.2,2.95);
\coordinate (eR) at ( 4.9,2.95);
\coordinate (pL) at (-5.2,-2.25);

\coordinate (piOut) at (4.9,0.75);

\def\xX{4.9}

\coordinate (xFFu) at (\xX,-0.80);
\coordinate (xFFd) at (\xX,-1.20);
\coordinate (xPDFu) at (\xX,-1.65);
\coordinate (xPDFm) at (\xX,-2.25);
\coordinate (xPDFd) at (\xX,-2.85);

\coordinate (PDFW)  at ($(PDF)+(-\rpdfx,0)$);
\coordinate (PDFNE) at ($(PDF)+(0.60,0.56)$);
\coordinate (PDFE)  at ($(PDF)+(\rpdfx,0)$);
\coordinate (PDFEu) at ($(PDF)+(1.00,0.20)$);
\coordinate (PDFEd) at ($(PDF)+(1.00,-0.20)$);

\coordinate (FFW)   at ($(FF)+(-\rff,0)$);
\coordinate (FFNE)  at ($(FF)+(0.66,0.34)$);
\coordinate (FFEup) at ($(FF)+(0.72,-0.12)$);
\coordinate (FFEdn) at ($(FF)+(0.70,-0.28)$);

\draw[fermion] (eL) -- (A);
\draw[fermion] (A) -- (eR);

\node at (-2.85,2.82) {$e(k)$};
\node at ( 2.65,2.82) {$e'(k')$};

\draw[photon] (A) -- (B);
\node[right] at ($(A)!0.50!(B)+(0.16,0.12)$) {$\gamma^{*}(q)$};

\draw ($(pL)+(0,0.10)$) -- ($(PDFW)+(0,0.10)$);
\draw[fermion] (pL) -- (PDFW);
\draw ($(pL)+(0,-0.10)$) -- ($(PDFW)+(0,-0.10)$);

\node at (-3.65,-1.82) {$P(p)$};

\draw (PDFNE) -- (B);
\node[right=-23pt, yshift=-0pt] at ($(PDFNE)!0.50!(B)$) {$xp$};

\draw (B) -- (FFW);

\draw[doublefermion] (FFNE) -- (piOut);
\node[above, xshift=-10pt, yshift=2pt] at ($(FFNE)!0.58!(piOut)$) {$h(P_h)$};

\draw (FFEup) -- (xFFu);
\draw (FFEdn) -- (xFFd);

\draw (PDFEu) -- (xPDFu);
\draw (PDFE)  -- (xPDFm);
\draw (PDFEd) -- (xPDFd);

\draw
  (5.18,-0.55)
  .. controls (5.48,-1.20) and (5.48,-2.45) ..
  (5.18,-3.10);

\node[right] at (5.42,-1.825) {$X$};

\draw[fill=white] (PDF) ellipse ({\rpdfx} and {\rpdfy});
\node at (PDF) {\shortstack{(TMD)\\PDF}};

\draw[fill=white] (FF) circle (\rff);
\node[yshift=-3pt] at (FF) {\shortstack{(TMD)\\FF}};

\end{tikzpicture} 
    \caption{Schematic of a SIDIS reaction in the single-photon-exchange approximation. At least one hadron is detected in the final state.}
\label{fig:SIDIS}
\end{figure}
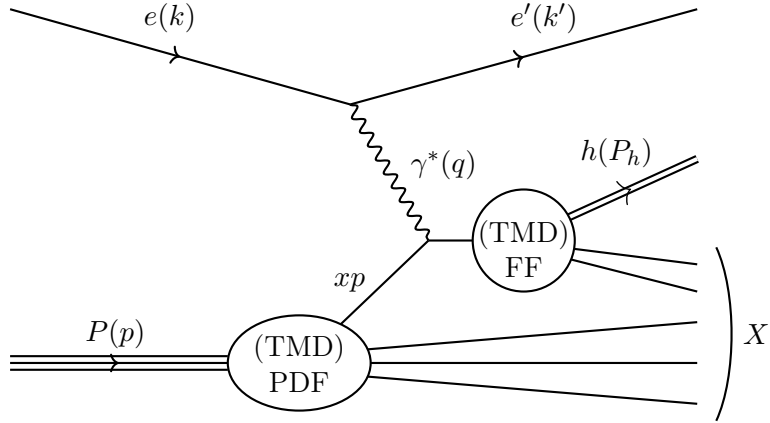

SIDIS processes, where at least one final-state hadron is detected in coincidence with the scattered lepton, as depicted in Fig.~\ref{fig:SIDIS}, provide a significantly richer landscape of information than inclusive DIS alone.
They offer improved sensitivity to the flavor of the struck quark. Moreover, while inclusive DIS primarily provides a one-dimensional (longitudinal) projection of the nucleon structure via collinear PDFs, SIDIS enables a more complete dynamical picture of the nucleon by granting sensitivity to transverse degrees of freedom. A scientific strength of the SIDIS program lies in its ability to translate specific experimental observables into a multidimensional map of the nucleon.
This is achieved through transverse-momentum-dependent (TMD) PDFs, which encode the three-dimensional momentum structure of the nucleon, in conjunction with TMD fragmentation functions (FFs), which describe the hadronization of a struck parton into the observed final-state hadron and carry sensitivity to its flavor, spin, and momentum.

An important quantity entering the description of SIDIS is the fractional momentum $z=(p \cdot P_h) / (p \cdot q)$, where $P_h$ and $p$ are the four-momenta of the final-state hadron and the incoming nucleon, respectively, and $q$ is the momentum transfer. 
For TMD physics, hadrons at intermediate-to-higher $z$ are of interest, as they exhibit stronger correlations with the flavor and spin of the fragmenting parton. Crucially, the species of the detected hadron must be correctly identified to exploit this flavor sensitivity, a capability that was limited at HERA but is a central design feature of ePIC.

Each SIDIS observable probes different aspects of QCD dynamics. Table~\ref{tab:sidis_capabilities} provides an overview of how SIDIS  measurements evolve with the early running conditions and the capabilities that enable them.
The EIC marks a transition for SIDIS from a qualitative era to a high-precision quantitative era. While previous experiments (like HERMES at DESY \cite{Airapetian:2004tw}, COMPASS at CERN~\cite{COMPASS:2008isr,COMPASS:2014bze}, and the 6/12\,GeV program at JLab~\cite{JeffersonLabHallA:2011ayy}) established the existence of three-dimensional effects such as the Sivers~\cite{Sivers:1989cc} and Collins~\cite{Collins:1992kk} asymmetries, the EIC provides several capabilities that were previously technically or kinematically inaccessible: 
\begin{itemize}
\item \textbf{Access to sea quarks and gluons via lower-$x$ SIDIS measurements.} The higher collision energies at the EIC extend the kinematic reach of previous fixed-target SIDIS experiments from the ``valence quark'' region ($x > 0.05$) to much smaller $x$ values, enabling the first three-dimensional momentum mapping of gluons and the quark ``sea''. 
\item \textbf{Polarized light ion beams.} A polarized $^3$He beam provides effective access to neutron spin structure through a source with nearly 90~\% polarization, free from the dilution effects inherent to fixed-target configurations.
\item \textbf{Unpolarized ion beams.} High-statistics measurements with a range of unpolarized beam species, from proton to uranium, enable systematic studies of nuclear PDFs and FFs individually for each ion species.
\item \textbf{Gluon-specific observables}. While single-hadron production in SIDIS provides sensitivity primarily to quark distributions, the back-to-back production of two hadrons (typically two pions) provides direct sensitivity to gluon distributions and, in collisions with heavy ions, to potential gluon saturation effects. The $4\pi$ hermeticity of the ePIC detector makes it well-suited for measurements of this kind.
\end{itemize}

\begin{table}[htbp]
\centering
\centering
\footnotesize
\renewcommand{\arraystretch}{1.8}
\begin{tabular}{L{0.15\textwidth} L{0.40\textwidth} L{0.35\textwidth}}
\hline
\textbf{Configuration} & \textbf{Representative measurements} & \textbf{Enabling capabilities} \\ 
\hline

$e+p$ running  & Multiplicities for FFs & Hadron ID ($\pi, K, p$) and $z$ coverage\\ 

$e+p$ running  & Multifold differential cross sections for unpolarized TMD PDFs & Needs previous capabilities + higher collected luminosity than in the  previous case \\ 

$e+p$ running with polarization & Azimuthal single-spin asymmetries for Sivers TMD PDF and Collins TMD FF & Needs previous capabilities + transversely polarized proton beam \\ 

$e+p$ running with polarization  & Double spin asymmetry measurements for sea quark helicities & Needs previous capabilities + longitudinally polarized proton and electron beams \\ 

$e+^3$He beam with polarization & TMD and helicity measurements on the neutron & Spectator tagging capabilities + polarized hadron and electron beams\\ 

$e+\mathrm{A}$ running with heavy nuclei & Di-hadron correlations for the onset of gluon saturation and non-linear QCD effects & Large acceptance (4$\pi$) tracking; and previous capabilities without the need for beam polarization\\

\hline
\end{tabular}
\caption{Overview of SIDIS measurements in the ePIC early science program grouped by the approximate running conditions and capabilities that enable them.}
\label{tab:sidis_capabilities}
\end{table}

The early years of physics running at ePIC already offer unique opportunities to advance our knowledge on several of the science pillars identified in the NAS report. 

\paragraph{Unpolarized TMD PDFs}
Even with an unpolarized nucleon beam, ePIC can provide flavor-separated TMD PDFs across a wide phase space in $x$ and $Q^2$, covering transverse momenta from the non-perturbative TMD PDF dominated region into the perturbative regime.
Measurements of unpolarized TMD PDFs are a primary requirement of the EIC science program, as they provide the essential baseline for extracting all other polarized TMD observables. Spin asymmetries used to isolate different polarized TMDs, such as the Sivers or Collins effects, are ratios in which the unpolarized TMD cross section appears in the denominator. Consequently, uncertainties in the unpolarized TMD baseline propagate directly into the extracted polarized functions, making high-precision characterization of the unpolarized distributions a prerequisite for quantitative spin-physics results.

The extraction of unpolarized TMD PDFs is intricately linked to our understanding of FFs. In SIDIS, the observed transverse momentum of the final hadron ($P_{h\perp}$) is a convolution of the intrinsic transverse momentum of the parton inside the nucleon and the transverse momentum generated during hadronization, the latter being encoded in the TMD FF. The ability of ePIC to perform measurements on both proton and nuclear targets will, for the first time, enable a simultaneous, high-precision characterization of both intrinsic partonic and hadronization contributions to $P_{h\perp}$.

In recent years, the extraction of unpolarized TMD PDFs and TMD FFs has reached a high level of accuracy, reaching N$^{4}$LL and N$^{3}$LL precision for, respectively, Drell-Yan and SIDIS production~\cite{MAPTMD22,Bacchetta:2024qre,Moos:2025sal,Moos:2023yfa}.
The MAPTMD24 extraction~\cite{Bacchetta:2024qre} incorporates a flavor-dependent parameterization, allowing $u$, $d$, and sea quarks to possess different intrinsic transverse momentum widths. Despite this theoretical progress, current global fits suffer from a lack of high-precision data in the low-$x$ and high-$Q^2$ regions. Most existing data come from low-energy fixed-target experiments (HERMES, COMPASS) and high-energy $Z$-boson production at the LHC, leaving a significant kinematic gap that limits our ability to test TMD evolution and flavor-dependent dynamics. Early ePIC data will bridge this gap in $x -Q^2$, providing substantial coverage during the early running stages.

\begin{figure} [htb!]
   \centering
\includegraphics[width=0.7\textwidth]{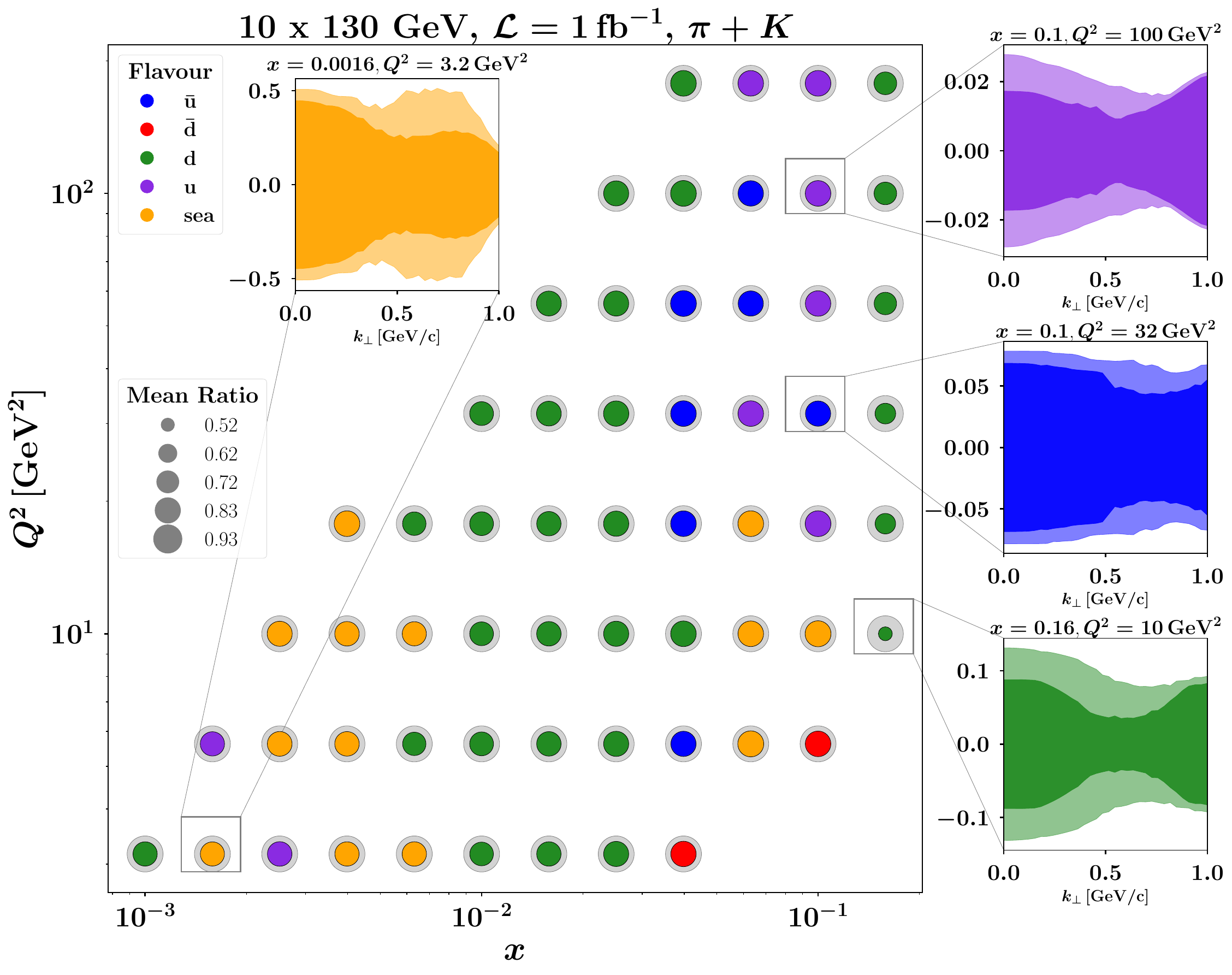}

\caption{ Impact of early SIDIS data on TMD PDFs by MAPTMD24~\cite{Bacchetta:2024qre} from 10\,GeV\,$\times$\,130\,GeV collisions and a collected luminosity of 1~fb$^{-1}$. Inner (outer) circles represent the maximum uncertainty averaged over all $k_T$ with (without) ePIC data, whereas colors indicate the most impacted quark flavors. Current uncertainties (lighter/gray shades) are compared to the expected improvements (darker shades) after incorporating ePIC pseudo-data. 
}
    \label{fig:MAP24}
    \end{figure}

An impact study was performed using ePIC pseudo-data based on the configuration expected during the early running period, for an integrated luminosity of 1~fb$^{-1}$. 
A 3.5~\% 
bin-by-bin systematic uncertainty and a 1.5~\% uncertainty from the luminosity determination are assigned to the pseudo-data. As shown in Fig.~\ref{fig:MAP24}, early data from ePIC are already expected to significantly reduce extraction uncertainties for both valence and sea quarks, establishing the essential baseline required for the full EIC physics program. A complementary study~\cite{Seidl:2022dmh} based on \cite{Moos:2023yfa} demonstrates how precise ePIC data will help discriminate between differing predictions from current phenomenological analyses, particularly regarding the poorly constrained non-perturbative component of the Collins-Soper kernel. This kernel governs how the transverse momentum distribution of a quark evolves as the hard scale $Q^2$ increases. Its precise determination is essential for robust TMD evolution, providing a more reliable baseline for future polarized and unpolarized TMD measurements.

\paragraph{Transverse-momentum-dependent observables: Sivers and Collins effects}

A primary goal of the SIDIS program at ePIC is to characterize the multidimensional partonic structure of the nucleon in momentum space through the extraction of TMD PDFs. Two flagship measurements in this program are the Sivers and Collins asymmetries, which have provided the first evidence for spin-orbit correlations and the transverse spin structure of the nucleon, and whose precise determination remains one of the central objectives of the EIC.

The Sivers function $f_{1T}^\perp$~\cite{Sivers:1989cc, Sivers:1990fh} is a leading-twist TMD PDF describing a fundamental correlation between the nucleon’s transverse spin and the intrinsic transverse momentum $k_T$ of its constituent quarks and gluons.
A non-zero Sivers effect requires a non-zero partonic orbital angular momentum. 
Its extraction at ePIC requires precise measurements of the Single-Spin Asymmetry $A_{UT}^{\sin(\phi_h - \phi_s)}$ in SIDIS, with a transversely polarized proton beam, where the azimuthal angles of the hadron transverse momentum and of the nucleon spin component are defined relative to the lepton scattering plane.
Extractions of the Sivers function have been performed in~\cite{Portela:2026wwn,Cammarota:2020qcw,Bury:2021sue,Bacchetta:2020gko,Echevarria:2020hpy}.

\begin{figure}[htb]
	\centering
	\includegraphics[width=0.50\textwidth,valign=c]{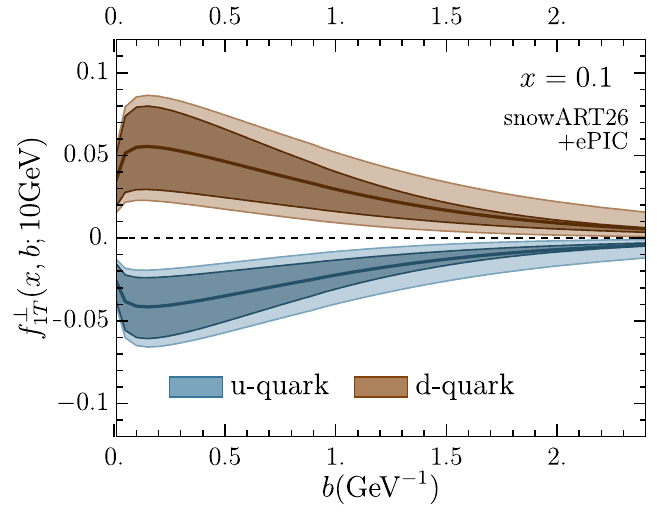}
	\hfill
	\includegraphics[width=0.45\textwidth,valign=c]{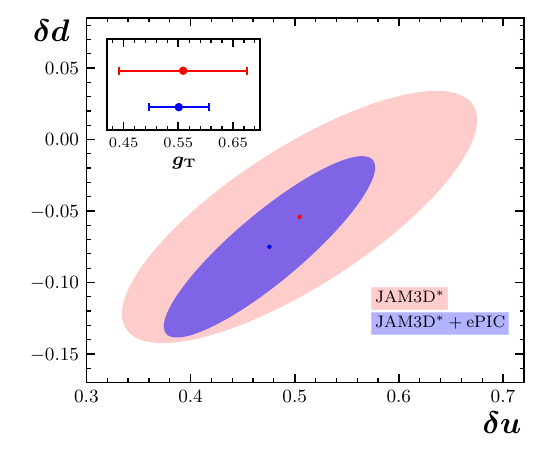}
	\caption{Left: Expected uncertainties of the $u$- and $d$-quark Sivers function as a function of the $b$ parameter conjugated to the quark momentum $k_T$ for $x=0.1$, based on \cite{Portela:2026wwn}. The light shaded uncertainties represent the current uncertainties while the dark shaded uncertainties represent the impact including the early EIC data. Right: Expected uncertainties on the up and down quark tensor charges when including the early EIC data based on \cite{Gamberg:2022kdb,Cocuzza:2023vqs}. The pink shaded area represents the current uncertainties while the purple region represents the uncertainty when including the early EIC data from e+p and e+$^3$He collisions.}
	\label{fig:SIDIS_SSA_maindoc}
\end{figure}

The Collins function $H_1^\perp$~\cite{Collins:1992kk} is a time-reversal-odd FF that describes the mechanism by which a transversely polarized quark fragments into an unpolarized final-state hadron.
The Collins effect serves as the ``analyzer'' needed to measure the nucleon’s transversity distribution $h_{1}$, which quantifies the net transverse polarization of quarks inside a transversely polarized nucleon. Transversity is one of three fundamental collinear PDFs, alongside unpolarized and helicity distributions, but remains the least well-known due to its chiral-odd nature. It can be accessed via the study of single-hadron production, through its coupling to the Collins TMD FF, and via the study of di-hadron production, through its coupling to the dihadron FF.

Because the quark’s spin is not directly observable, the Collins FF serves as the spin analyzer, transforming the spin information into a measurable azimuthal modulation in the distribution of final-state hadrons. An important connection to transversity is through its moments, the tensor charges, which can also be obtained from Lattice QCD~\cite{FlavourLatticeAveragingGroupFLAG:2024oxs}. Possible differences between the two  could hint at new physics that manifests itself via a tensor interaction~\cite{Courtoy:2015haa}. 

The ePIC early science program will deliver unprecedented multi-dimensionality in these measurements at intermediate-to-low $x$. By simultaneously binning in $z$, $P_{h\perp}$, $x$, and $Q^2$, it becomes possible to disentangle contributions from the initial-state TMD PDFs and the final-state fragmentation, and to map their interplay across a wide kinematic range. Collins asymmetries are also discussed in the context of fragmentation within jets in Sec.~\ref{secmain:hf}.

The availability of transversely polarized beams will enable ePIC to begin this three-dimensional imaging program. Even with the luminosities achieved in the early running period, the statistical precision for pion and kaon asymmetries will be sufficient to significantly reduce existing uncertainties in global TMD PDF fits.
An impact study has been prepared using the $e+p$~10\,GeV\,$\times$\,130\,GeV and 10\,$\times$\,250\,GeV data sets, as well as tagged 10\,GeV\,$\times$\,166\,GeV~$e+^3\text{He}$ data sets, scaled to 1.25, 0.5 and 0.75 fb$^{-1}$, respectively. A conservative systematic uncertainty, taken as the bin-dependent  difference between the generated and reconstructed asymmetries, and a 1.5~\% scale uncertainty from the polarization determination are considered. The resulting pseudo-data, which achieve sub-percent-level statistical precision over a large fraction of the available phase space, were then analyzed following the methodology of previous impact studies \cite{Gamberg:2021lgx,Seidl:2022dhh}. 
 Figure~\ref{fig:SIDIS_SSA_maindoc} shows the expected precision with which the Sivers functions and tensor charges can be extracted at ePIC during the early running period. As can be seen, these early data are expected to advance studies beyond the one-dimensional longitudinal picture of nucleon structure, contributing to a more comprehensive dynamical understanding of how mass, spin, and confinement emerge from QCD.
 
\paragraph{Helicity measurements}
While parton helicities and their contributions to the nucleon spin have been discussed in Sec.~\ref{secmain:inclusive}, the flavor sensitivity of SIDIS is particularly valuable for extracting the helicities of sea quark flavors. 
Measurements of W-boson production at RHIC have established that the light quark sea is polarized and asymmetric between anti-up and anti-down flavors \cite{PHENIX:2015ade,PHENIX:2018wuz,STAR:2018fty}. These investigations need to be extended to lower momentum fractions to evaluate their total contributions to the spin sum rule through their integrals. A special role in these measurements falls to the strange quark helicities. As of now, HERMES~\cite{HERMES:2004zsh} and COMPASS \cite{COMPASS:2010hwr} have not seen any indication of a non-zero polarization, while global fits that use the hyperon decay constant and SU(3)$_\mathrm{F}$ find a negative total contribution. 
Impact studies have been performed for the early running stage of ePIC using $e+p$ data sets at 10\,GeV\,$\times$\,130\,GeV and 10\,GeV\,$\times$\,250\,GeV, scaled to accumulated luminosities of 0.5 and 1.25 fb$^{-1}$, respectively. A bin-dependent systematic uncertainty, taken as the difference between the generated and reconstructed asymmetries, and a scale uncertainty of 1.5~\% per beam polarization are included. 

\begin{figure}[htb]
    \centering
    \includegraphics[width=0.95\linewidth]
   {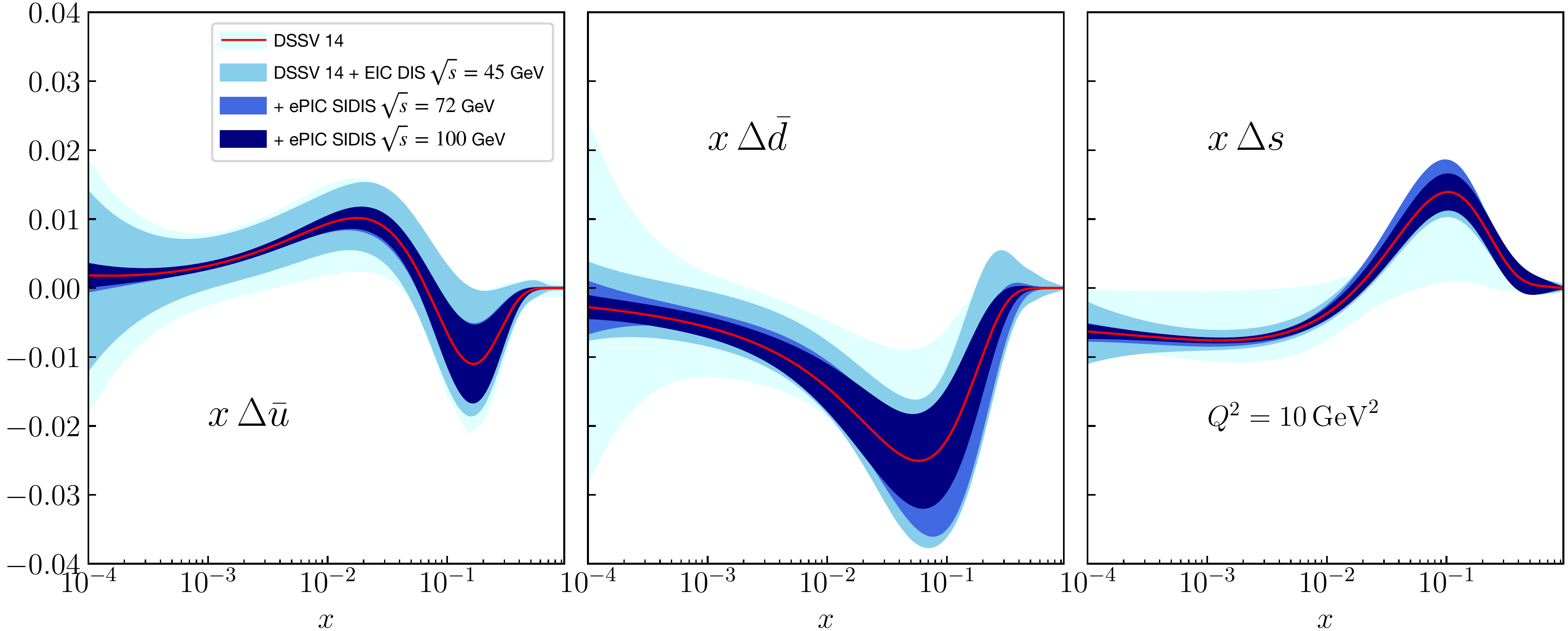}
    
    \caption{Expected impact of the semi-inclusive double-spin asymmetry measurements on the sea-quark helicities as a function of $x$. The light blue shaded regions represent the current knowledge, based on \cite{deFlorian:2014yva}, and the slightly darker shaded area (DSSV 14 + EIC DIS $\sqrt{s}=45$ GeV) includes a set of 10 fb$^{-1}$ of pseudo-EIC inclusive DIS data at a center-of-mass energy of $\sqrt{45}$~GeV, added in order to perform the reweighting with the SIDIS data. The darker shaded regions represent the uncertainties when including in addition  separately early ePIC data at the two different collision energies of 10\,GeV $\times$ 130\,GeV and 10\,GeV $\times$ 250\,GeV.}
\label{fig:ALL_SIDIS}
\end{figure}
Figure~\ref{fig:ALL_SIDIS} shows the expected impact from the early EIC data on the sea quark helicities  based on the analysis performed by \cite{Borsa:2020lsz}.
As can be seen from the figure, an extensive kinematic coverage can be achieved in which the uncertainties of all helicities can be determined precisely. These early data will be particularly relevant for the determination of the strange quark helicity distributions. Not shown here are expected impacts for the valence quark helicities that are already better constrained.

\paragraph{Nuclear modifications and fragmentation}

A central goal of the EIC is to understand how the dense nuclear environment modifies the structure and interactions of partons relative to those in a free proton or neutron. One of the most direct approaches is to study modifications of the hadronic final state in the presence of a nuclear medium, a phenomenon that remains poorly understood in cold nuclear matter physics~\cite{Qiu:2019sfj,JET:2013cls,Qin:2015srf,Blaizot:2015lma,Majumder:2010qh,KunnawalkamElayavalli:2017hxo,He:2018xjv}.

At hadron-hadron colliders such as the LHC and RHIC, nPDFs can be studied in proton-nucleus collisions~\cite{Eskola:2022vaf,AbdulKhalek:2022ihb}. 
SIDIS processes in electron+ion collisions provide a particularly clean avenue for exploring nPDFs and fragmentation in the nuclear medium. Measuring hadron multiplicities enhances the sensitivity to nuclear fragmentation functions (nFFs), since the dependence on nPDFs largely cancels in the ratio. These studies were pioneered at HERMES~\cite{HERMES:2007plz}, which, despite a restricted kinematic coverage, enabled the first direct extraction of nFFs~\cite{Sassot:2009sh}. No comparable measurements have been performed since, leaving nFFs poorly constrained to this day. 
The broader kinematic coverage of the EIC, combined with the statistical precision expected from early ePIC running, will significantly advance the characterization of nFFs and the corresponding nPDFs.

\begin{figure} [htbp]
   \centering
\includegraphics[width=0.49\textwidth]{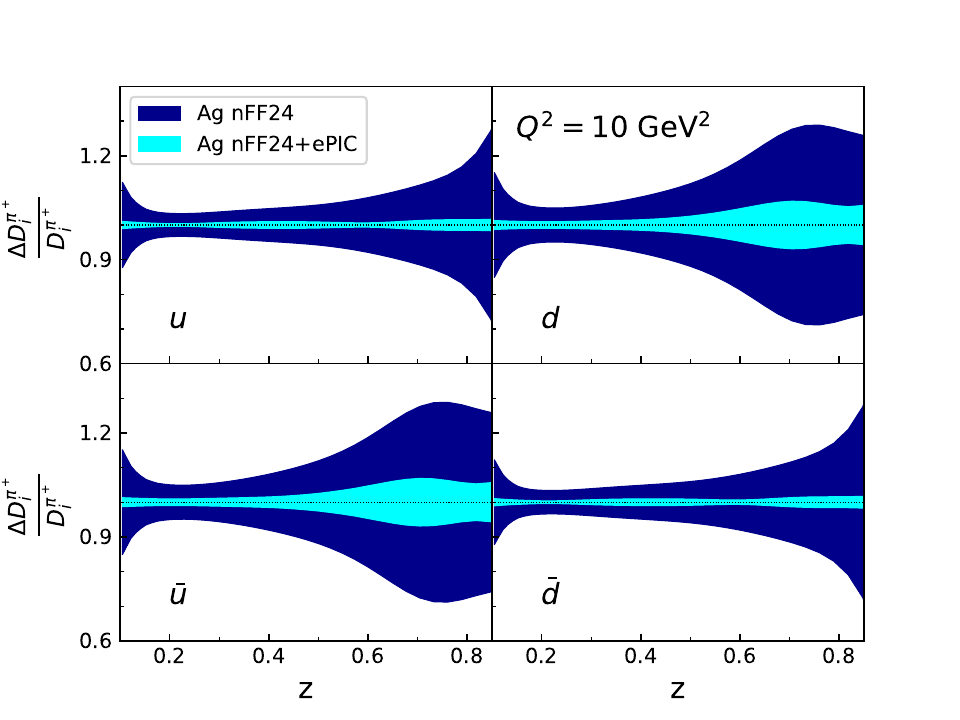}
\includegraphics[width=0.49\textwidth]
{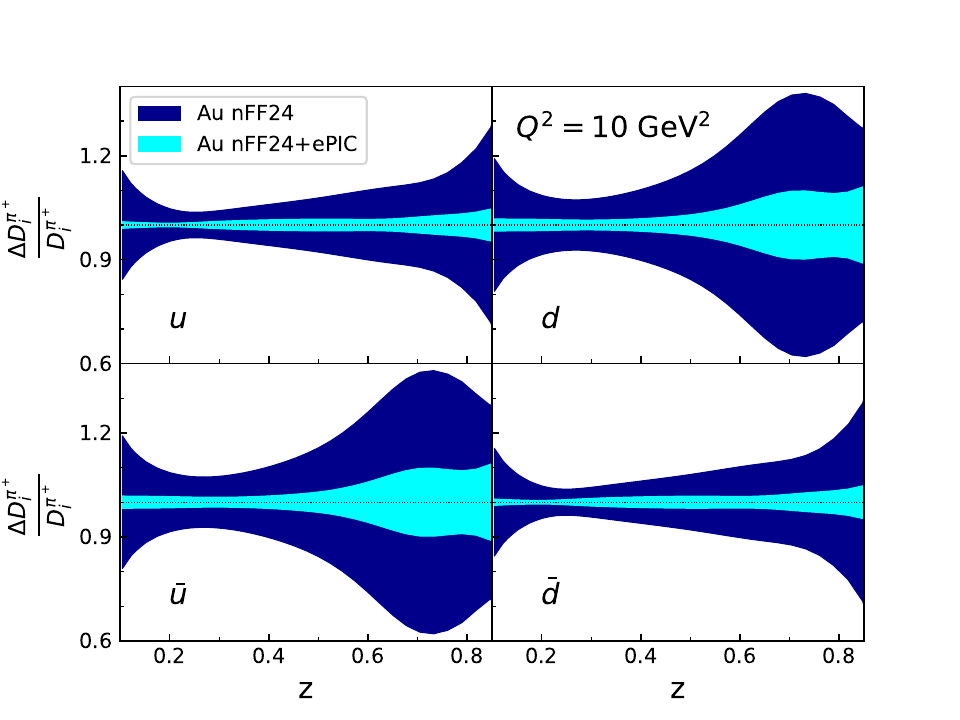}
\caption{Current relative uncertainties on nFFs~\cite{Doradau:2024wli} (dark-blue) and relative nFF uncertainties from the reweighting study performed here, as described in the text (cyan). Left: for Ag. Right: for Au.}
    \label{fig:SIDIS_FF}
\end{figure}

An impact study for the determination of nFFs based on simulated early ePIC data was performed using a reweighting technique and a set of newly extracted nFFs~\cite{Doradau:2024wli}, where the latter are based on the A-dependent parameterization of available nuclear data. The present study considers the early
$e+\text{Ag}$ and $e+\text{Au}$ running periods at 10\,GeV\,$\times$\,100\,GeV. Because the study uses the reweighting technique and because only a very limited amount of data exists  for the study of nFFs, it was necessary to restrict the simulation sample drastically. Here, 5~\% of the bins from the simulated data corresponding to an integrated luminosity of 1~fb$^{-1}$ are used. The result of the study is shown in Fig.~\ref{fig:SIDIS_FF}, where the relative uncertainty on the positive pion nFF is shown without (dark-blue bands) and with (cyan bands) the inclusion of the ePIC data considered here. As can be seen, a large reduction in uncertainty is expected with the addition of the early run ePIC data. 

Vacuum FFs can also be improved significantly from the $e+p$ collision data alone. The combination of different final-state hadron types enables a substantial reduction in the uncertainties on the FFs for all flavors and over a wide range of momentum fraction $z$.

\begin{figure} [htbp]
   \centering
\includegraphics[width=0.49\textwidth]{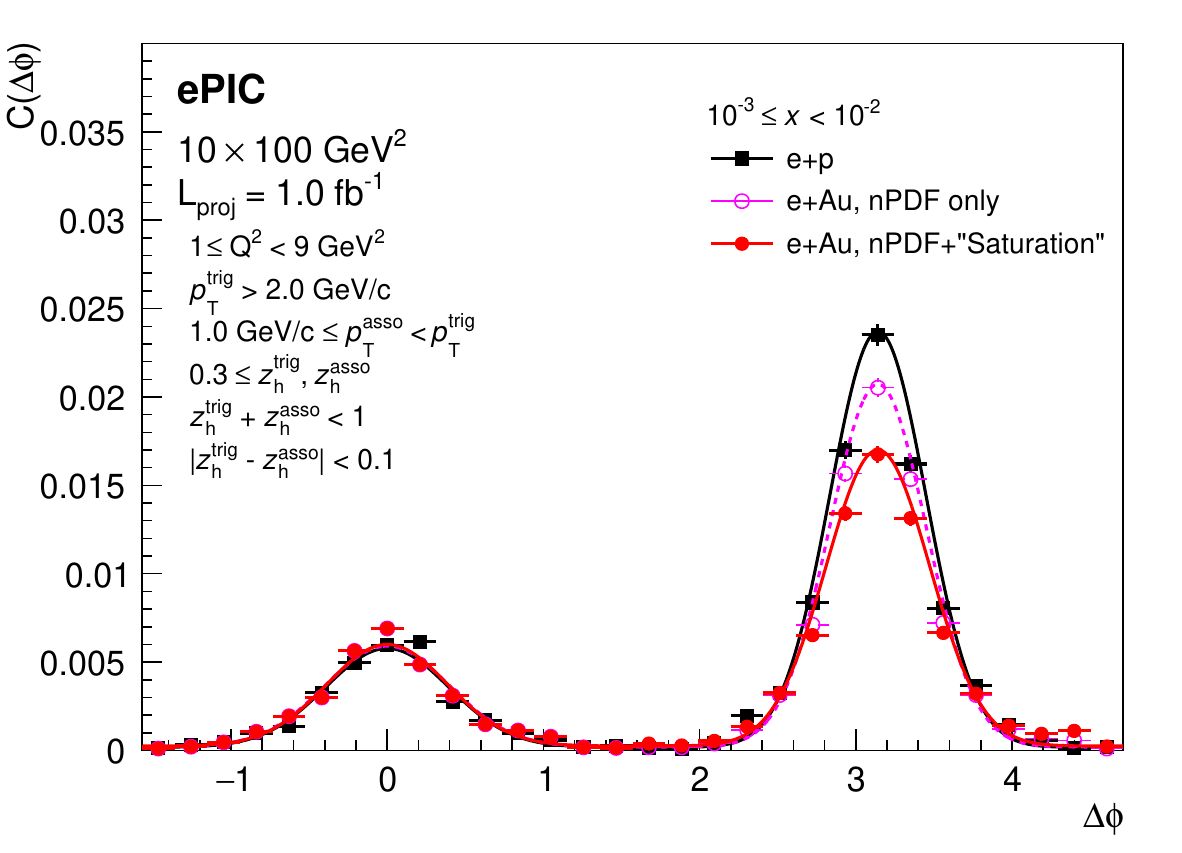}
\includegraphics[width=0.49\textwidth]{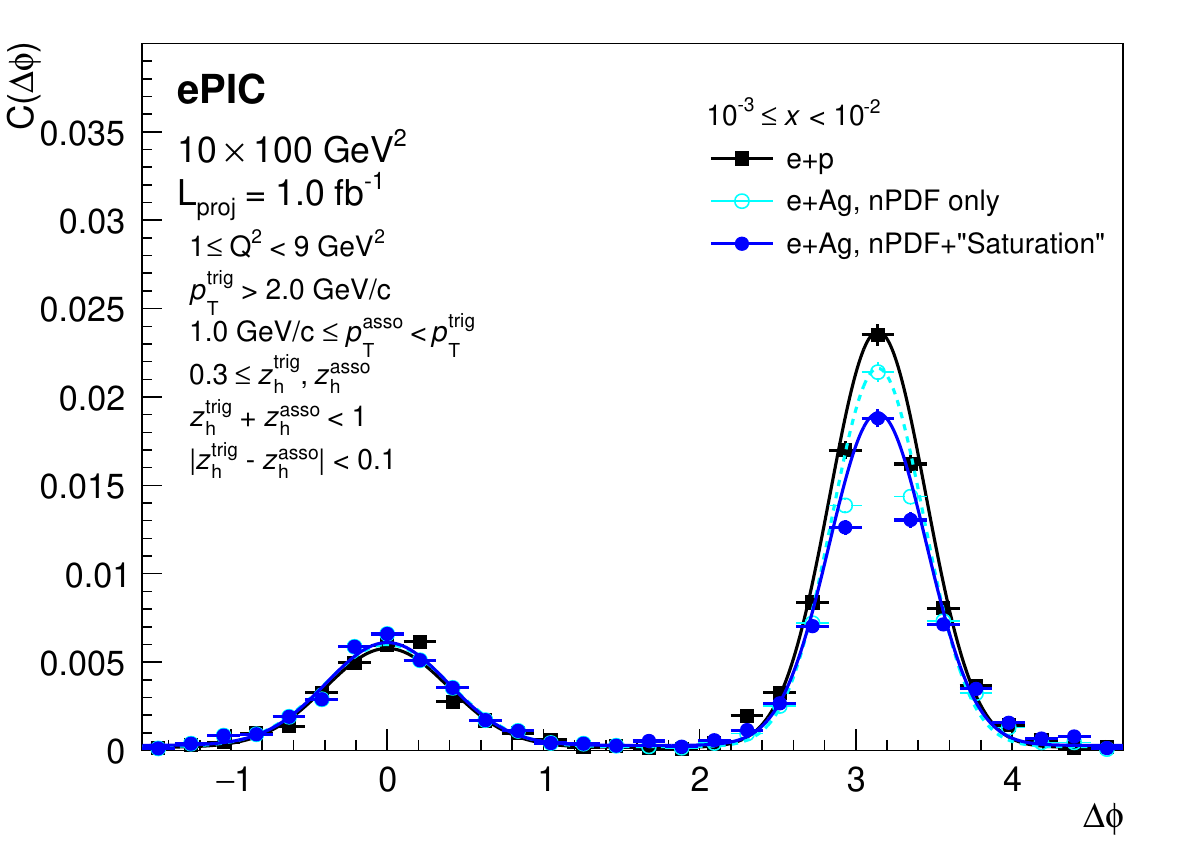}
\caption{Correlation functions of $e+p$ (black)
and $e+\mathrm{A}$ without “saturation weight”  (light color) and with “saturation weight”  (dark color) in the range $10^{-3} \le x < 10^{-2}$. Left: $e+\mathrm{Au}$. Right: $e+\mathrm{Ag}$. While this figure is not based on full simulations, no significant changes are expected due to the high efficiency and resolution of the ePIC tracking system.}
    \label{fig:sat2}
\end{figure}

\paragraph{Transition into gluon saturation regime}
The production of two back-to-back, high-transverse-momentum, charged hadrons in SIDIS processes from electron-nucleus collisions provides sensitivity to non-linear QCD effects arising from the high gluon density in the nuclear medium. These effects can be probed through the measurement of azimuthal angular correlations between near- and away-side peaks in the di-hadron distributions.
In the presence of saturation, the away-side peak becomes decorrelated, and the yield of back-to-back pairs is reduced in comparison to $e+p$ collisions at the same energy~\cite{Zheng:2014vka}. The use of di-hadrons instead of jets extends the kinematic reach to lower momentum fractions, where nonlinear effects due to the onset of gluon saturation are expected to be more pronounced.
An impact study was performed using $e+p$, $e+\mathrm{Ag}$ and $e+\mathrm{Au}$ data sets at 10~GeV $\times$ 100~GeV, each corresponding to an integrated luminosity of 1~fb$^{-1}$. Nuclear modifications are taken into account through the nPDF set EPPS16~\cite{Eskola:2016oht}, while an additional event-level reweighting includes non-linear QCD effects. 
Figure~\ref{fig:sat2} presents the results for $e+\mathrm{Au}$ and $e+\mathrm{Ag}$, respectively, compared to the $e+p$ baseline for $10^{-3} \le x < 10^{-2}$. A clear away-side peak suppression is visible, demonstrating that non-linear QCD effects can be probed through di-hadron correlation analysis within the expected luminosities of the early stages of EIC running.

\paragraph{Summary}
The early semi-inclusive program at ePIC marks the transition from a one-dimensional picture of nucleon structure to a comprehensive three-dimensional tomographic mapping of its internal dynamics. 
By measuring hadron multiplicities and TMD observables, ePIC will provide high-precision constraints on fragmentation functions and unpolarized TMD PDFs, establishing the essential baseline for all subsequent spin-physics extractions.

With the introduction of polarized beams, ePIC will substantially advance our understanding of spin contributions by sea quarks to the spin sum rule. The measurements of the Sivers and Collins asymmetries will access spin-orbit and spin-spin correlations and provide insights into the tensor charges of the nucleon.  Early investigations into di-hadron correlations in electron-nucleus collisions will probe the high-density gluonic regime, seeking the first signatures of gluon saturation. Collectively, these early measurements position ePIC to deliver world-leading insights into the emergence of nucleon mass, spin, and confinement from the earliest years of operation.


\section{Exclusive, Diffractive and Tagged Measurements}
\label{secmain:exclusive}

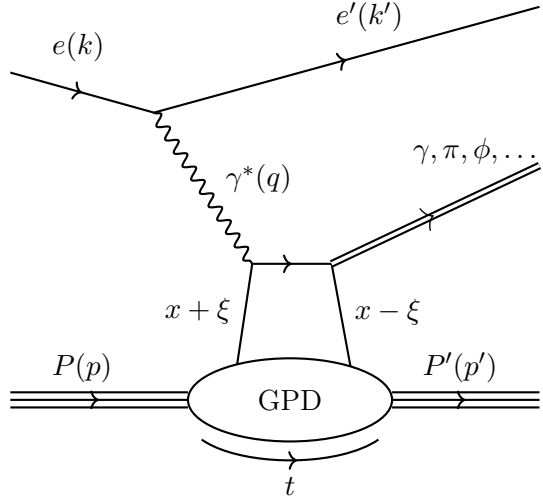
\begin{figure}[tbh!]
\centering
\begin{tikzpicture}[scale=1.0, line width=0.8pt]

\tikzset{
  fermion/.style={
    postaction={decorate},
    decoration={markings, mark=at position 0.5 with {\arrow{>}}}
  },
  arcwitharrow/.style={
    postaction={decorate},
    decoration={markings, mark=at position 0.54 with {\arrow{>}}}
  },
  photon/.style={
    decorate,
    decoration={snake, amplitude=2pt, segment length=6pt}
  },
  doublefermion/.style={
    double,
    double distance=1.3pt,
    postaction={decorate},
    decoration={markings, mark=at position 0.5 with {\arrow[scale=0.65]{>}}}
  }
}

\coordinate (A) at (-0.2,1.55);     
\coordinate (B) at (1.10,-0.45);    
\coordinate (B2) at (2.15,-0.45);   

\coordinate (C) at (1.60,-2.25);    
\def\a{1.35}                        
\def\b{0.55}                        

\coordinate (eL) at (-2.10,2.08);   
\coordinate (eR) at ( 4.90,2.95);   
\coordinate (pL) at (-2.10,-2.25);  
\coordinate (pR) at ( 4.90,-2.25);  

\coordinate (out) at (4.90,0.85);

\coordinate (GL) at ($(C)+(-\a,0)$);
\coordinate (GR) at ($(C)+(\a,0)$);

\coordinate (L) at ($(C)+(-0.70,0.47)$);
\coordinate (R) at ($(C)+( 0.80,0.44)$);

\draw[fermion] (eL) -- (A);
\draw[fermion] (A) -- (eR);

\node at (-1.20,2.42) {$e(k)$};
\node at ( 2.65,2.82) {$e'(k')$};

\draw[photon] (A) -- (B);
\node[right] at ($(A)!0.50!(B)+(0.16,0.12)$) {$\gamma^{*}(q)$};

\draw[fermion] (B) -- (B2);

\draw (B)  -- (L);
\draw (B2) -- (R);

\node at ($(B)!0.50!(L)+(-0.62,+0.05)$) {$x+\xi$};
\node at ($(B2)!0.50!(R)+(0.62,+0.05)$) {$x-\xi$};

\draw ($(pL)+(0,0.10)$) -- ($(GL)+(0,0.10)$);
\draw[fermion] (pL) -- (GL);
\draw ($(pL)+(0,-0.10)$) -- ($(GL)+(0,-0.10)$);

\node at (-1.15,-1.82) {$P(p)$};

\draw ($(GR)+(0,0.10)$) -- ($(pR)+(0,0.10)$);
\draw[fermion] (GR) -- (pR);
\draw ($(GR)+(0,-0.10)$) -- ($(pR)+(0,-0.10)$);

\node at (3.85,-1.82) {$P'(p')$};

\draw[fill=white] (C) ellipse [x radius=\a, y radius=\b];
\node at (C) {GPD};

\pgfmathsetmacro{\offset}{0.25}
\coordinate (T) at ($(C)+(0,-\offset)$);

  \draw[arcwitharrow]
  ($(T)+({cos(210)*\a},{sin(210)*\b})$)
  arc (210:330:{\a} and {\b});

\node[yshift=-7pt] at ($(T)+(0,-0.65)$) {$t$};

\draw[doublefermion] (B2) -- (out);
\node at (4.05,1.05) {$\gamma,\pi,\phi,\ldots$};

\end{tikzpicture}
\caption{Schematic diagram of a generic hard exclusive process in lepton--nucleon scattering. A virtual photon emitted by the electron interacts with the nucleon, producing an exclusive final-state photon or meson while the recoiling nucleon carries the squared momentum transfer $t$. In the GPD formalism, $x$ denotes the average longitudinal momentum fraction of the
active parton, while $\xi$ is the skewness variable, which quantifies the longitudinal momentum transfer to the nucleon.}
\label{fig:excldiag}
\end{figure}

Diffractive processes in lepton--hadron scattering have been instrumental in establishing the modern QCD description of hadronic structure, providing unique sensitivity to both the longitudinal momentum and transverse spatial distributions of partons.
At HERA, diffractive deep-inelastic scattering was found to be a sizable component of the total virtual-photon--proton cross section, with ZEUS measuring a diffractive-to-total ratio of \(15.8~\%\) at \(Q^2=4~\mathrm{GeV}^2\), decreasing to \(5.0~\%\) at \(Q^2=190~\mathrm{GeV}^2\)~\cite{Chekanov:2008fh}. The H1 and ZEUS diffractive measurements further showed that diffractive DIS is compatible with a leading-twist QCD description based on collinear factorization~\cite{Aktas:2006hy,Chekanov:2008fh,Collins:1997sr}.

Among diffractive processes, exclusive processes, like deeply virtual Compton scattering (DVCS) and deeply virtual meson production~\cite{Mueller:1994ses, Ji:1996ek,Radyushkin:1997ki}, provide access to generalized parton distributions (GPDs). Exclusive processes are reactions in which the full final state is determined. 
A schematic of a generic exclusive reaction is shown in Fig.~\ref{fig:excldiag}: the electron emits a virtual photon, which interacts with the nucleon and produces an exclusive final-state photon or meson. The squared four-momentum transfer between the initial and final nucleon is denoted by $t$.
First measurements of DVCS and exclusive vector meson production at HERA demonstrated sensitivity to the transverse spatial distribution of partons in the small-$x$ regime~\cite{Aaron:2009ac}, including gluon-dominated dynamics probed through vector meson channels~\cite{Chekanov:2007zr}.

The fixed-target GPD program was subsequently developed in parallel at HERMES and JLab, with complementary experimental capabilities. Early DVCS measurements at HERMES established beam-spin asymmetries as a key observable~\cite{Airapetian:2001yk}, and were followed by a broad program exploiting beam-charge, beam-spin, and target polarization observables on both proton and nuclear targets~\cite{Airapetian:2008aa,Airapetian:2009aa}. At JLab, high luminosity enabled increasingly precise and multidimensional measurements of DVCS cross sections and asymmetries, providing stringent constraints in the valence region~\cite{Stepanyan:2001sm,MunozCamacho:2006hx,Jo:2015ema,Defurne:2015kxq,Defurne:2017paw,JeffersonLabHallA:2022pnx}. Together, these programs have established the experimental foundations for extracting Compton form factors and constraining GPDs~\cite{Kumericki:2016ehc}. Measurements at COMPASS have provided complementary information at intermediate energies, in particular through DVCS cross sections and the extraction of transverse spatial scales~\cite{Adolph:2017hne}. Despite these advances, the experimental knowledge remains uneven: while the valence region is increasingly well constrained, the gluon-dominated small-$x$ regime lacks the precision and multidimensional coverage required for quantitative spatial imaging of sea quarks and gluons.

By the time the EIC begins operation, the state of the art will also be shaped by collider measurements of exclusive processes at high energies. 
Ultra-peripheral collisions at RHIC and the LHC have enabled studies of exclusive vector meson photoproduction. While measurements at hadron-hadron colliders lack the precision of DIS experiments and only provide limited kinematic coverage, because they are restricted to photoproduction, they do provide the advantage of reaching the very low-$x$ region and, as such, contribute to our knowledge of gluon distributions and their spatial structure~\cite{Baltz:2007kq, CMS:2025lsm, CMS:2025oxg}. Measurements of coherent $J/\psi$ production in heavy-ion collisions~\cite{Abelev:2012ba, CMS:2023snh} and in proton--proton interactions~\cite{Aaij:2013jxj} have been complemented by higher-mass quarkonium channels such as $\Upsilon$~\cite{CMS:2026soy, CMS:2018bbk}, as well as more recent measurements extending the kinematic reach~\cite{Acharya:2021bnz}. Recent RHIC data provide additional evidence for strong nuclear effects in exclusive quarkonium production~\cite{Abdulhamid:2024jdq}, highlighting the sensitivity of these observables to nuclear shadowing and gluon saturation. In parallel, theoretical developments have significantly improved the quantitative description of exclusive processes, including global analyses of DVCS observables~\cite{Kumericki:2016ehc}, dipole-based descriptions of vector meson production~\cite{Kowalski:2003hm}, saturation-based approaches~\cite{Rezaeian:2012ji}, and the meaning of coherence in the Good-Walker paradigm~\cite{Klein:2023GW}. The inclusion of higher-order corrections in exclusive vector meson production has further reduced theoretical uncertainties in the small-$x$ regime~\cite{Mantysaari:2021dvmNLO,Mantysaari:2022dvmFull}. 
 
The ePIC early science program will build on this foundation by turning the broad physics landscape of exclusive, diffractive, and tagged reactions into a staged set of measurements matched to the beam species, luminosities, polarization configurations, and detector capabilities expected during the first years of EIC running. The measurements are summarized in Table~\ref{tab:exclusive_diffractive_tagging_staging}. The table is organized by approximate running conditions rather than by physics topic alone, because the early reach of this program is determined not only by the underlying cross sections but also by the availability of proton or nuclear beams, polarization, spectator tagging, nuclear-breakup vetoes, and fully commissioned far-forward instrumentation.

\begin{table}[bth!]
\centering
\footnotesize
\renewcommand{\arraystretch}{1.8}
\begin{tabular}{L{0.15\textwidth} L{0.40\textwidth} L{0.35\textwidth}}
\hline
\textbf{Configuration} &
\textbf{Representative measurements} &
\textbf{Enabling capabilities} \\
\hline

$e+\mathrm{A}$ with medium-mass nuclei &
Rapidity-gap diffraction; coherent-enhanced exclusive $\rho/\phi$ samples;
initial exclusive $J/\psi\rightarrow \ell^+\ell^-$ candidate studies &
Unpolarized $eA$ beams; central electron tracking and PID; low-multiplicity and rapidity-gap selections\\

$e+p$ running &
DVCS, $e+p\rightarrow e'p'\gamma$; exclusive vector-meson production, including $J/\psi$, $\rho$, and $\phi$; diffractive DIS with leading-proton or rapidity-gap selection &
Higher luminosity; proton beams; initial far-forward proton detection; broad central and forward acceptance \\

$e+\text{D}$  &
Spectator-tagged DIS on deuterium; tagged neutron structure; tagged studies of nuclear binding and off-shell effects &
Deuteron beams; far-forward proton and neutron tagging; control of spectator kinematics \\

$e+p$ running with polarization &
Beam-spin and target-spin asymmetries in DVCS and exclusive meson production; spin-dependent GPD observables; kaon structure via $\Lambda$ tagging&
Longitudinal and/or transverse proton polarization; polarized electron beam; good azimuthal acceptance\\

$e+$Au running &
Coherent and incoherent vector-meson production in gold; coherent-to-incoherent ratios; nuclear diffractive DIS &
Heavy-ion beams; improved luminosity; robust nuclear-breakup vetoes; precision $t$ reconstruction and unfolding \\

$e+^3$He with polarization &
Double spectator tagging in $^3$He; spin-dependent tagged neutron measurements &
Polarized $^3$He beams; forward spectator detection; mature light-ion reconstruction \\


\hline
\end{tabular}
\caption{Staged overview of exclusive, diffractive and tagging measurements in the ePIC early science program, grouped by the approximate running conditions and detector capabilities that enable them.}
\label{tab:exclusive_diffractive_tagging_staging}
\end{table}

The first opportunities arise in nuclear running, where events with a rapidity-gap in the final state can already isolate diffractive topologies, and where coherent enhancement makes light vector-meson production a natural early channel. These measurements provide an initial path toward nuclear gluon imaging and establish the event-selection, exclusivity, and background-control methods that will be needed for the later heavy-nucleus program. 

With higher-luminosity proton running, the program expands to the benchmark exclusive reactions: DVCS, exclusive vector-meson production, and diffractive DIS with either leading-proton or rapidity-gap selection. These channels provide complementary access to quark and gluon GPDs, diffractive parton dynamics, and the transverse spatial structure of the proton.

Light-ion beams add a distinct tagging program that is unique to the EIC. Spectator tagging in deuterium and, later, polarized $^3$He enables measurements of neutron structure with controlled nuclear corrections, including spin-dependent observables once the relevant polarization configurations become available. The sequence in Table~\ref{tab:exclusive_diffractive_tagging_staging} therefore links three experimental themes that are often discussed separately: exclusive proton imaging, nuclear diffraction and gluon-density imaging, and tagged measurements of nucleon structure in light nuclei.

Across these stages, the common experimental requirement is the ability to establish exclusivity and reconstruct the relevant exclusive kinematics, including both the momentum transfer $t$ and the azimuthal angle between the lepton and hadron planes. The recoiling proton can be measured by far-forward spectrometers. 
Diffractive measurements can also be selected through large rapidity gaps or forward baryon tagging. In both cases, the $t$ dependence of the cross section is a central observable: it connects the measured final-state kinematics to the transverse spatial distribution of quarks and gluons in protons and nuclei.


The following subsections illustrate this staged strategy through concrete early measurements, beginning with DVCS as the cleanest benchmark channel for proton imaging, followed by exclusive vector-meson production, diffractive DIS, and tagged measurements in light nuclei. 

\paragraph{Deeply Virtual Compton Scattering (DVCS)}

DVCS offers the theoretically cleanest access to generalized parton distributions and therefore to tomographic imaging of the proton. The final-state topology is experimentally simple: one scattered electron, one high-energy photon, and a forward proton. Simulations including realistic detector response show high reconstruction efficiency, excellent control of backgrounds, and robust determination of the squared momentum transfer $t$ at the proton vertex, using either the measured forward proton or the over-constrained lepton--photon final state. With the luminosities foreseen in the first polarized proton runs, ePIC will rapidly provide differential cross sections and beam-spin asymmetries over a broad kinematic domain, establishing one of the earliest benchmark channels for the exclusive program.

\begin{figure}[bht!]
\centering
\includegraphics[width=0.45\linewidth,valign=c]{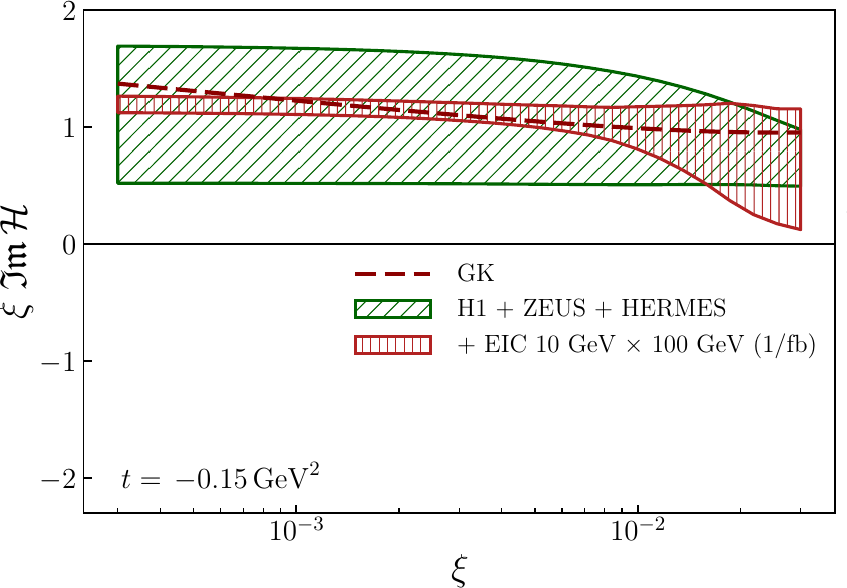}\hfill
\includegraphics[width=0.43\linewidth,valign=c]{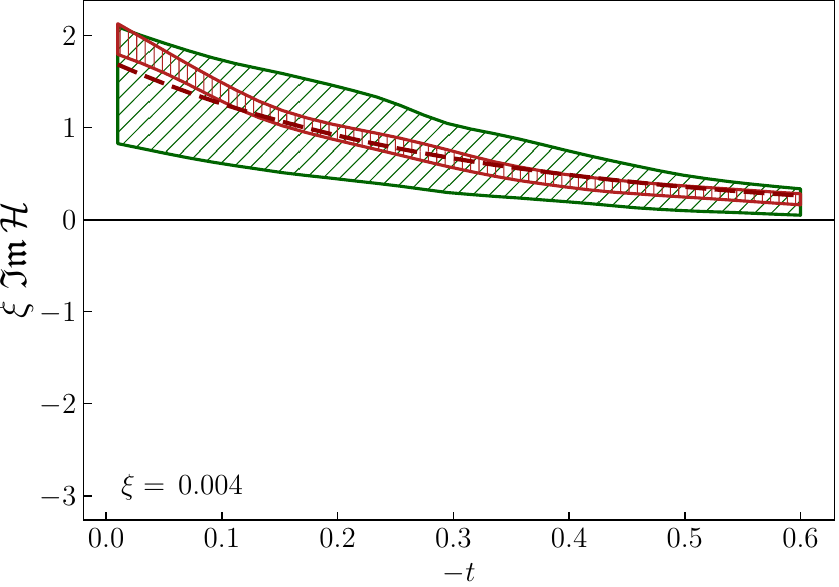}

\caption{
Left: $\mathrm{Im}\,\mathcal{H}$  as a function of $\xi$. Right: $\mathrm{Im}\,\mathcal{H}$  as a function of $-t$. Both are extracted by training an ensemble of neural nets on HERA data only (green slanted dashes) and additionally on simulated EIC data (red vertical dashes) at $Q^2=4$~GeV$^2$.
Goloskokov-Kroll (GK) model values~\cite{Goloskokov2005, Goloskokov2009} are plotted for comparison (red dashed line).}
\label{fig:exclusive}
\end{figure}

The physics impact of such measurements goes beyond demonstrating the reconstruction of exclusive final states. As illustrated in Fig.~\ref{fig:exclusive}, projected DVCS measurements at EIC kinematics substantially improve the extraction of some of the Compton form factors (CFFs), the experimentally accessible convolutions of GPDs entering the DVCS amplitude. The figure shows the expected constraints on the imaginary part of CFF $\mathcal{H}$ 
as a function of the skewness variable $\xi$, which quantifies the longitudinal momentum transfer to the proton in the GPD formalism, and of the squared momentum transfer $t$. The comparison between fits based on existing HERA and HERMES data alone and fits including simulated ePIC beam-spin-asymmetry pseudo-data demonstrates a strong reduction of the uncertainty bands. The leading potential systematic effects, namely QED radiative corrections and neutral-pion background contamination, were studied following the procedures described in Ref.~\cite{fy8y-bjc9} and found to be negligible compared with the projected fit uncertainties.

This shows that early ePIC DVCS measurements will already provide quantitative constraints on the CFFs most directly connected to proton imaging and, through GPDs, to the quark contribution to the proton spin and orbital angular momentum. In this sense, DVCS provides a direct link between an experimentally clean early measurement and the broader EIC goals of spatial imaging and spin decomposition of the nucleon.

\begin{figure}[bth]
\raisebox{-0.5\height}{%
\includegraphics[width=0.541\linewidth]{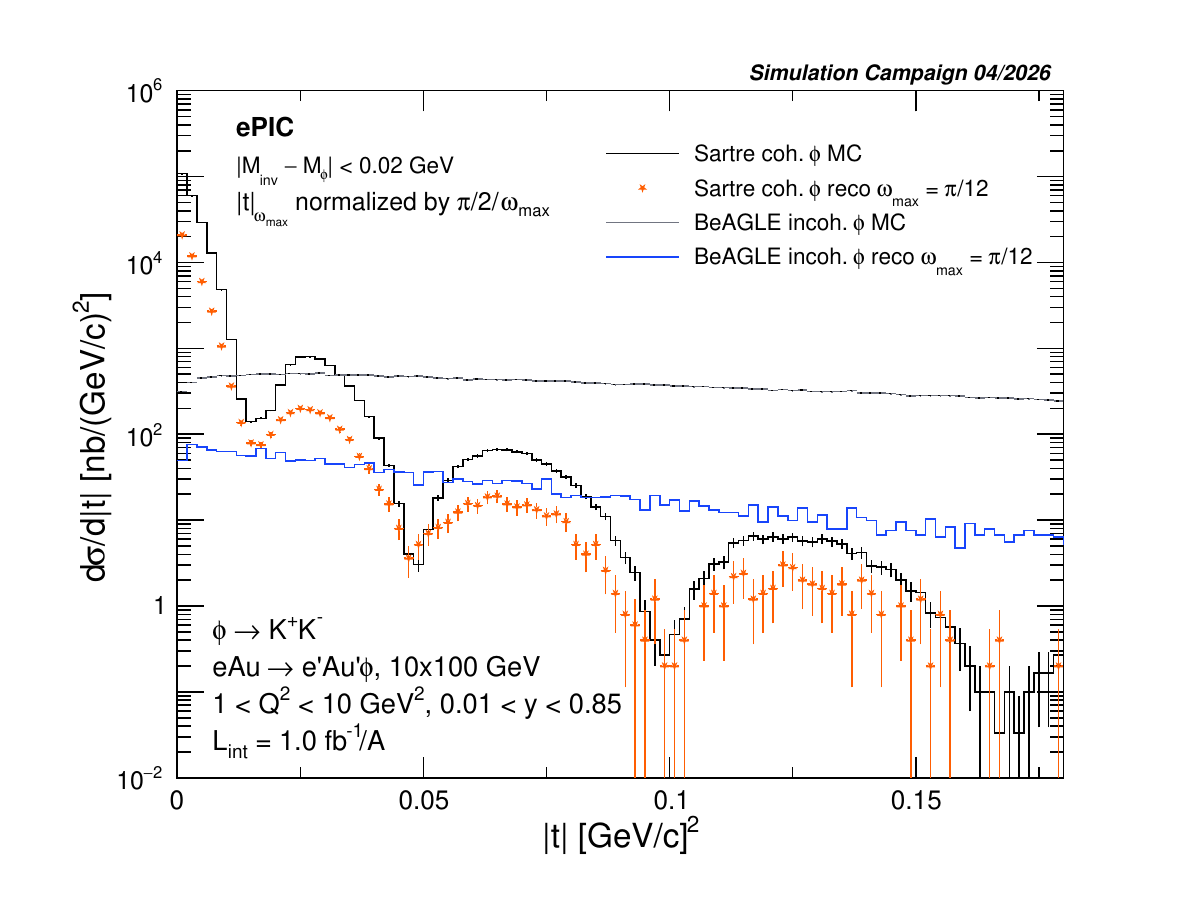}
}
\hfill
\raisebox{-0.5\height}{%
\includegraphics[width=0.45\linewidth]{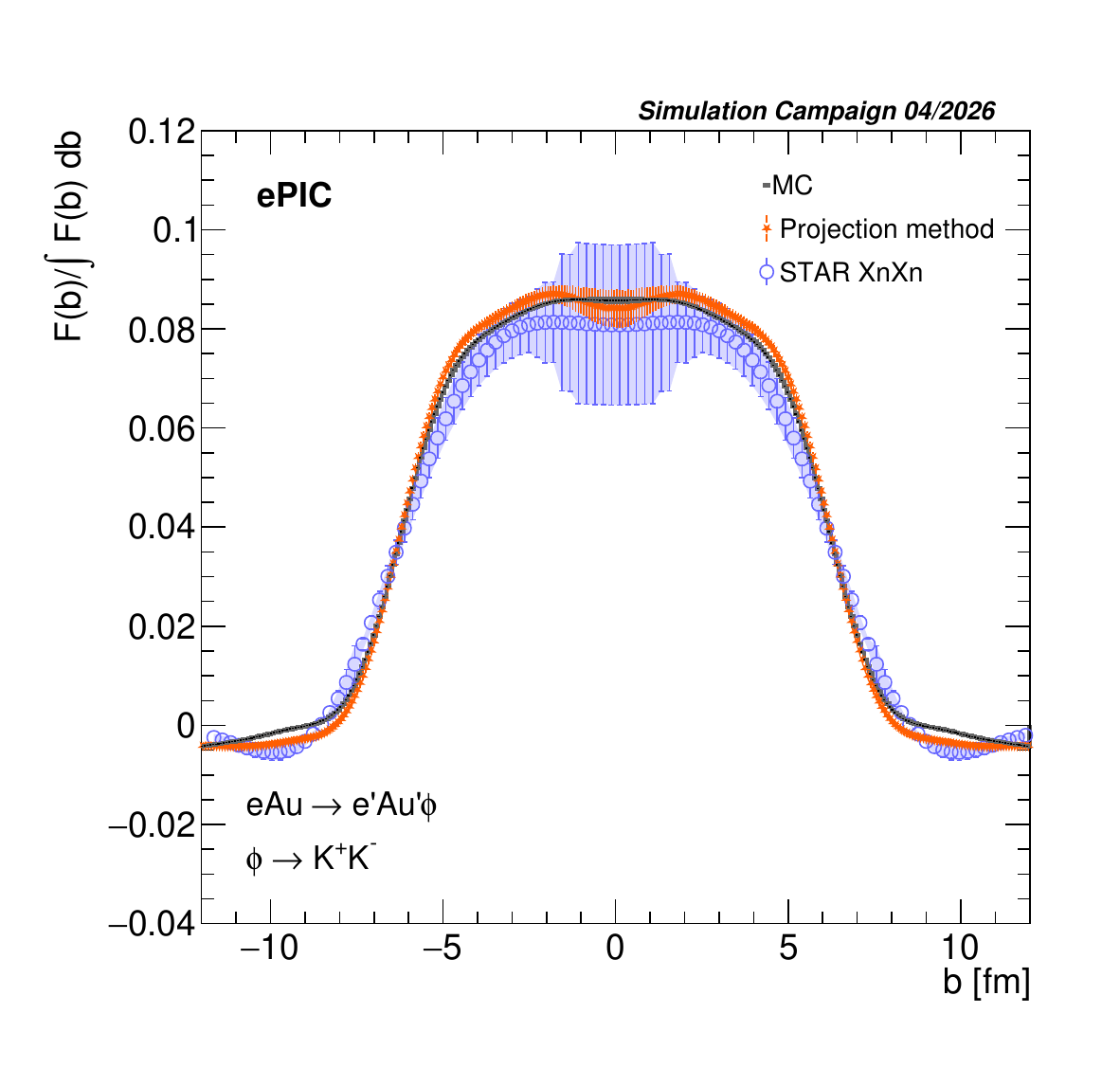}
}
\caption{\label{figmain:diff_phi}
Left: Generated (continuous, black line) and reconstructed (orange points) $|t|$ distribution for the selection of exclusive $\phi$ production, with $\phi\to K^+K^-$, in $e+\mathrm{Au}$ collisions. The suppression (blue line) of the generated (gray line) incoherent contribution is also shown. The reconstructed result uses the projected-$|t|$ method with a wedge selection $\omega<\omega_{\max}=\pi/12$ around the direction $\hat n$ normal to the electron scattering plane; the distribution is normalized by the accepted angular phase-space fraction. Right: Fourier--Bessel transform of the reconstructed $|t|$ distribution, illustrating the resulting transverse spatial gluon profile in the nucleus. The ePIC projection result is compared with the MC truth and with the STAR $\mathrm{Xn}+\mathrm{Xn}$ result from coherent $\rho^0$ photoproduction in ultra-peripheral $\mathrm{Au}+\mathrm{Au}$ collisions~\cite{Adamczyk:2017vfu}.} 
\end{figure}

\paragraph{Exclusive vector-meson production}
Production of heavy quarkonia, especially $J/\psi$, probes the transverse spatial distribution and fluctuations of gluons. For an intact proton, coherent production constrains the average transverse profile, whereas proton-dissociative production is sensitive to fluctuations. The charm-quark mass provides a comparatively hard scale, making the channel directly sensitive to gluon dynamics, while the clean dilepton signature leads to excellent signal--to--background ratios. Early measurements of the $t$ dependence will enable gluon imaging of the proton. Measurements of $\Upsilon$ production provide access to an even harder scale, with the EIC expected to resolve all three S-wave bottomonium states, $\Upsilon(1S)$, $\Upsilon(2S)$, and $\Upsilon(3S)$, in contrast to the more limited separation achievable at HERA or the LHC. Light vector mesons ($\rho$, $\phi$) offer higher statistics and, in nuclei (Fig.~\ref{figmain:diff_phi}), sensitivity to coherence effects and nuclear geometry.

Coherent diffractive vector-meson production probes the transverse gluon distribution in nuclei through the $t$ dependence of the cross section, whose Fourier--Bessel transform maps the impact-parameter profile. Figure~\ref{figmain:diff_phi} shows diffractive $\phi$ production using the projection method of Ref.~\cite{KESLER2026140585}, where the momentum transfer is projected onto the direction normal to the electron scattering plane. This minimizes sensitivity to electron momentum resolution, preserving the diffractive minima. Restricting the projection to events within a wedge of opening angle $\omega<\pi/12$ (where $\omega$ is measured relative to the projection axis) further improves the $t$ resolution, with normalization accounting for the reduced acceptance. The dominant residual systematic arises from incoherent $\phi$ production, where DIS and misidentified $\rho$ backgrounds are strongly suppressed.

As shown in Fig.~\ref{figmain:diff_phi}, the projected-$t$ method accurately reproduces the MC $|t|$ spectrum and preserves the diffractive structure with realistic ePIC reconstruction. The Fourier–-Bessel transform obtained from the ePIC projection has much better precision than that obtained from the STAR Xn+Xn measurement from coherent $\rho^0$ photoproduction in ultra-peripheral Au+Au collisions~\cite{Adamczyk:2017vfu}.
These results demonstrate that ePIC can achieve robust imaging of the nuclear gluon distribution during early $e+\mathrm{A}$ running with moderate integrated luminosity.

\paragraph{Diffractive DIS and rapidity--gap measurements}
Large rapidity gaps and forward baryon tagging can be identified with high efficiency using the wide acceptance of ePIC.  
In $e+p$, early data will map diffractive structure functions with precision exceeding that of HERA in many regions.  
In the early $e+\mathrm{A}$ running, the ePIC detector will already allow us to distinguish coherent and incoherent regimes, opening the path toward imaging of gluon densities and their event--by--event fluctuations.

\paragraph{Spectator tagging in light nuclei}
Tagging recoil protons in deuteron or $^{3}\mathrm{He}$ beams isolates scattering from nearly on-shell neutrons and dramatically reduces nuclear uncertainties. In deuterium, measurements at very low spectator momentum enable the use of pole-extrapolation techniques~\cite{Sargsian:2005rm,Jentsch:2021qdp}, providing access to the free-neutron structure function with minimal nuclear corrections. In $^{3}\mathrm{He}$, spectator tagging provides a complementary path to neutron spin observables once polarized light-ion beams are available. These measurements rely on the far-forward instrumentation and represent a unique EIC capability~\cite{Jentsch:2021qdp}.

The physics impact of spectator-tagged neutron measurements goes beyond the direct improvement of $F_2^n$ itself. Precision free-neutron data constrain the isospin decomposition of nucleon structure, including the $d/u$ ratio and sea-quark distributions, and improve the flavor separation entering global PDF analyses~\cite{Li:2023yda,Accardi:2026hdv}. Through QCD evolution and global-fit correlations, these constraints also propagate to the sea-quark and gluon sectors at higher scales. Equally important, the combination of free-proton and free-neutron information provides a controlled $p+n$ baseline for nuclear ratios such as $D/(p+n)$, measured in the same collider environment. This is essential for testing nuclear shadowing in the simplest nucleus, where the expected effect is at the percent level and therefore requires neutron-structure uncertainties of comparable precision. Such measurements provide a benchmark for shadowing calculations in light nuclei and strengthen the baseline needed to identify saturation effects at smaller $x$ or in heavier nuclei.

\begin{figure}[bth]
\centering
\includegraphics[width=0.4\linewidth]
{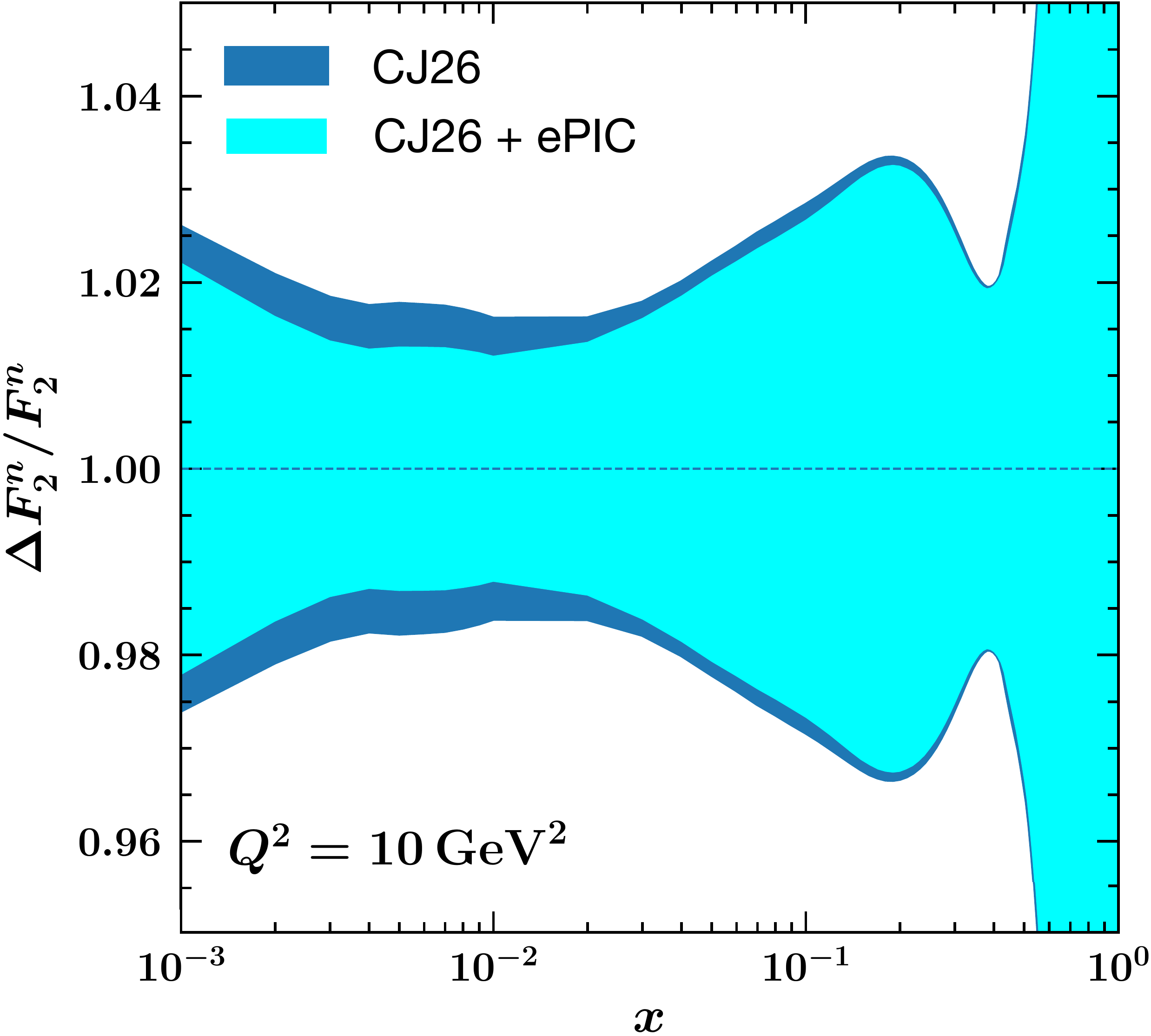}
\caption{\label{figmain:f2n}
Projected impact of early ePIC spectator-tagged deuteron data on the neutron structure function $F_2^n$ in the CJ26 framework~\cite{Accardi:2026hdv}. The upper panel shows the relative uncertainty band for $F_2^n$ at $Q^2=10~\mathrm{GeV}^2$ before and after including ePIC pseudo-data sample corresponding to $\rm 1~fb^{-1}$. 
}
\end{figure}

Figure~\ref{figmain:f2n} illustrates the projected impact of early ePIC spectator-tagged data on the neutron structure function in the CJ26 framework~\cite{Accardi:2026hdv}. At $Q^2 = 10~\mathrm{GeV}^2$, the inclusion of ePIC pseudo-data reduces the relative uncertainty on $F_2^n$ over a broad range in $x$. Around $x \simeq 5\times 10^{-3}$, the uncertainty decreases from about $2\,\%$ to about $1.5\,\%$, corresponding to a reduction of roughly $25$--$30\,\%$. While the absolute precision is already high before including ePIC data, the additional reduction is important because it brings the uncertainty closer to the scale required for quantitative tests of deuteron shadowing and for robust flavor-separated PDF constraints.

\paragraph{Meson structure with forward tagging}
Forward-tagged exclusive channels provide an additional early opportunity to study meson structure. Processes such as \(e p \to e'\pi^+ n\) give access to the structure of the pion and provide complementary information for flavor separation in GPD studies. In this channel, detection of the forward neutron in the Zero Degree Calorimeter (ZDC) tags the pion-exchange process and enables an extraction of the pion form factor, \(F_\pi\).

\begin{figure}[htb]
\centering
\includegraphics[width=0.8\textwidth]{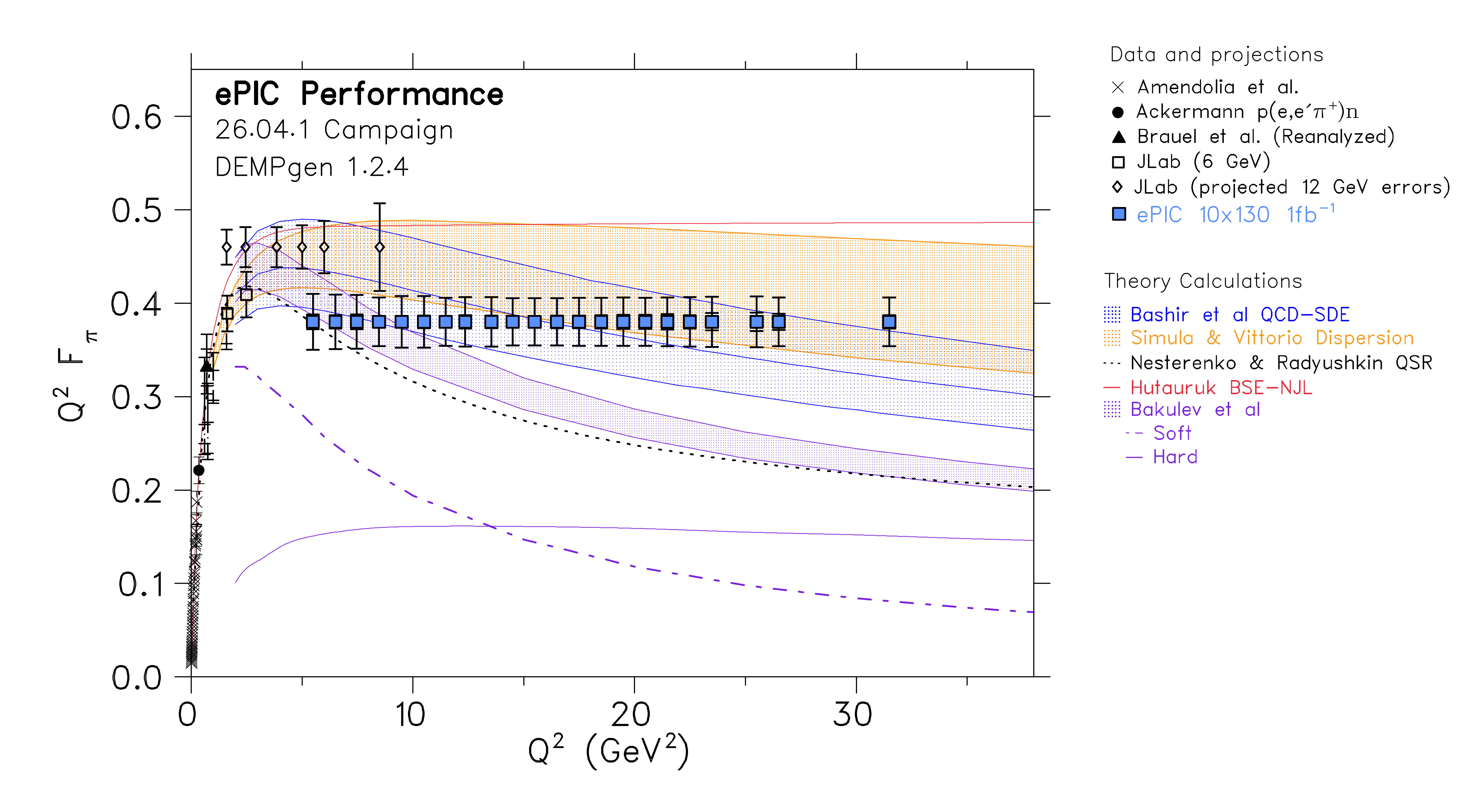}
\caption{Existing data and projected uncertainties for future data on the pion form factor from JLab and ePIC (blue), in comparison to a variety of hadronic structure models. The projected ePIC data have double error bars, where the inner error bar includes only statistical and assumed point-to-point systematic errors, and the outer error bar includes also the correlated systematic errors. For most of the ePIC data points, the statistical errors are smaller than the marker and only the total uncertainty is visible.}
\label{fig:pion_FF}
\end{figure}

Figure~\ref{fig:pion_FF} shows the projected uncertainties for \(Q^2 \mathrm{F}_\pi\), compared with existing data, future JLab projections, and a representative set of hadronic-structure calculations. These projections include statistical and assumed systematic contributions. In the current estimate, the dominant systematic effect is taken to be an overall scale uncertainty of about 12~\%, pending a dedicated detector-level systematic study. The projected precision demonstrates that an early ePIC measurement can provide meaningful constraints on pion structure in collider kinematics, thereby extending the exclusive and tagged program beyond nucleon imaging alone. 

Access to partonic meson structure is also available through tagged semi-inclusive DIS measurements, which enable pion and kaon structure-function studies. The far-forward instrumentation is crucial for these measurements, for example through \(\Lambda\) tagging to facilitate kaon structure-function extractions.

\paragraph{Summary}
Exclusive and diffractive reactions possess several distinctive characteristics: they feature low occupancies and benefit from kinematic closure tests.  Key observables such as exponential $t$ slopes, coherent/incoherent ratios, and spin asymmetries can reach physics relevance while simultaneously validating detector alignment, calibration, and reconstruction strategies.
Realistic simulations incorporating detector response, reconstruction, and background estimates demonstrate that clean signals and precise $t$ measurements are achievable from the outset. 
Completing a full GPD program with exclusive reactions would require a significantly larger data sample than the one envisioned in the early running. Nevertheless, Fig.~\ref{fig:exclusive} shows that a significant impact on some CFFs is already possible with early data. The spectra illustrated in Fig.~\ref{figmain:diff_phi}~(left) provide the essential input for spatial imaging through Fourier transformation techniques. 
Figure~\ref{figmain:f2n} illustrates how early ePIC deuteron data improve the extraction of $F_2^n$, providing the free-neutron input needed to isolate nuclear effects in deuterium through controlled $D/(p+n)$ ratios. The pion form-factor projection in Fig.~\ref{fig:pion_FF} provides a complementary meson-structure benchmark, demonstrating that forward-tagged exclusive measurements can already test hadronic-structure models during early running.
These results demonstrate that ePIC can deliver competitive and distinctive exclusive, diffractive, and tagged measurements during early running, while establishing the forward-reconstruction, background-control, and unfolding techniques required for the full EIC program.

\section{Heavy flavor and jet measurements}
\label{secmain:hf}

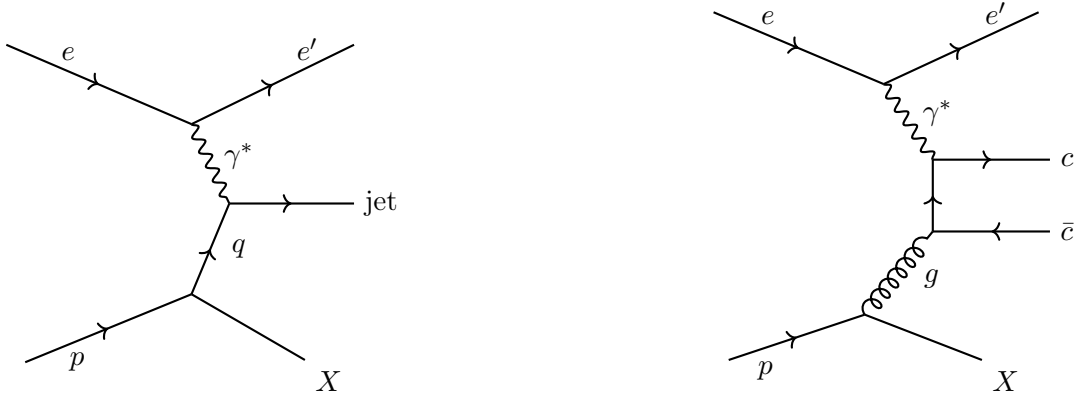
\begin{figure}[htbp]
\centering
\begin{subfigure}[t]{0.48\textwidth}
\centering
\begin{tikzpicture}[scale=1.0, line width=0.8pt]
\tikzset{
  fermion/.style={
    postaction={decorate},
    decoration={markings, mark=at position 0.5 with {\arrow{>}}}
  }
}
\coordinate (A) at (-0.15,1.10);
\draw[fermion] (-2.6,2.15) -- (A)
  node[midway, above left=5pt] {$e$};
\draw[fermion] (A) -- (2.0,2.15)
  node[midway, above right=5pt] {$e'$};
\coordinate (B) at (0.35,0.05);
\draw[decorate, decoration={snake, amplitude=2pt, segment length=6pt}]
  (A) -- (B);
\node at (0.35,0.68) {$~~\gamma^*$};
\draw[fermion] (B) -- (2.0,0.05);
\node[right] at (2.0,0.05) {jet};
\coordinate (D) at (-0.15,-1.15);
\draw[fermion] (-2.35,-2.05) -- (D)
  node[midway, below left=5pt] {$p$};
\draw[fermion] (D) -- (B);
\path (D) -- (B) node[midway, right, xshift=4pt] {$q$};
\draw (D) -- (1.35,-2.02);
\node[below right] at (1.35,-2.02) {$X$};
\end{tikzpicture}
\end{subfigure}
\hfill
\begin{subfigure}[t]{0.48\textwidth}
\centering
\begin{tikzpicture}[scale=1.0, line width=0.8pt]
\tikzset{
  fermion/.style={
    postaction={decorate},
    decoration={markings, mark=at position 0.5 with {\arrow{>}}}
  },
  gluon/.style={
    decorate,
    decoration={coil, aspect=0.75, segment length=5pt, amplitude=3pt}
  }
}
\coordinate (A) at (-0.1,1.10);
\draw[fermion] (-2.35,2.05) -- (A)
  node[midway, above left=5pt] {$e$};
\draw[fermion] (A) -- (1.95,2.05)
  node[midway, above right=5pt] {$e'$};
\coordinate (B) at (0.55,0.10);
\draw[decorate, decoration={snake, amplitude=2pt, segment length=6pt}]
  (A) -- (B);
\node at (0.55,0.72) {$~\gamma^{*}$};
\draw[fermion] (B) -- (2.10,0.10);
\node[right] at (2.10,0.10) {$c$};
\coordinate (C) at (0.55,-0.85);
\draw[fermion] (2.10,-0.85) -- (C);
\node[right] at (2.10,-0.85) {$\bar{c}$};
\draw[fermion] (C) -- (B);
\coordinate (D) at (-0.35,-1.95);
\draw[gluon] (D) -- (C);
\path (D) -- (C) node[midway, right, xshift=5.5pt, yshift=-2pt] {$g$};
\draw[fermion] (-2.15,-2.55) -- (D)
  node[midway, below left=5pt] {$p$};
\draw (D) -- (1.20,-2.55);
\node[below right] at (1.20,-2.55) {$X$};
\end{tikzpicture}
\end{subfigure}
\caption{Illustration of production processes for inclusive jets (left) and charm quarks (right) at leading order in $e+p$ collisions.}
\label{fig:prod-jets-hf}
\end{figure}

Measurements of jets and heavy-flavor hadrons (collectively, ``hard probes'') constitute an integral component of the early ePIC scientific program. As illustrated in Fig.~\ref{fig:prod-jets-hf}, jet and heavy-flavor production, at leading order in $\alpha_S$, couple directly to quarks and gluons, respectively, thus constraining PDFs and nPDFs. Compared to inclusive, semi-inclusive, and exclusive measurements, hard probes provide a uniquely differential window on QCD dynamics: they resolve the kinematics of the hard scattering, and allow the study of subsequent parton shower evolution and hadronization in both vacuum and cold nuclear matter. Many observables, with clean reconstruction signatures and robust theoretical interpretations (see Table~\ref{tab:jethf_capabilities}), can be measured with sufficient statistical precision in the early running period to
produce impactful results. 


\begin{table}[htbp]
\centering
\centering
\footnotesize
\renewcommand{\arraystretch}{1.8}
\begin{tabular}{L{0.15\textwidth} L{0.40\textwidth} L{0.35\textwidth}}
\hline
\textbf{Configuration} & \textbf{Representative measurements} & \textbf{Enabling capabilities} \\ 
\hline

$e+p$ running & Jet FFs; $\Lambda^+_c/\mathrm{D}^0$ & Particle ID ($\pi, K, p$), primary and secondary vertexing, tracking and calorimetry, high luminosity\\ 

$e+p$ running with polarization & Hadron-in-jet Collins FF & Same as above with transversely polarized proton beam \\ 

$e+A$ running with heavy nuclei & Jet $R_{eA}$; $\mathrm{D}^0$-in-jet $R_{eA}$ & Same as above\\

\hline
\end{tabular}
\caption{Overview of measurements related to jets and heavy flavor in the ePIC early science program grouped by the approximate running conditions and capabilities that enable them.
}
\label{tab:jethf_capabilities}
\end{table}

\paragraph{Jets in Cold Nuclear Matter} Impacts of cold nuclear matter on jets have been
investigated extensively through measurements of hadrons with high transverse momenta and jet-related observables in $p + \mathrm{A}$ and $d + \mathrm{A}$ collisions at RHIC and the LHC~\cite{ALICE:2012mj,ALICE:2017svf,CMS:2016xef,CMS:2025jbv,ATLAS:2014cpa,ATLAS:2016xpn,ATLAS:2022iyq,ATLAS:2023zfx,PHENIX:2023dxl,STAR:2014qsy,STAR:2024nwm}. Overall, inclusive jet and high-$p_T$ hadron measurements are broadly described by calculations incorporating nuclear PDFs, although some observables indicate that additional cold nuclear matter effects beyond nPDF modifications may also play a role. 
Similarly, charm quark production in a cold nuclear environment has also been studied in detail both experimentally and theoretically (see~\cite{Arleo:2025oos} and references therein). In fixed-target collisions, no significant nuclear effects on $D$ meson production are observed~\cite{WA82:1992yoq,FermilabE769:1992tiy,E769:1993igi,E789:1994nhc,BEATRICE:1996bnh,HERA-B:2007rfd}. At the LHC, D meson production in $p+\rm{Pb}$ collisions has been measured~\cite{ALICE:2019fhe,LHCb:2017yua,LHCb:2023kqs,LHCb:2022dmh}. Compared to $p+p$ collisions, strong suppression of the yields is observed at forward rapidity and low transverse momenta. At the EIC, the cleaner lepton--nucleus environment will provide strong sensitivity to cold nuclear matter effects in $x$--$Q^{2}$ kinematic regions complementary to those accessed at RHIC and the LHC. 

Inclusive jet electro-production in $e+\mathrm{A}$ collisions provides a direct probe of QCD dynamics in nuclear matter, with sensitivity to both initial-state nuclear effects, such as modifications of nPDFs, and final-state effects, including parton energy loss.

\begin{figure}[htbp]
\centering
\raisebox{-0.5\height}{%
\includegraphics[width=0.43\textwidth]{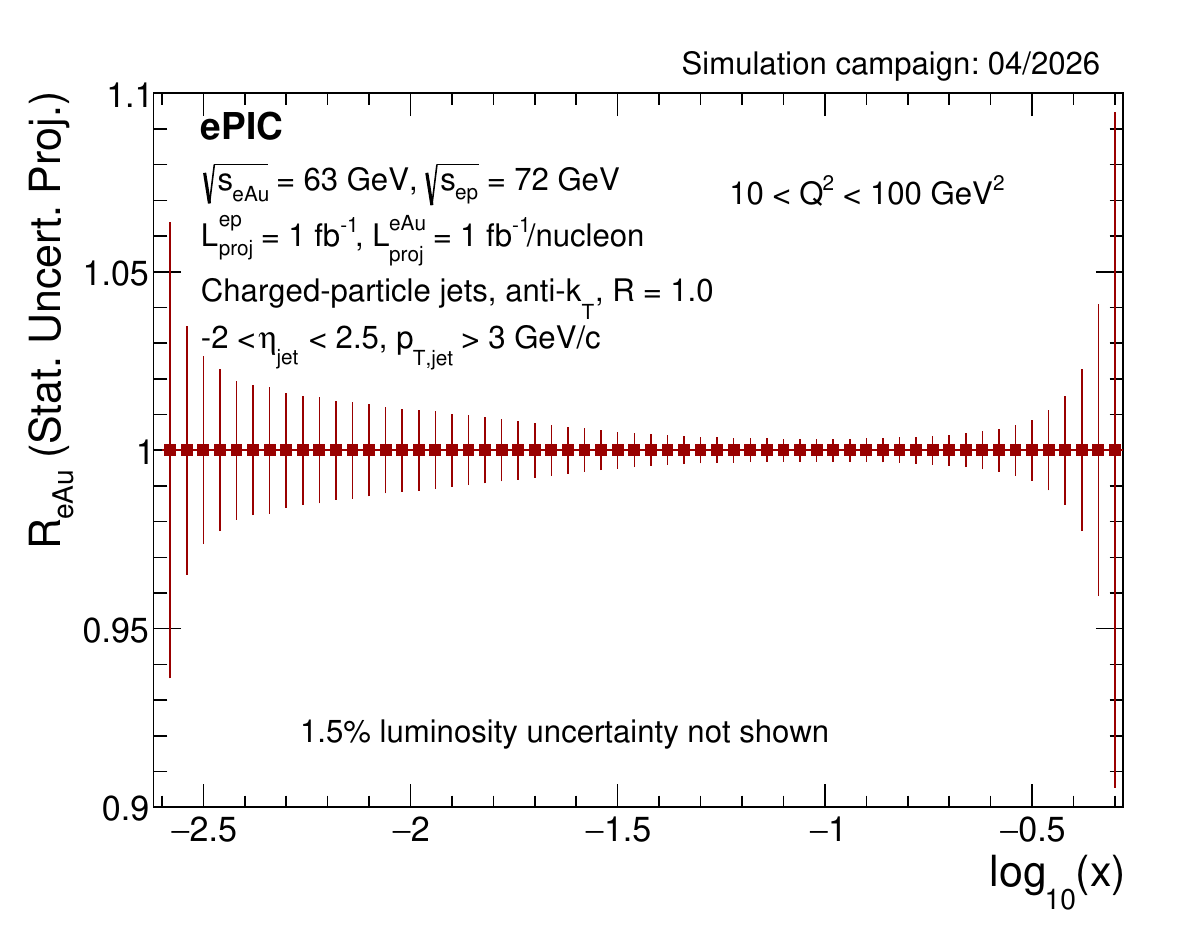}
}
\raisebox{-0.5\height}{%
\includegraphics[width=0.54\textwidth]{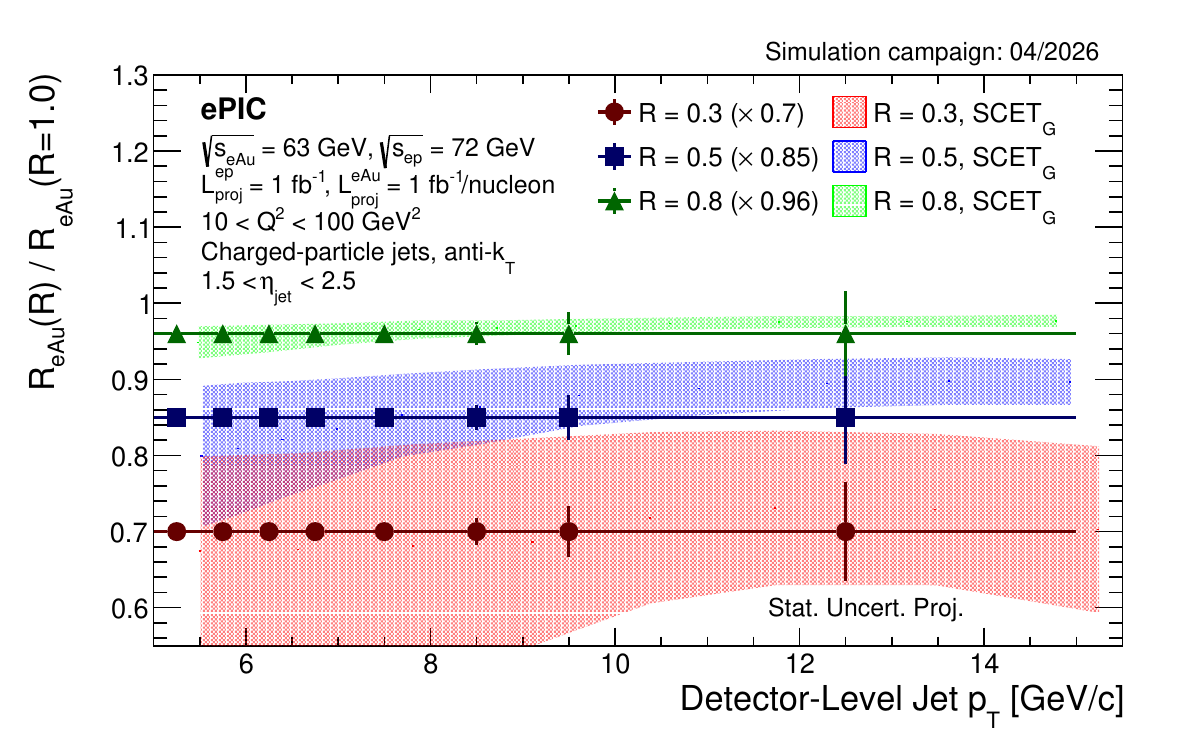}
}
\caption{Left: projected statistical precision for inclusive charged jet $R_{eA}$ for $R=1.0$ as a function of $\log(x)$. Right: projected statistical precision for inclusive charged jet $R_{eA}$ for $R=0.3$, $0.5$, and $0.8$ as a function of jet transverse momentum. The beam energies are  10\,GeV\,$\times$\,100\,GeV for $e+\rm{Au}$ collisions and 10\,GeV\,$\times$\,130\,GeV for $e+p$ collisions. In the right panel, the data points are vertically shifted for clarity. The theoretical calculations~\cite{Li:2020rqj} are shown as shaded bands. SCET denotes soft-collinear effective theory.}
\label{fig:jets}
\end{figure}

To disentangle these initial- and final-state contributions, jets with different radii ($R$) will be measured in both $e+p$ and $e+\mathrm{A}$ collisions. Large-radius jets are expected to capture most of the medium-modified jet shower, since radiation induced by parton propagation through the nuclear environment is expected to be redistributed within a broader angular region. Consequently, measuring the inclusive jet
cross-section ratio between $e + \mathrm{A}$ and $e + p$ collisions, denoted $R_{eA}$, at large jet radii provides enhanced sensitivity to nPDF effects because of its correlation with parton kinematics and its reduced sensitivity to final-state effects. The projected statistical precision for $R_{eA}$ of $R=1.0$ charged-jets in 10\,GeV\,$\times$\,100\,GeV $e+\rm{Au}$ collisions and 10\,GeV\,$\times$\,130\,GeV $e+p$ collisions is shown in the left panel of Fig.~\ref{fig:jets}, where the expected statistical error is generally less than 2~\% in the range of $-2.3<\log(x)<-0.5$.

In contrast, the ratio of $R_{eA}$ measured for a smaller jet radius to that obtained for $R=1.0$ directly probes cold nuclear matter effects on jet propagation and shower dynamics. In this ratio, initial-state nuclear effects largely cancel, thereby enhancing sensitivity to modifications arising from final-state interactions~\cite{Li:2020rqj}. A corresponding precision projection is shown in the right panel of Fig.~\ref{fig:jets} as a function of jet transverse momentum in the forward rapidity region, $1.5 < \eta_{\rm jet} < 2.5$, where medium-induced effects are expected to be larger than those at midrapidity and backward rapidities~\cite{Li:2020rqj}. For visual clarity, the data points corresponding to different small jet radii are shifted vertically. The figure also includes theoretical predictions incorporating parton energy loss in cold nuclear matter~\cite{Li:2020rqj}. With the anticipated EIC luminosities during early running, these measurements are expected to achieve sufficient statistical precision to quantitatively constrain final-state nuclear effects.

\paragraph{Charm Hadronization in Vacuum and Nuclear Matter}
Heavy-flavor production, particularly open-charm production, plays a unique role in addressing the gluonic structure of the nucleon. This topic therefore links to the first NAS science pillar, since gluons account for the dominant fraction of the nucleon's momentum, and therefore the dynamical generation of its mass, and charm provides a particularly clean probe of gluon dynamics.

\begin{figure}[htbp]
\centering
\raisebox{-0.5\height}{%
\includegraphics[width=0.52\textwidth]{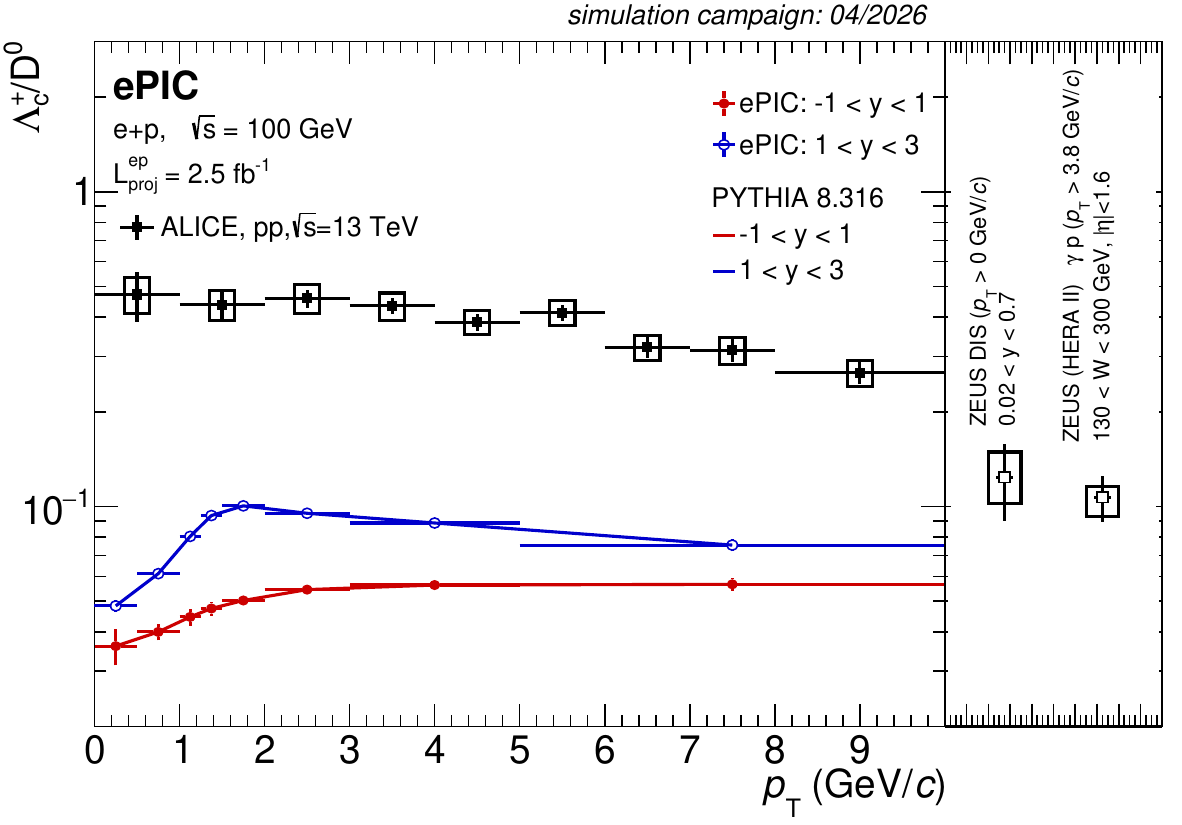}
}
\raisebox{-0.5\height}{%
\includegraphics[width=0.45\textwidth]{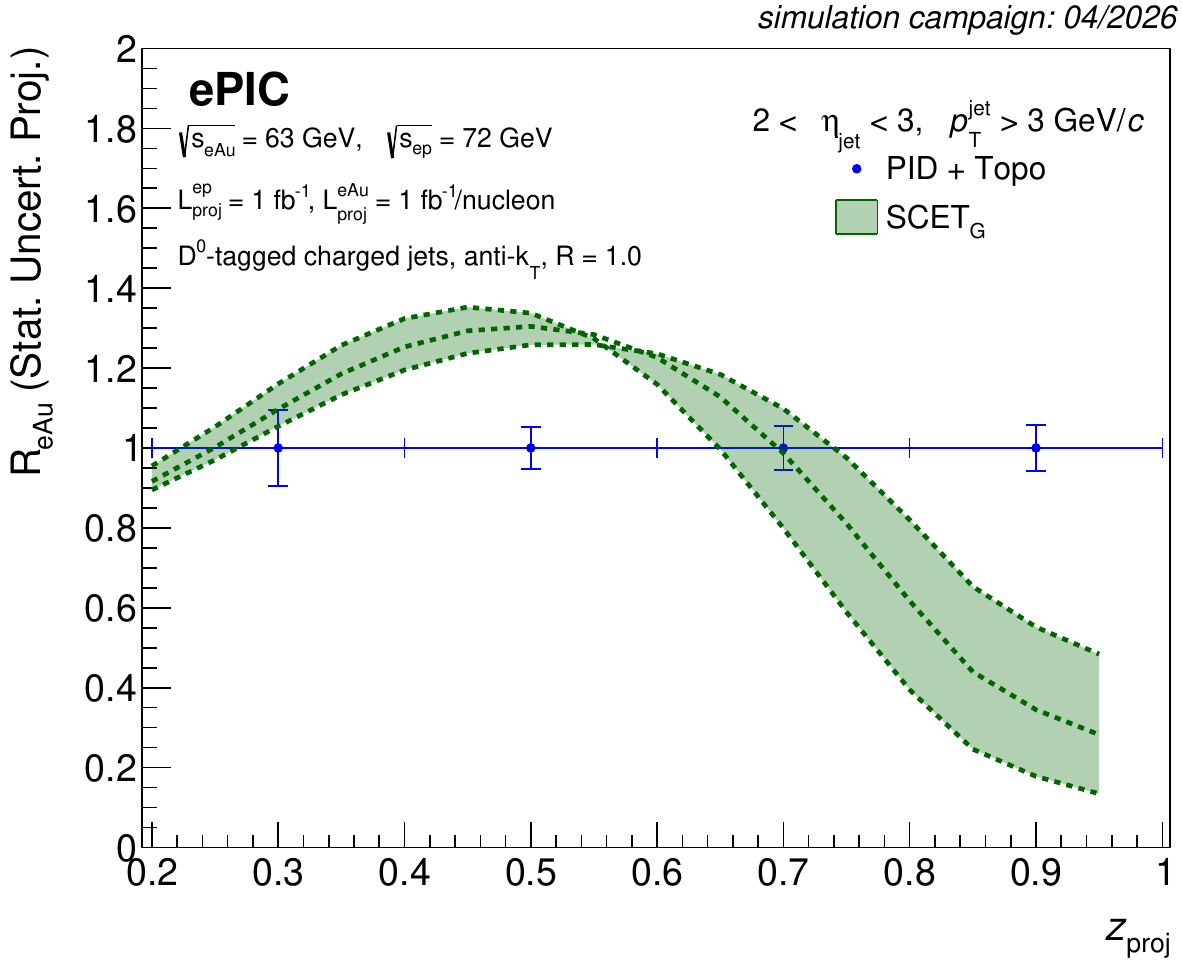}
}

\caption{Left: projected inclusive $\Lambda^+_{c}$/$D^{0}$ ratio as a function of $p_{\rm T}$ at mid ($-1<y<1$) and forward ($1<y<3$) rapidities in 10\,GeV\,$\times$\,250\,GeV $e+p$ collisions. The error bars correspond to the statistical precision for an integrated luminosity of $\rm 2.5 ~fb^{-1}$, while the curves represent PYTHIA 8.316 predictions. The projected ePIC measurements are shown together with existing ALICE~\cite{ALICE:2023sgl} and ZEUS measurements~\cite{ZEUS:2013fws,ZEUS:2010cic} for comparison. Right: statistical projection (blue solid markers) for the measurements of modifications to the $D^{0}$ fragmentation function within charged jets in $e+\rm{Au}$ collisions relative to $e+p$ collisions, in the forward region ($2<\eta_{jet}<3$), assuming an integrated luminosity of $\rm 1 ~fb^{-1}$. The projection is compared with the theoretical predictions (dotted filled bands) based on soft-collinear effective theory with Glauber gluons ($\rm SCET_G$) ~\cite{Li:2020zbk,Li:2025lxr}.}
\label{fig:hf}
\end{figure}

Beyond constraining gluon distributions, charm observables also provide a powerful tool for studying hadronization mechanisms in both vacuum and nuclear environments. The fragmentation of heavy quarks differs from that of light flavors, since a
heavy-flavor meson retains a larger fraction of the original heavy quark's momentum; thus open-charm measurements enable detailed investigations of how heavy quarks fragment. Recent measurements at the LHC~\cite{ALICE:2023sgl} have shown that the ratio of charmed-baryon to meson ($\Lambda^+_{c}$/$\mathrm{D}^{0}$) in $p+p$ collisions is significantly higher than that in $e^{+} + e^{-}$ collisions at LEP and at B factories (see~\cite{Lisovyi:2015uqa} and references therein), challenging the long-assumed universality of the fragmentation function among different collision systems. Measurements of $\Lambda^+_{c}$/$\mathrm{D}^{0}$ in the intermediate $e+p$ collisions provide critical insights into this puzzle. Such measurements have been carried out in $e+p$ collisions at HERA~\cite{ALICE:2020wla,ZEUS:2013fws,ZEUS:2010cic}, which, however, have only integrated $p_{\rm T}$ results with large uncertainties. With the anticipated luminosities during the early science period, a multi-differential measurement of inclusive $\Lambda^+_{c}$/$\mathrm{D}^{0}$ production as a function of $p_{\rm T}$ can be performed. The projected precision achievable for 10\,GeV\,$\times$\,250\,GeV $e+p$ collisions is shown in the left panel of Fig.~\ref{fig:hf} for two different rapidity intervals. The projection is obtained using particle identification and topological selection optimized with a Boosted Decision Tree (BDT) binary classifier \cite{barioglio_2022_7014886}, and is compared with ALICE $p+p$ data at $\sqrt{s}$ = 13~TeV and ZEUS data at $\sqrt{s}$ = 320~GeV. Data points are placed at the predicted values by PYTHIA~\cite{Bierlich:2022pfr} with the vertical bars indicating statistical uncertainties.

Heavy-flavor quarks also provide a sensitive probe of how hadronization may be modified by nuclear matter. Estimates based on the uncertainty principle~\cite{Adil:2006ra,Sharma:2012dy} suggest that the hadron formation time is inversely proportional to the meson mass squared. As a result, heavy mesons such as the $\mathrm{D}^{0}$ are more likely than light mesons to hadronize inside the nucleus in $e+\mathrm{A}$ collisions. The influence of the nuclear medium can be quantified by comparing fragmentation functions measured in $e+\mathrm{A}$ collisions with those in $e+p$ collisions. Such measurements are sensitive to both parton energy loss and possible modifications of the hadronization process itself. The reconstruction of $\mathrm{D}^{0}$ mesons is performed via their hadronic decay channel ($\mathrm{D}^{0} \rightarrow K^- \pi^+$) using particle identification and topological selection criteria. 
The projected statistical precision for the measurement of $R_{eA}$ for $\mathrm{D}^{0}$ mesons within charged-particle jets in 10\,GeV\,$\times$\,100\,GeV $e+\rm{Au}$ collisions and 10\,GeV\,$\times$\,130\,GeV $e+p$ collisions, shown as a function of the longitudinal momentum fraction $z_{\rm proj}= (\vec{p}_{\mathrm{D}^0} \cdot \vec{p}_{\rm jet}) / |\vec{p}_{\rm jet}|^2$ in the right panel of Fig.~\ref{fig:hf}, demonstrates this capability. With the integrated luminosities anticipated during the early years of EIC operation, the measurement is expected to achieve a statistical precision of $\sim 5-10~\%$ over a broad range of $z_{\rm proj}$ at forward rapidity. Theoretical predictions incorporating parton energy loss~\cite{Li:2020zbk,Li:2025lxr} are also shown in Fig.~\ref{fig:hf} for comparison. In the rapidity interval $2<y<3$, the projected experimental precision is sufficient to resolve the predicted nuclear modifications.

\paragraph{Spin Sensitivity}

\hyphenation{trans-verse-mo-men-tum-de-pen-dent}
With polarized proton beams, measurements of identified hadrons within jets provide access to transverse-momentum-dependent fragmentation functions and to quark transversity distributions via Collins-type asymmetries. While heavy-flavor production is in principle sensitive to gluon helicity through photon–gluon fusion, its impact in early running is expected to be limited by statistical precision and should be viewed as complementary to, rather than competitive with, constraints from inclusive DIS scaling violations. Although the ultimate precision program will require higher luminosities and center-of-mass energies, early running already permits statistically meaningful spin asymmetry measurements in selected kinematic regions. These will expand the kinematic reach beyond that accessible at previous facilities and contribute to the broader program of constraining  $\Delta g(x)$ at moderate $x$.

\paragraph{Summary} 
Jet and heavy-flavor measurements combine strong sensitivity to gluon dynamics with high statistical precision from the early EIC operational era and robust theoretical interpretability. The projected measurements span a range of key observables. These include the inclusive charged-jet $R_{eA}$ for large and small jet radii, which together disentangle nPDF effects from final-state parton energy loss in cold nuclear matter, the $\Lambda^+_c/D^0$ ratio as a function of $p_T$ in $e+p$ collisions, which will provide new insights into the universality of charm hadronization, and the $D^0$-in-jet $R_{eA}$ as a function of $z_{\rm proj}$, which probes modifications of heavy-quark fragmentation in the nuclear environment. With polarized beams, hadron-in-jet Collins asymmetries and photon--gluon fusion processes additionally provide early sensitivity to quark transversity and gluon helicity $\Delta g(x)$. Together, these hard-probe measurements are expected to provide some of the earliest and clearest demonstrations of the QCD science program enabled by the EIC, while simultaneously establishing analysis methodologies that will be used in the full EIC physics program.

\section{Limitations of the early science configuration}
\label{sec:limitations}

This report showcases how the early years of ePIC running will deliver world-class measurements addressing all three NAS science pillars. It is equally important to state explicitly 
which elements of the EIC science case identified in the NAS report~\cite{NAP25171} can only be realized with the full design energy, luminosity and polarization capabilities. 
The early program described in this report is the first stage of the EIC physics mission, not a substitute for it. 

\paragraph{Kinematic reach}
The baseline 9~GeV electron beam limits the $e+p$ center-of-mass energy to $\sqrt{s} = 100$~GeV, compared with $\sqrt{s} = 140$~GeV at the design configuration (18\,GeV\,$\times$\,275\,GeV). The accessible range in $Q^2$ at fixed $x$
is correspondingly reduced, shortening the DGLAP evolution lever arm that underpins precision PDF and $\alpha_S$ extractions, and the lowest reachable $x$ at perturbative $Q^2$ is roughly a factor of four higher than at full energy. The very low-$x$ frontier in both $e+p$ and $e+\mathrm{A}$ collisions therefore remains reachable only at the full-energy EIC.

\paragraph{Luminosity-limited precision}
Per-species integrated luminosities of 1.0--2.5~fb$^{-1}$ during the
early years, while already exceeding the HERA lifetime total, fall short of the ${\gtrsim}10$~fb$^{-1}$ per year expected for full EIC operations. This restricts the fine multi-dimensional binning in $(x, Q^2, z, P_{h\perp}, t)$ on which model-independent extractions of GPDs, TMD PDFs, and FFs depend. Quantitative tomographic imaging of the nucleon through GPDs -- a flagship EIC deliverable -- can be initiated, but not completed, with early science statistics. Spin asymmetries at $x \lesssim 10^{-3}$, where the gluon helicity contribution dominates the proton spin sum rule, remain statistics-limited, as do rare exclusive channels such as near-threshold quarkonium production, which carries unique sensitivity to the gluonic gravitational form factors and hence to the origin of the nucleon mass.

\paragraph{Dense gluonic matter and the saturation regime}
The saturation scale $Q_s$ grows with the mass number of the nucleus, $Q_s^2 \propto A^{1/3}$, which is precisely why heavy ions such as gold are the probe of choice for dense gluonic matter: the large gluon density in a gold nucleus makes it an efficient ``amplifier'' of non-linear QCD effects.
However, even with this nuclear enhancement, the kinematic reach of the early configuration, a 9~GeV electron beam on a 100~GeV/nucleon gold-ion beam, does not extend deeply enough into the low-$x$, perturbative-$Q^2$ corner of phase space where the saturated regime, $Q^2 \lesssim Q_s^2(x, A)$, is expected to open up. Pushing firmly into the saturation regime, even with the heaviest beams, requires the 18~GeV electron beam of the upgraded EIC, which extends
the per-nucleon center-of-mass energy from $\sqrt{s} \approx 60$~GeV to $\sqrt{s} \approx 90$~GeV, and
correspondingly lowers the accessible $x$ at fixed $Q^2$. The di-hadron decorrelation program (Sec.~\ref{secmain:sidis}) and the diffractive gluon-imaging program (Sec.~\ref{secmain:exclusive}) will both
deliver benchmark measurements in the early years, characterizing the approach toward the saturated regime, but a definitive discovery and characterization of gluon saturation requires the full EIC energy reach, high-luminosity $e+\mathrm{A}$ running, and the complete suite of nuclear species.

\paragraph{Hard probes and charged-current DIS}
The jet transverse-momentum reach at early energies extends only to $p_T \sim 12$--15~GeV/$c$ in the rapidity regions most sensitive to nuclear modifications. 
The full-energy configuration substantially extends this reach, enabling jet substructure studies, improved separation of initial- and final-state nuclear effects, and access to the kinematic domain where higher-order QCD calculations are most robust.
Charged-current DIS, which provides unique flavor separation in the valence-quark sector, 
is not accessible with early luminosities and energies and is deferred entirely to full EIC operations. 
\section{Summary}
\label{sec:summary}

The early operational years of the EIC will already deliver a broad and coherent set of measurements that take the first decisive steps toward answering the three central questions identified in the 2018 NAS report~~\cite{NAP25171}. The cross-cutting connection between representative ePIC early science measurements and the three NAS science pillars has been summarized in Table~\ref{tab:nas-pillars-exe} in the Executive Summary.

The studies presented in Secs.~\ref{secmain:inclusive}--\ref{secmain:hf} demonstrate that ePIC can make world-leading measurements bearing on these questions with realistic early running beam configurations, luminosities, polarization scenarios, and detector capabilities, as summarized in Table~\ref{tab:early-science-matrix}. The measurement program is staged by capability: inclusive DIS provides the earliest high-statistics measurements, SIDIS and jet/heavy-flavor channels expand the program as hadron reconstruction and particle identification mature, and exclusive, diffractive, and tagged measurements fully exploit the progressive commissioning of far-forward instrumentation.

The early ePIC program will also provide reference measurements with impact on interpreting heavy-ion measurements at RHIC and the LHC. 
By constraining nuclear PDFs, gluon spatial distributions, parton propagation, heavy-flavor production, and hadronization in cold QCD matter, early electron–nucleus data will provide clean benchmarks for interpreting heavy-ion measurements.

The discussion below follows the same logic, organized by NAS pillar, and highlights how the measurements presented in this report collectively address each question while also clarifying which parts of the full NAS science program require the ultimate EIC luminosity and energy.

\paragraph*{Origin of Nucleon Mass: Gluon Dynamics and Hadron Structure}
The first NAS pillar is approached by measurements that determine how quarks and, especially, gluons generate the internal structure and dynamics of hadrons. Inclusive DIS, introduced through the neutral current topology in Fig.~\ref{fig:InclusiveDIS}, provides the most direct early route to high-precision reduced cross sections. In Sec.~\ref{secmain:inclusive}, the projected nuclear PDF impact shown in Fig.~\ref{fig:inclusive_R_Au} demonstrates that early $e+\mathrm{A}$ data will already constrain nuclear modifications of sea quarks, valence quarks, and gluons. The proton structure function program, illustrated by the early science kinematic coverage for $F_L$ in Fig.~\ref{inclusive-phase-space-FL}, the proton PDF impact in Fig.~\ref{fig:inclusive_pdf}, and the simultaneous $\alpha_S$ extraction in Fig.~\ref{fig:inclusive-alphaS}, constrains the one-dimensional momentum-space description of quarks and gluons in the proton. 
The spectator-tagging projection in Fig.~\ref{figmain:f2n} further extends this structure-function program to the neutron, showing how early deuteron data can improve the extraction of $F_2^n$ and strengthen flavor-separated constraints on nucleon PDFs.

The same mass pillar is extended beyond collinear PDFs by the multi-dimensional measurements discussed in Secs.~\ref{secmain:sidis} and~\ref{secmain:exclusive}. SIDIS, whose basic topology is shown in Fig.~\ref{fig:SIDIS}, provides access to transverse-momentum-dependent structure and to the fragmentation process. The projected impacts on unpolarized TMD PDFs in Fig.~\ref{fig:MAP24} and on 
nuclear fragmentation functions in Fig.~\ref{fig:SIDIS_FF} connect, respectively, the 3D quark distributions and the emergence of hadrons from quarks and gluons to measurements that can begin in the early running period. Exclusive and diffractive reactions provide the complementary spatial picture. The generic hard-exclusive topology in Fig.~\ref{fig:excldiag} and the projected CFF/GPD impact in Fig.~\ref{fig:exclusive} illustrate how early DVCS and exclusive channels initiate the tomographic imaging program. 
Together with the forward-tagged pion form-factor projection in Fig.~\ref{fig:pion_FF}, and with pion/kaon structure-function measurements enabled by tagged reactions, these observables connect the early ePIC program to the QCD mechanisms underlying hadron mass and confinement.

Heavy flavor and jets add a third handle on the same question. As shown schematically in Fig.~\ref{fig:prod-jets-hf}, charm production proceeds through photon--gluon fusion at leading order, providing direct sensitivity to gluon distributions in the proton and in nuclei. The projected $\Lambda_c/\mathrm{D}^0$ precision in Fig.~\ref{fig:hf} further extends the mass pillar program from gluon structure to the dynamics of charm hadronization, testing whether fragmentation and baryon formation in the clean DIS environment differ from those observed in hadronic collisions.

Although the complete nucleon-mass and tomography program will require the full EIC luminosity and extended beam-energy lever arm,  the early measurements already provide complementary constraints on quark and gluon momentum distributions, multidimensional imaging observables, and heavy-flavor and jet channels sensitive to gluon dynamics and hadron formation.

\paragraph*{Origin of Nucleon Spin: Helicity, Orbital Motion, and Spin--Orbit Correlations}
The early running period takes the first concrete steps toward the second NAS pillar through the polarized-beam measurements enabled during this time. Inclusive double-spin asymmetries in Sec.~\ref{secmain:inclusive} provide the foundation.
The projected impact on $g_1^{p,n}$ in Fig.~\ref{fig:inclusive_g1-main} demonstrates that early polarized running will substantially expand the kinematic reach of spin-structure measurements and improve the precision of spin structure global QCD analyses, particularly toward low $x$.

SIDIS provides the essential next layer by correlating the nucleon spin with the flavor, momentum, and transverse motion of identified final-state hadrons. The Sivers and Collins projections in Fig.~\ref{fig:SIDIS_SSA_maindoc} show how early transversely polarized proton running can access spin--orbit correlations and quark transversity. The semi-inclusive helicity projections in Fig.~\ref{fig:ALL_SIDIS} complement inclusive measurements by improving flavor separation, especially for sea-quark helicities. Exclusive and tagged measurements add a further connection to orbital angular momentum through GPDs and to neutron spin through spectator-tagged light nuclei. In particular, the spectator-tagging program discussed in Sec.~\ref{secmain:exclusive} improves the extraction of neutron spin observables from light nuclear targets by constraining nuclear effects such as Fermi motion, binding, and final-state interactions.

Jets and heavy flavor, discussed in Sec.~\ref{secmain:hf}, provide additional spin sensitivity through gluon-dominated hard subprocesses and hadron-in-jet observables. The jet and charm production mechanisms in Fig.~\ref{fig:prod-jets-hf} indicate the channels through which early data can begin to constrain $\Delta g(x)$ and spin-dependent fragmentation observables. 

Again, the ultimate precision spin program will still require the full EIC luminosity and complete polarization capabilities; however the early measurements already provide a coherent set of complementary observables for separating quark helicity, gluon helicity, transversity, and orbital-motion effects.

\paragraph*{Emergent Properties of Dense Systems of Gluons}
Measurements with nuclei offer the most direct path toward the third NAS pillar. Inclusive nuclear DIS in Sec.~\ref{secmain:inclusive} provides the first clean collider access to nuclear structure functions and nPDFs over a broad kinematic range. The nuclear PDF projection in Fig.~\ref{fig:inclusive_R_Au} shows that early $e+\mathrm{Au}$ data alone can already reach competitive sensitivity to the poorly known nuclear gluon distribution, while the $F_L$ kinematic coverage in Fig.~\ref{inclusive-phase-space-FL} illustrates the importance of the EIC lever arm for gluon-sensitive structure function measurements. 

The most direct spatial probe of dense gluonic matter comes from exclusive and diffractive scattering in nuclei. In Sec.~\ref{secmain:exclusive}, coherent diffractive vector-meson production is shown to provide access to the transverse spatial distribution of gluons through the measured $t$ spectrum. Figure~\ref{figmain:diff_phi} demonstrates that, using realistic ePIC reconstruction and the projected-$t$ method, the diffractive structure in coherent $\phi\to K^+K^-$ production in $e+\mathrm{Au}$ can be recovered and transformed into a transverse gluon profile, provided adequate control of the background contribution.

SIDIS and jets and heavy flavor provide complementary sensitivity to the same nuclear dynamics. The SIDIS nuclear fragmentation projection in Fig.~\ref{fig:SIDIS_FF} tests how hadron formation is modified in cold nuclear matter, while the di-hadron and low-$x$ program described in Sec.~\ref{secmain:sidis} targets saturation-sensitive azimuthal decorrelations. In Sec.~\ref{secmain:hf}, inclusive charged-jet nuclear modification measurements, shown in Fig.~\ref{fig:jets}, and $\mathrm{D}^0$-in-jet nuclear modification studies, shown in Fig.~\ref{fig:hf}, probe how parton showers, heavy-quark propagation, and hadronization are modified in nuclei. These observables connect the collinear nuclear structure measured in inclusive DIS to the real-time propagation and fragmentation of colored partons in cold nuclear matter.

A definitive study of dense gluonic matter and the possible onset of saturation will necessitate the full EIC energy, luminosity, and nuclear-species reach. Despite this, the early measurements already provide a connected set of observables: nuclear structure functions and nPDFs, coherent vector-meson production, SIDIS nuclear effects, and jet and heavy-flavor modification in cold nuclear matter.

\paragraph*{Scientific Opportunities Requiring the Full EIC}
Achieving the EIC science goals in their full scope, including the central questions highlighted by the NAS report, requires capabilities beyond those available during the early science phase, as
detailed in Sec.~\ref{sec:limitations}. The early science configuration does not reach the very low-$x$ domain where the saturation scale can be mapped, does not provide the luminosity needed for quantitative GPD tomography or for closing the spin sum rule at
small $x$, defers charged-current DIS and the 18~GeV program entirely, and places key hard-probe and near-threshold observables at the edge of practicality. None of these limitations diminishes the early program; rather, they define its proper role. The early years deliver a first set of world-leading measurements within each pillar while validating the experimental and analysis infrastructure of the experiment. Nevertheless, the answers to the NAS questions in their full depth, from the discovery and characterization of gluon saturation to the complete spin and mass decomposition of the nucleon, are deliverables of the full EIC. In its full configuration, the EIC will be transformational for our understanding of hadronic structure and QCD. Preserving the path to the complete machine capabilities is therefore not an optional enhancement of the program described here: it is the condition for its scientific completion.

\paragraph*{Conclusions}
From the first collisions, ePIC will deliver unique and competitive measurements while establishing the experimental and analytical foundations needed to fulfill the whole scientific promise of the Electron–Ion Collider. The complete realization of the science program will require the full EIC capabilities, including higher luminosity and extended energy coverage. However, the early ePIC program is not simply the beginning of the EIC operations --- it is the beginning of a new era of precision QCD.
\clearpage
\phantomsection
\section*{Acknowledgments}
\label{sec:acknowledgments}

We are grateful to Wim Cosyn -- Florida International University, Nicole d'Hose -- CEA-Saclay, Marco Mirazita -- Laboratori Nazionali di Frascati, and Rithya Kunnawalkam Elayavalli -- Vanderbilt University, for having served as external reviewers of this manuscript.


The ePIC Collaboration gratefully acknowledges the support of the following funding agencies and organizations:
the High Education and Science Committee of the Republic of Armenia (Armenia);
the Research Foundation--Flanders (FWO) (Belgium);
the Natural Sciences and Engineering Research Council of Canada (NSERC) and the Canada Foundation for Innovation (CFI) (Canada);
the National Natural Science Foundation of China (NSFC) (China);
the Ministry of Education, Youth and Sports (MEYS) (Czech Republic);
the Agence Nationale de la Recherche (ANR) and the Centre National de la Recherche Scientifique/Institut National de Physique Nucléaire et de Physique des Particules (CNRS/IN2P3) (France);
the German Federal Ministry of Education and Research (BMBF), the Helmholtz Association, and the Deutsche Forschungsgemeinschaft (DFG) (Germany);
the Hungarian National Research, Development and Innovation Office (NKFIH) (Hungary);
the Department of Atomic Energy (DAE), the Department of Science and Technology (DST), and the Ministry of Education (India);
the Israel Science Foundation (ISF) (Israel);
the Istituto Nazionale di Fisica Nucleare (INFN) (Italy);
the Ministry of Education, Culture, Sports, Science and Technology (MEXT) and the Japan Society for the Promotion of Science (JSPS) (Japan);
the Ministry of Higher Education and Scientific Research (Jordan);
the National Research Foundation of Korea (NRF) (Republic of Korea);
the Centre National pour la Recherche Scientifique et Technique (CNRST) (Morocco);
the Research Council of Norway (Norway);
the National Science Centre (NCN) (Poland);
the Slovenian Research Agency (Slovenia);
the Department of Science and Innovation (DSI) and the National Research Foundation (NRF) (South Africa);
the National Science and Technology Council (NSTC) and the Ministry of Education (MOE) (Taiwan);
the Ministry of Education and Science of Ukraine (Ukraine);
the Science and Technology Facilities Council (STFC), UK Research and Innovation (UKRI) (United Kingdom);
the U.S. Department of Energy, Office of Science, Office of Nuclear Physics, and the National Science Foundation (United States of America);
and the U.S. Department of Energy, National Nuclear Security Administration, Office of Defense Nuclear Nonproliferation Research \& Development by the Nuclear Science and Security Consortium.

Individuals have received support from
the French ANR under the grant ANR-24-CE31-5571 (``ROAD4EIC'') (France);
the R\&D program at UPES, Dehradun (India);
Future Artificial Intelligence Research (FAIR) funded by the NextGenerationEU program (Italy);
the Center for Frontiers in Nuclear Science (CFNS) funded through the Simons Foundation, Stony Brook University (United States of America);
the National Science Centre (Poland) under the grant 2024/53/B/ST2/03257; and the grant 2025/58/E/ST2/00045).


\addcontentsline{toc}{section}{Acknowledgments}
\clearpage
\appendix
\phantomsection
\addcontentsline{toc}{section}{Appendix: Charge Letter}
\section*{Appendix: Charge Letter}
\label{app:charge-letter}
 
This Early Science Report was prepared by the ePIC Collaboration in response to the
charge letter reproduced below, sent on June 13, 2025 by David Dean (Deputy Director,
JLab) and Abhay Deshpande (Associate Laboratory Director for Nuclear and
Particle Physics, BNL).
 

\centering
\includegraphics[width=0.93\linewidth]{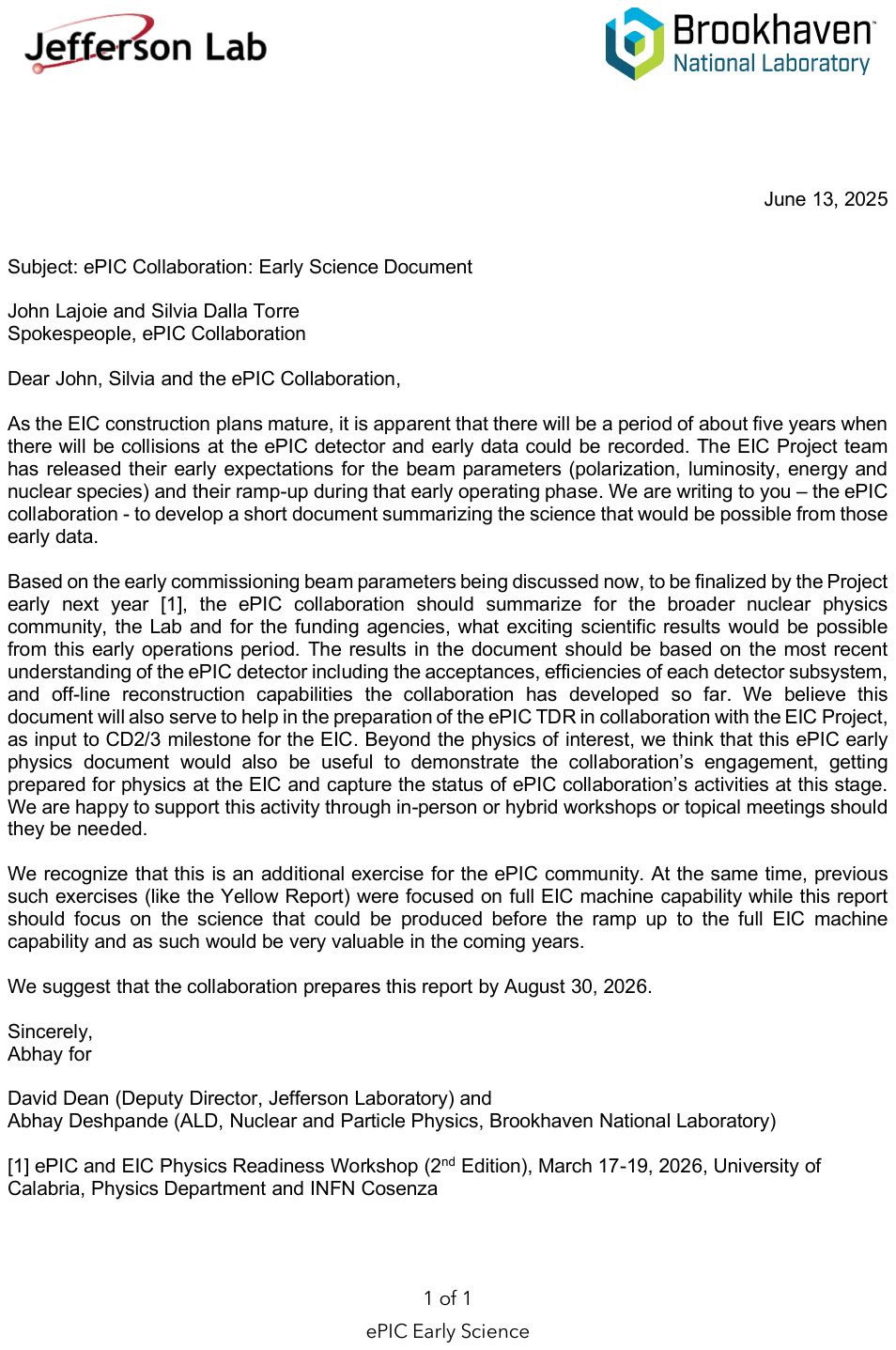}

\clearpage
\nolinenumbers
\phantomsection
\addcontentsline{toc}{section}{References}

\setlength{\bibsep}{0pt plus 0.3ex}
\renewcommand{\bibfont}{\small}

\bibliographystyle{h-physrev3}
\bibliography{reference}
\end{document}